\documentclass[12pt,nofootinbib]{article}

\usepackage[dvips]{graphicx}
\usepackage{amssymb}

\usepackage{amsmath}
\usepackage{epsfig}

\numberwithin{equation}{section}

\def\beq{\begin{equation}}
\def\eeq{\end{equation}}
\def\be{\begin{equation}}
\def\ee{\end{equation}}
\def\bea{\begin{eqnarray}}
\def\eea{\end{eqnarray}}
\def\d{{\rm d}}
\def\F{\mathcal{F}}
\def\hr{\hat{r}}
\def\FF{\mathbb{F}}

\DeclareRobustCommand{\SkipTocEntry}[4]{}

\begin{document}

\begin{titlepage}

\setcounter{page}{1} \baselineskip=15.5pt \thispagestyle{empty}

\begin{flushright}
PUPT-2237 \\
ITEP-TH/22-07
\end{flushright}
\vfil

\bigskip\
\begin{center}
{\LARGE Towards an Explicit Model of D-brane Inflation}
\vskip 15pt
\end{center}

\vspace{0.5cm}
\begin{center}
{\large Daniel Baumann,$^{1}$ Anatoly Dymarsky,$^{1}$ }
\vskip 3pt
{\large Igor R. Klebanov,$^{1,2}$ and Liam McAllister$^{1}$}
\end{center}

\begin{center}
\textit{$^{1}$Department of Physics,
Princeton University, Princeton, NJ 08544}\\
\textit{$^{2}$Princeton Center for Theoretical Physics,
Princeton University, Princeton, NJ 08544}
\end{center} \vfil

\noindent

We present a detailed analysis of an explicit model of warped D-brane inflation, incorporating the effects of moduli stabilization.  We consider the potential for D3-brane motion in a warped conifold background that includes fluxes and holomorphically-embedded D7-branes involved in moduli stabilization.  Although the D7-branes significantly modify the inflaton potential, they do not correct the quadratic term in the potential, and hence do not cause a uniform change in the slow-roll parameter $\eta$. Nevertheless, we present a simple example based on the Kuperstein embedding of D7-branes, $z_1=$ constant, in which the potential can be fine-tuned to be sufficiently flat for inflation.  To derive this result, it is essential to incorporate the fact that the compactification volume changes slightly as the D3-brane moves. We stress that the compactification geometry dictates certain relationships among the parameters in the inflaton Lagrangian, and these microscopic constraints impose severe restrictions on the space of possible models.  We note that the shape of the final inflaton potential differs from projections given in earlier studies: in configurations where inflation occurs, it does so near an inflection point.  Finally, we comment on the difficulty of making precise cosmological predictions in this scenario.  This is the companion paper to \cite{ShortPaper}.

\vfil
\begin{flushleft}
\today
\end{flushleft}

\end{titlepage}

\newpage
\tableofcontents

\newpage
\section{Introduction}
\label{sec:intro}

Inflation \cite{Inflation} provides an explanation for certain observed properties of our universe, from large-scale
isotropy and homogeneity to the small-scale anisotropies in the cosmic microwave background (CMB).  Recent observational advances
\cite{Observations} have begun to constrain the space of inflationary models, and in the next decade it should be possible to exclude a large fraction of current models. This is a striking opportunity to extract clues about physics at extremely high
energy scales. At the same time, there is a pressing need to develop a more fundamental understanding of the physics of inflation.
String theory, as a quantum theory of Planck-scale physics, is a promising candidate for the fundamental theory underlying the inflationary paradigm.

In this paper we will be concerned with making progress towards an {\it{explicit}} model of inflation in string theory, by which we mean a model in which the fields and parameters in the low-energy Lagrangian are derived from the data of a string compactification.  In such a scenario, questions about the structure of the inflaton potential can be resolved by concrete string theory computations.  This should be contrasted with string-inspired models constructed directly in effective field theory, for which naturalness is the final arbiter of the form of the potential.  We will not quite reach the ambitious goal of an entirely explicit model of inflation derived from string theory, and indeed one main point of this paper is that truly explicit models of string inflation can be rather intricate, involving multiple nontrivial microscopic constraints that are surprising from the low-energy perspective.

We will work in the setting of D-brane inflation \cite{Dvali,otherbraneinflation,KKLMMT}, a promising framework that has attracted considerable interest, but in which concrete, working models remain scarce.\footnote{For other approaches to string inflation, see {\it{e.g.}} \cite{otherinflation}, as well as \cite{reviews} and references therein.}  In this approach to string inflation the inflaton field is identified with the location of a mobile D-brane, usually a D3-brane, in the compactification manifold.  The proposal \cite{Dvali} is that a weak force on this brane, {\it{e.g.}} from a distant antibrane, could give rise to a relatively flat inflaton potential.  In practice, this is difficult to achieve: the Coulomb potential of a brane-antibrane pair in an unwarped compactification is generically too steep for inflation.  In \cite{KKLMMT}, it was observed that a brane-antibrane pair separated along a warped direction enjoys a much flatter potential.  At the same time, it was recognized that moduli stabilization, which is essential for a successful cosmological model in string theory, gives rise to corrections to the inflaton potential.  Some of these corrections arise from complicated properties of the compactification and have been computed only recently \cite{BDKMMM}.\footnote{For important earlier work, see \cite{Ganor,BHK,GM}.}

The attitude taken in most of the literature on the subject (see {\it{e.g.}}
\cite{KKLMMT,CloserTowards,reviews}) is that because of
the vast number and complexity of string vacua, in some nonzero fraction of them it should be
the case that the corrections to the inflaton potential cancel to high precision,
leaving a suitable inflationary model.
However, there is no guarantee that this hope can be realized:
for example, the correction terms may invariably have the same sign
so that no cancellation can ever occur.
Moreover, the precise nature of the cancellation will affect the detailed predictions of the model.
In this paper we will systematically address the question of whether or not this hope of fine-tuned
cancellation can in fact be realized. The new ingredient that makes this investigation possible is
our recent result, obtained in collaboration with J. Maldacena and A. Murugan \cite{BDKMMM},
for the one-loop correction to the volume-stabilizing nonperturbative superpotential.
This effect is due to the interaction \cite{BHK,GM} between the inflationary D3-brane
and the moduli-stabilizing wrapped branes, {\it{i.e.}} D7-branes
wrapping a four-cycle within the Calabi-Yau threefold.\footnote{
Alternatively, one could consider Euclidean D3-branes wrapping this four-cycle. Their effect on the nonperturbative
superpotential is very similar to that of the D7-branes.}
The location of these wrapped branes therefore becomes a crucial parameter
in the D3-brane potential.

With this theoretical input we ask whether the known correction terms to the inflaton potential
can indeed cancel for specific values of the
microphysical input parameters.  To investigate this, we study D3-brane motion in
warped conifold backgrounds \cite{KS,KT} for a large class of wrapped brane embeddings.
By identifying radial trajectories that are stable in the angular directions,
we integrate out the angular degrees of freedom and arrive at an effective two-field potential
for the inflaton -- corresponding to the radial direction in the throat -- and the
compactification volume.  Because we work in a framework with explicitly stabilized moduli,
the compactification volume has a positive mass-squared.  However, this mass is not
so large that the volume remains fixed at a single value during inflation.
Instead, the minimum of the potential for the volume shifts as the D3-brane moves:
the compact space shrinks slightly as the D3-brane falls down the throat.
Properly incorporating this phenomenon leads to a nontrivial change in the effective single-field
inflaton potential.  Thus, we find that an approximation that keeps the volume fixed at its KKLT \cite{KKLT} minimum during inflation is
not sufficiently accurate for a D-brane inflation model.
Our improved approximation is that the volume changes adiabatically,
remaining in an inflaton-dependent minimum, as the D3-brane moves.

Equipped with the effective single-field potential, we ask whether the trajectories
that are stable in the angular directions can enjoy flat potentials. For
the large class of holomorphic wrapped brane embeddings described in \cite{ACR}, we find trajectories
that are too steep to permit inflation, even with an arbitrary amount of fine-tuning of the
compactification parameters: the functional form of the leading corrections to the potential makes
fine-tuning impossible.  (Our conclusions are consistent with those reached in
\cite{Burgess} for the special case of the Ouyang embedding \cite{Ouyang}.)
This illustrates a key virtue of the explicit, top-down approach:
by direct computation we can refute the very reasonable expectation that fine-tuning is generically possible.

Undeterred by this no-go result for a wide class of D3-brane trajectories,
we devote a large portion of this paper to showing that in a particularly simple and symmetric
embedding due to Kuperstein \cite{Kuperstein}, the stable trajectory is {\it{not}} necessarily steep.
We then establish that for fine-tuned values of the microphysical parameters, a
D3-brane following this stable trajectory leads to sustained slow-roll inflation near an
inflection point of the potential.  Finally, we derive nontrivial constraints, due to the consistency of
the embedding of the warped throat in a flux compactification,
that relate the microscopic parameters of the inflaton Lagrangian.
These constraints sharply restrict the parameter space of the model,
and in fact exclude most, but not all, of the parameter space that permits sustained inflation.

Our result provides evidence for the existence of successful warped D-brane inflation models based on concrete microscopic data of a flux compactification.  However, we emphasize that inflation is non-generic in this class of D-brane models.  In fact, because of the geometric constraints, it is surprisingly difficult, though not impossible, to achieve inflation.

The outline of this paper is as follows: in \S\ref{sec:potential}
we briefly review D-brane inflation in warped backgrounds.
We then provide results for the complete D3-brane potential in a warped throat
region of the compactification. This includes an important
correction to the volume-stabilizing nonperturbative
superpotential first computed in \cite{BDKMMM}. In
\S\ref{sec:Kuperstein} we present a detailed study of a simple
example, the case of the Kuperstein embedding \cite{Kuperstein} of
the wrapped branes. By integrating out the complex K\"ahler
modulus and the angular positions of the D3-brane, we derive
effective single-field potentials for different brane
trajectories. In \S\ref{sec:Analysis} we then prove the existence of a stable inflationary
trajectory, but also discuss important constraints on microscopic parameters that make inflation challenging to achieve. In \S\ref{sec:ACR} we comment on generalizations to other embeddings.
We then take the opportunity, in \S\ref{sec:discussion}, to make some general remarks about the problem of relating string compactification data to the low energy Lagrangian.
We conclude in \S\ref{sec:Conclusion}.

In order to make the body of the paper more readable, we have relegated a number of more technical results to a series of appendices.  Although most of these results are new, they could be omitted on a first reading.
Appendix \ref{sec:conifold} gives details of the conifold geometry and of the
supergravity F-term potential. In Appendix \ref{sec:reduc} we dimensionally reduce the ten-dimensional string action and derive microscopic constraints on the inflaton field range and the warped four-cycle volume.  In particular, we explain how to generalize the field range bound of \cite{BM} to a compactification with a nontrivial breathing mode.  We also derive a new constraint that relates the field range and the volume of the wrapped four-cycle. Appendix \ref{sec:stability} provides more technical aspects of the proof that the inflationary trajectory is stable against angular fluctuations. In Appendix \ref{sec:sigma} we derive the dependence of the compactification volume on the D3-brane position.  This is an important improvement on the typical approach of keeping the volume fixed as the D3-brane moves.  Appendix \ref{sec:magnetic} addresses the concern that induced magnetic fields on the D7-branes could affect the D3-brane potential.    Finally, Appendix \ref{sec:symbols} collects the definitions of all variables used in this work.

A condensed presentation of some of the key results of this paper has appeared in \cite{ShortPaper}.


\section{D3-brane Potential in Warped Backgrounds}
\label{sec:potential}

In warped brane--antibrane inflation \cite{KKLMMT}, one considers
the motion of a D3-brane towards an anti-D3-brane that sits at the
bottom of a warped throat region of a stabilized flux
compactification.
The inflaton field is the separation between the brane and the antibrane.

\subsection{The compactification}

Our setting is a flux compactification \cite{GKP} (see \cite{FluxReview} for a review) of type IIB
string theory on an orientifold of a Calabi-Yau threefold (or an F-theory compactification on a Calabi-Yau fourfold).  We suppose that
the fluxes are chosen so that the internal space has a warped throat region.  As a simple, concrete example of this local geometry, we consider
the warped deformed conifold \cite{KS}.  The deformed conifold is a subspace of complex dimension three in
$\mathbb{C}^4$ defined by the constraint equation \beq
\label{equ:Cconstraint} \sum_{i=1}^4 z_i^2 = \varepsilon^2\, ,
\eeq where $\{z_i, i = 1,2,3,4\}$ are complex coordinates in
$\mathbb{C}^4$.  The deformation parameter $\varepsilon$ can be made real by an appropriate phase rotation.
The region
relevant to our modeling of D-brane inflation lies far from the
bottom of the throat, where the right hand side of (\ref{equ:Cconstraint}) can be ignored and the metric
of the deformed conifold is well-approximated by that of the
singular conifold,
\beq \label{conicform}
\d s^2_6 = \d \hr^2 + \hr^2 \d s_{T^{1,1}}^2\, ,
\eeq
where $\d s_{T^{1,1}}^2$ is the metric of the Einstein manifold $T^{1,1}$, the base of the cone (see Appendix A).
This Calabi-Yau metric is obtained from the K\"ahler potential
\cite{Candelas}
\beq \label{Kpot}
k = \frac{3}{2} \left(
\sum_{i=1}^4 |z_i|^2 \right)^{2/3} = \frac{3}{2} r^2 = \hr^2 \, . \eeq

The warping is achieved by turning on $M$ units of $F_3$ flux through the
A-cycle of the deformed conifold (the three-sphere at the bottom) and
$-K$ units of $H_3$ flux through the dual B-cycle. The resulting warped deformed
conifold background is given in
\cite{KS,HKO}. If $\hat r_{\rm UV}$ is the maximum radial coordinate where the
throat is glued into a compact manifold, then for
$\varepsilon^{2/3} \ll \hat r \ll \hat r_{\rm UV}$
the background
is well-approximated by the warped conifold \cite{KT}
\beq \label{warpedansatz}
\d s_{10}^2 = h^{-1/2}(\hr) \d s_4^2+ h^{1/2}(\hr) \d s_6^2\, ,
\eeq
with the warp factor \cite{KT,HKO}
\beq \label{equ:KTh}
h(\hr) = \frac{L^4}{\hr^4} \ln \frac{\hr}{\varepsilon^{2/3}}\, ,\qquad
L^4 = \frac{81}{8} (g_s M{\alpha^{\prime}})^2\, .
\eeq
We also have \cite{GKP,HKO}
\beq
 \ln \frac{\hr_{\rm UV}}
{\varepsilon^{2/3}} \approx \frac{2 \pi K}{3 g_s M}\, .
\eeq
The scale of supersymmetry breaking associated with an anti-D3-brane at the
bottom of the throat is $D=2T_3 h_0^{-1}$, where $h_0$ is the warp factor there.
The approximation (\ref{equ:KTh}) is not accurate enough to determine the
warp factor at the bottom of the throat; its value is \cite{KS,HKO}
\beq \label{warptip}
h_0 = a_0 (g_s M \alpha')^2 2^{2/3} \varepsilon^{-8/3} \ ,\qquad a_0 \approx 0.71805\, ,
\eeq
which is approximately $h_0 \approx e^{8\pi K/3g_s M}$ \cite{GKP}.

Following \cite{KKLT}, we require that all the closed string
moduli are stabilized,\footnote{This condition is necessary for a
realistic model, and amounts to a nontrivial selection criterion
on the space of compactifications.} by a combination of fluxes and
nonperturbative effects.  Each nonperturbative effect may arise
either from Euclidean D3-branes wrapping a four-cycle, or from
strong gauge dynamics, such as gaugino condensation, on a stack of
$n >1$ D7-branes wrapping a four-cycle. Finally, as in
\cite{BDKMMM}, we require that at least one of the four-cycles
bearing nonperturbative effects descends a finite distance into
the warped throat.  For simplicity of presentation we will refer
to the nonperturbative effects on this cycle as originating on
D7-branes, but all our results apply equally well to the case in which
Euclidean D3-branes are responsible for this effect.

\begin{figure}[htbp]
    \centering
        \includegraphics[width=0.70\textwidth]{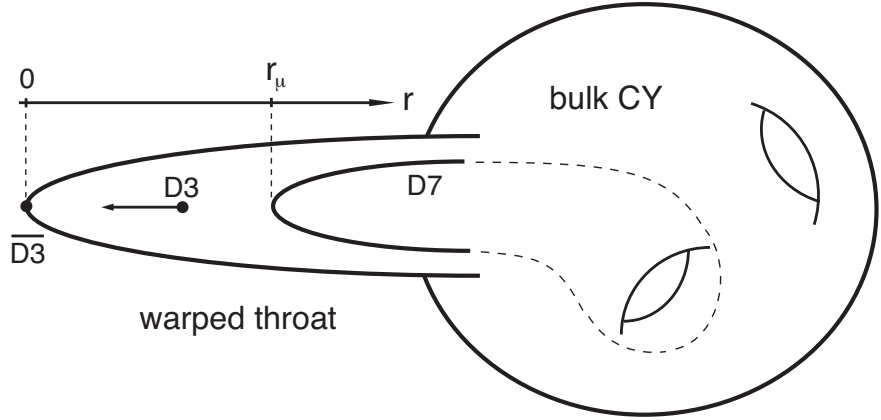}
    \caption{Cartoon of an embedded stack of D7-branes wrapping a four-cycle $\Sigma_4$, and a mobile D3-brane, in a warped throat region of a compact Calabi-Yau.  The D3-brane feels a force from the D7-branes and from
an anti-D3-brane at the tip of the throat.}
    \label{fig:throat}
\end{figure}

An embedding is specified by the number $n= N_{D7}>1$ of D7-branes and the minimal radial coordinate
$r_{\mu}$ reached by the D7-branes.  The stabilization of $r_{\mu}$ is a potentially confusing issue,
so we pause to explain it.  In the construction \cite{Kuperstein} of supersymmetric wrapped
D7-branes in the noncompact KS throat, $r_{\mu}$ is a free parameter.  One might therefore think
that the wrapped D7-branes are not stabilized, and that there is a massless field corresponding
to changes in $r_{\mu}$.  However, in the F-theory picture, $r_{\mu}$ is determined by the
complex structure of the fourfold.  For generic choices of four-form fluxes, this complex
structure is entirely fixed \cite{Munich} (see also \cite{Gorlich}), just as
the threefold complex structure is fixed in type IIB compactifications with generic
three-form fluxes \cite{GKP}.  Moreover, the scale of the associated mass terms
(see {\it{e.g.}} \cite{Svrcek}), $ m_{\rm flux} \sim \frac{\alpha^{\prime}}{\sqrt{V_6}} $,
with $V_6$ the volume of the compact space, is considerably higher than the (warped) energy scale associated with the brane--antibrane pair under consideration.  Hence, for our purposes the D7-brane moduli are massive enough to be ignored.  Next, the stabilized value of $r_{\mu}$ is determined by the fluxes in the bulk of the fourfold.  In a generic compactification the number of choices of such fluxes is vast, so we expect that for a given compactification and for any desired value $r_{\mu}^{\star}$, there exist choices of flux that fix the D7-brane to a location $r_{\mu} \approx r_{\mu}^{\star}$.

\subsection{D3-brane potential from moduli stabilization}

The effect of moduli stabilization on the D3-brane is captured by
the F-term potential of ${\cal N}=1$ supergravity, \beq
\label{equ:VF} V_F = e^{\kappa^2 {\cal K}} \Bigl[ D_\Sigma W
{\cal K}^{\Sigma \overline{\Omega}} \overline{D_\Omega W}- 3
\kappa^2 W \overline{W}\Bigr]\, , \quad \quad \kappa^2 =
M_P^{-2} \equiv 8 \pi G \, ,\eeq where $\{Z^\Sigma\} \equiv
\{\rho, z_\alpha; \alpha=1,2,3 \}$ and $D_\Sigma W = \partial_\Sigma W + \kappa^2 (\partial_\Sigma {\cal K}) W$. The combined K\"ahler
potential for the volume modulus, $\rho$, and the three open string
moduli (D3-brane positions), $z_\alpha$, is of the form postulated
by DeWolfe and Giddings  \cite{DeWG}\footnote{In \cite{wsb} it was suggested that this result may receive corrections in strongly-warped scenarios.  However, the proposed corrections do not affect the metric on the K\"ahler moduli space, and thus are irrelevant for most of the considerations presented here.  However, a truly thorough search for possible effects of such corrections on our analysis must await a more complete understanding of the structure of corrections to the K\"ahler potential.}
\beq \label{equ:KKK} \kappa^2
{\cal K}(\rho,\overline{\rho},z_\alpha,\overline{z_\alpha})  =
- 3 \log [\rho + \bar \rho - \gamma k(z_\alpha, \overline{z_\alpha}) ]
\equiv - 3 \log U \, ,
\eeq
where in general $k(z_\alpha, \overline{z_\alpha})$ denotes the K\"ahler
potential of the Calabi-Yau manifold.
The normalization constant $\gamma$ in (\ref{equ:KKK}) is derived in Appendix \ref{sec:reduc} and may be expressed as
\beq
\label{equ:gamma}
\gamma \equiv \frac{\sigma_0}{3} \frac{T_3}{M_P^2}\, ,
\eeq
where
$2 \sigma_0 \equiv 2 \sigma_\star(0) = \rho_\star(0) + \bar \rho_\star(0)$
is the stabilized value of the K\"ahler modulus
when the D3-brane is near the tip of the throat.

The K\"ahler metric ${\cal K}_{\Omega \overline{\Sigma} } \equiv
{\cal K}_{, \Omega \overline{\Sigma} }$ assumes the form
\beq  {\cal
K}_{\Omega \overline{\Sigma} }  = \frac{3}{\kappa^2 U^2} \left( \begin{array}{c|c} 1 & -
\gamma k_{\overline{\beta}} \\ \hline
- \gamma k_\alpha & U \gamma k_{\alpha \overline{\beta}} + \gamma^2 k_\alpha k_{\overline{\beta}} \\
\end{array}\right)\, ,
\eeq
where
$k_{\alpha \overline{\beta}} \equiv \partial_\alpha
\partial_{\overline{\beta}} k$ is the Calabi-Yau metric, and $k_\alpha\equiv k_{,\alpha}$.
The problem of finding the inverse metric,
${\cal K}^{\Delta \overline{\Gamma}} {\cal K}_{\overline{\Gamma} \Omega} = \delta_{~\Omega}^{\Delta}$,
was solved in \cite{Burgess}:
\beq  \label{equ:inverse_alpha} {\cal K}^{\Delta
\overline{\Gamma}}  = \frac{\kappa^2 U}{3}  \left(
\begin{array}{c|c} U + \gamma k_{\gamma} k^{\gamma
\overline{\delta}} k_{\overline{\delta}} & k_\gamma k^{ \gamma \overline{\beta}}  \\ \hline
k^{ \alpha \overline{\delta} } k_{\overline{\delta}}
& \frac{1}{\gamma} k^{\alpha \overline{\beta}}\\
\end{array}\right)\, .
\eeq
After some calculation,
these results lead to the F-term potential
\bea V_F(\rho, z_\alpha) &=&
\frac{\kappa^2}{3 U^2} \Biggl[ \bigg (\rho + \overline{\rho}+
\gamma ( k_{\gamma} k^{\gamma
\overline{\delta}} k_{\overline{\delta}}-k) \bigg )
|W_{,\rho}|^2 - 3 (\overline{W} W_{, \rho} + c.c.) \nonumber \\
&& \hspace{1.5cm} + \underbrace{ ( k^{\alpha \overline{\delta}}k_{\overline{\delta}}
\overline{W_{,\rho}}  W_{,\alpha} + c.c.) +
\frac{1}{\gamma} k^{\alpha \overline{\beta}} W_{,\alpha}
\overline{W_{,\beta}} }_{\Delta V_F} \Biggr]\, .
\label{equ:Fterm}
\eea
The label $\Delta
V_F$ has isolated the terms in F-term potential (\ref{equ:Fterm})
that arise exclusively from the dependence of the nonperturbative
superpotential on the brane position \cite{BDKMMM}.  The remainder
of (\ref{equ:Fterm}) is the standard KKLT F-term potential \cite{KKLT}.

The
superpotential $W$ is the sum of the constant Gukov-Vafa-Witten
flux superpotential \cite{GVW}, $W_{\rm flux} = \int G \wedge
\Omega \equiv W_0$,
and a term from nonperturbative effects, $W_{\rm np} = A(z_\alpha) e^{-a \rho}$,
\beq
\label{equ:W}
 W(\rho,z_\alpha) = W_0 + A(z_\alpha) e^{- a \rho}\, , \quad \quad \quad
a \equiv \frac{2 \pi}{n}\, .
\eeq
By a choice of phases we can arrange that $W_0$ is real and negative.
The nonperturbative term
$W_{\rm np}$ arises either from strong gauge dynamics on a stack of $n>1$ D7-branes
or from Euclidean D3-branes (with $n =1$ ).  We assume that either sort of brane
supersymmetrically wraps a four-cycle in the warped throat that
is specified by a holomorphic embedding equation $f(z_\alpha)=0$.
The warped volume of the four-cycle governs the magnitude of the nonperturbative
effect, by affecting the gauge coupling on the D7-branes
(equivalently, the action of Euclidean D3-branes) wrapping this four-cycle.
The presence of a D3-brane gives rise to a perturbation to the warp factor, and
this leads to a correction to the warped four-cycle volume.
This correction depends on the D3-brane position and is responsible for the prefactor $A(z_\alpha)$ \cite{GM}.
In \cite{BDKMMM}, in collaboration with J. Maldacena and A. Murugan, we computed the
D3-brane backreaction on the warped four-cycle volume.  This gave the result\footnote{The D3-brane-independent factor $A_0$ in (\ref{equ:A})
arises from threshold corrections that depend on the complex structure moduli.
This quantity is not known except in special cases, but is a
relatively unimportant constant in our scenario, because the
complex structure moduli are stabilized by the flux background, and because,
as we shall see, $A_0$ appears in the final potential only as an overall constant prefactor.}
\beq \label{equ:A} A(z_\alpha) = A_0 \left(\frac{f(z_\alpha)}{f(0)} \right)^{1/n}\, . \eeq
See \cite{BDKMMM} for a derivation of this result, and for a more complete
discussion of the setup, which we have only briefly reviewed here.

\subsection{Potential in the warped conifold throat}

In this section we apply the general formulae of the previous
section to the case of a D3-brane moving in a warped deformed conifold.
We will assume that both the mobile
D3-brane and the fixed D7-branes are located far enough from the tip that
the deformation parameter $\varepsilon$ may be neglected.
If we use
$z_\alpha = \{z_1, z_2, z_3 \}$ as the three independent variables, the conifold constraint allows us to express
$z_4 =\pm i (\sum_{\alpha=1}^3 z_\alpha^2)^{1/2}$.
Using this basis, and the K\"ahler potential (\ref{Kpot}), we obtain the conifold metric
\bea
k_{\alpha\overline{\beta}}
&=& \frac{3}{2}{\partial^2\over \partial z_\alpha \partial \overline{z_\beta}}\left(
\sum_{\gamma=1}^3 |z_\gamma|^2+\left|\sum_{\gamma=1}^3 z_\gamma^2\right|\right)^{2/3}\, \\
&=& {1\over r} \bigg [ \delta_{\alpha\overline{\beta}}+{z_\alpha
\overline{z_\beta}\over |z_4|^2}- {1\over 3 r^3}\left ( z_\alpha
\overline{z_\beta} + z_\beta \overline{z_\alpha}- {z_4\over
\overline{z_4}} \overline{z_\alpha} \overline{z_\beta}
-{\overline{z_4}\over z_4} z_\alpha z_\beta \right ) \bigg ]\, .
\eea Its inverse assumes the simple form \beq \label{equ:kalphabeta}
k^{\alpha\overline{\beta}}= r \bigg [
\delta^{\alpha\overline{\beta}}+ \frac{1}{2} {z_\alpha
\overline{z_\beta}\over  r^3}  - {z_\beta \overline{z_\alpha}\over
r^3} \bigg ]\, . \eeq Expression (\ref{equ:Fterm}) for the F-term
potential simplifies significantly when we substitute
(\ref{equ:kalphabeta}) and note that
\beq U(\rho, r)= \rho
+\overline{\rho} - {3\gamma\over 2} r^2\ ,\qquad k^{\gamma
\overline{\delta}}k_{\overline{\delta}}={3\over 2}{z_\gamma}\, ,
\quad k_{\gamma} k^{\gamma \overline{\delta}}
k_{\overline{\delta}} =k\, .  \eeq

The remaining term in the potential is the contribution of an anti-D3-brane
at the tip of the conifold, including its Coulomb interaction with the mobile D3-brane
\cite{KKLMMT}
\beq \label{equ:Dterm} V_D(\rho, r)= \frac{D(r)}{U^2(\rho,r)}\, ,
\qquad D(r) \equiv D \left[ 1
- \frac{3 D}{16 \pi^2} \frac{1}{(T_3 r^2)^2} \right] + D_{\rm other}
\approx D + D_{\rm other}\, , \eeq
where $D\equiv 2 h_0^{-1} T_3$ is twice the warped D3-brane tension at the
tip\footnote{A similar potential for a mobile D3-brane arises if instead of
including the antibrane we generalize the throat background \cite{DKS}.}
and $D_{\rm other}$ represents a possible contribution from distant sources of supersymmetry breaking, {\it{e.g.}} in other throats.

The complete inflaton potential is then the sum of the F-term
potential from moduli stabilization, plus the contribution of the antibrane,
\beq \label{equ:Vtot} V = V_F(\rho,z_\alpha) + V_D(\rho,r) \, . \eeq

The canonical inflaton $\phi$ is proportional to $r$, the radial location of the D3-brane
(see \S\ref{sec:bound} for details).  Using (\ref{equ:Vtot})
to compute the slow-roll parameter \beq  \eta \equiv M_P^2 \frac{V_{,\phi\phi}}{V} \, , \eeq
we find \beq
\label{equ:eta}
 \eta = \frac{2}{3} + \Delta \eta(\phi) \, , \eeq
where $\Delta \eta$ arises from the dependence of the superpotential on $\phi$.
If $A$ were a constant independent of $\phi$, slow-roll inflation would be impossible \cite{KKLMMT}, because in that case
$\eta = \frac{2}{3}$.  In this paper, using the explicit result of
\cite{BDKMMM} for $A(\phi)$, we will compute $\Delta \eta$ and
determine whether the full potential can be flat enough for inflation.
Note that the sign of $\Delta \eta$, while crucial, is not obvious {\it{a priori}}.

\section{Case Study: Kuperstein Embedding}
\label{sec:Kuperstein}

Let us consider a particularly simple and symmetric
holomorphic embedding due to Kuperstein \cite{Kuperstein},
which is defined by the algebraic equation
\beq
\label{equ:KupEmbed} f(z_1) = \mu - z_1 = 0\, , \eeq
where, without loss of generality, we will consider the case in which $\mu \in \mathbb{R}^+$.
This embedding preserves an $SO(3)$ subgroup of the $SO(4)$ global symmetry acting on the
$z_i$ coordinates of the deformed conifold.
Kuperstein showed that this is a supersymmetric embedding not just
for the singular conifold, but also in the full warped deformed
conifold background with three-form fluxes.  (For comparison,
the embeddings of \cite{Ouyang,ACR} have so far been studied explicitly only
in the $AdS_{5}\times T^{1,1}$ background).
Adding just a mobile D3-brane does not break supersymmetry
in the case of the non-compact throat. Therefore, the interaction
between the D3-branes and D7-branes must vanish in that limit. When the throat is embedded in a compactification,
the D3-brane potential can receive a contribution from the nonperturbative superpotential (\ref{equ:W}).

The inflaton potential $V(\rho, r ,z_i)$ is in general a complicated function of the
K\"ahler modulus and of the radial and
angular coordinates of the D3-brane. In this section we systematically integrate out all fields except the radial coordinate, leading to an effective single-field potential for the radial inflaton.

\subsection{Multi-field potential}
\label{sec:KupersteinPotential}

\addtocontents{toc}{\SkipTocEntry}
\subsubsection{F-term potential}

By the results of \cite{BDKMMM}, equation (\ref{equ:KupEmbed})
implies \beq \label{equ:AA} A(z_1) = A_0 \left( 1 -
\frac{z_1}{\mu} \right)^{1/n} \, , \eeq and the F-term potential
(\ref{equ:Fterm}) is \bea \label{equ:FKup} V_F &=& \frac{\kappa^2
a |A(z_1)|^2 e^{-a(\rho + \bar{\rho})}}{3 U(\rho,r)^2} \Biggl[
\Bigl(a(\rho+\bar{\rho}) +6\Bigr)+ 6\, W_0\,  {\rm Re} \Bigl( \frac{e^{ a \rho} }{A(z_1)} \Bigr) \nonumber \\
&&  \hspace{3.7cm} - \ 3 \, {\rm Re}(\alpha_{z_1} z_1) + \frac{r}{
a  \gamma} \Bigl( 1 - \frac{|z_1|^2}{2 r^3} \Bigr)
|\alpha_{z_1}|^2 \Biggr]\, , \eea where \beq \alpha_{z_1} \equiv
\frac{A_{z_1}}{A} = - \frac{1}{n(\mu - z_1)}\, , \eeq and \beq
{\rm Re}(\alpha_{z_1} z_1) = -\frac{1}{2 n}
\frac{\mu (z_1 + \overline{z_1}) - 2 |z_1|^2}{|\mu-z_1|^2}\, .
\eeq
Note that the potential (\ref{equ:FKup}) depends only on $r$,
$z_1$,
and $\rho$. Therefore, it is invariant under the $SO(3)$ that acts
on $z_2, z_3, z_4$.

\addtocontents{toc}{\SkipTocEntry}
\subsubsection{Angular degrees of freedom}

\subsubsection*{\sl Imaginary part of the K\"ahler modulus}

\noindent
First, to reduce the complexity of the multi-field potential, we integrate out
the imaginary part of the K\"ahler modulus. Setting $\rho \equiv
\sigma + i \tau$, equation (\ref{equ:FKup}) becomes
\bea
\label{equ:FKup2}
V_F &=& \frac{\kappa^2 a |A|^2 e^{-2
a\sigma}}{3 U^2} \Biggl[
\Bigl(2 a\sigma +6\Bigr)+ 6\, W_0 e^{a \sigma} \, \underline{ \underline{ {\rm Re} \Bigl(  \frac{e^{i a \tau}}{A}  \Bigr) }} \nonumber \\
&&  \hspace{3cm} - \ 3\, {\rm Re}(\alpha_{z_1} z_1) +
\frac{r}{ a \gamma} \Bigl( 1 - \frac{|z_1|^2}{2 r^3} \Bigr)
|\alpha_{z_1}|^2 \Biggr]\, . \eea We see that only the underlined term depends on $\tau$, and the potential for $\tau$ is minimized when this term is as small as possible.
Because $W_0$ is negative, integrating out $\tau(z_1,r) = {\rm Im }(\rho)$ then amounts to the replacement \beq
\label{equ:tau} \frac{e^{i a \tau} }{A} \to \frac{1}{|A|}\, . \eeq
Notice that this is {\it not} the same as setting $\tau \equiv 0$.
In particular, $\tau(z_1,r)$ might be a complicated function, but
all we need to know is (\ref{equ:tau}).

\subsubsection*{\sl Angular directions}

\noindent  The D3-brane position is described by the radial coordinate $r$ and five angles $\Psi_i$ on the base of the cone. The angles are periodic coordinates on a
compact space, so the potential in $\Psi_i$ is either constant or else has discrete minima at some values $\Psi_i^\star$.
We are interested in trajectories that are stable in the angular directions, so that
the motion occurs purely along the radial direction.

We can therefore reduce the number of degrees of freedom by fixing the angular coordinates to the positions that minimize the potential.
In Appendix \ref{sec:stability} we show that for any Kuperstein-like embedding $f(z_1)=0$, these extrema in the angular directions occur only for trajectories
satisfying \beq \label{equ:KupersteinStable}
z_1 = \pm \frac{r^{3/2}}{\sqrt{2}}\, ,
\quad \quad \Leftrightarrow \quad \quad \frac{\partial V}{
\partial \Psi_i} = 0\, .
\eeq
Furthermore, in Appendix \ref{sec:stability} we examine the matrix of second derivatives,
$\frac{\partial^2 V}{
\partial\Psi_i
\partial\Psi_j}$, and find the conditions under which these extrema are stable
minima.  For the present discussion we only need one result from that
section: for small $r$, the trajectory (\ref{equ:KupersteinStable})
is stable against angular fluctuations for negative $z_1$ and unstable for positive $z_1$.
In Appendix \ref{sec:reduc} we show that the canonical inflaton field $\phi$ is well-approximated by a constant rescaling of the radial coordinate $r$,
\beq
\phi^2 \equiv T_3 \hr^2 = \frac{3}{2} T_3 r^2\, .
\eeq
An important parameter of the brane potential is the minimal radial coordinate of the D7-brane embedding \cite{BDKMMM, Kuperstein},
$r_\mu^3 \equiv 2 \mu^2$, or $\phi_\mu^2 = \frac{3}{2} T_3 (2 \mu^2)^{2/3}$.
The potential along the trajectory (\ref{equ:KupersteinStable}) may then be written as
\beq
V(\phi, \sigma) = V_F(\phi, \sigma) + V_D(\phi, \sigma)\, ,
\eeq
where
\bea
\label{equ:VFTwoField}
 V_F(\phi,\sigma) & =& \frac{\kappa^2 a |A_0|^2}{3} \frac{\exp (-2 a \sigma)}{ U(\phi,\sigma)^2}  g(\phi)^{2/n} \Biggl[
2 a \sigma +6- 6 \exp (a \sigma ) \frac{|W_0|}{|A_0|} \frac{1}{g(\phi)^{1/n}}  \nonumber \\
&& \hspace{3cm}
+ \frac{3}{n} \left( c \frac{\phi}{\phi_\mu}
  \pm \left(  \frac{\phi}{\phi_\mu}\right)^{3/2}  - \left(\frac{\phi}{\phi_\mu} \right)^3  \right) \frac{1}{g(\phi)^{2}} \Biggr]\, , \\
\label{equ:VDTwoField}
V_D(\phi, \sigma) & = &  \frac{D(\phi)}{U(\phi,\sigma)^2} \, ,
\eea
and
\beq
g(\phi) \equiv \left|1 \mp \left(\frac{\phi}{\phi_\mu}\right)^{3/2} \right| \, , \qquad U(\phi, \sigma) \equiv 2 \sigma -  \frac{\sigma_0}{3} \frac{\phi^2}{M_P^2}\, .
\eeq
Here we have introduced the constant
\beq
c \equiv \frac{1}{4 \pi\gamma r_\mu^2} = \frac{9}{4n \, a \sigma_0 \frac{\phi_\mu^2}{M_P^2}}\, .
\eeq
This two-field potential is the input for our numerical study in \S\ref{sec:numerics}.

\subsection{Effective single-field potential}
\label{sec:SingleField}

\subsubsection*{\sl Real part of the K\"ahler modulus}

Having reduced the potential to a function of two real fields, $\phi$ and $\sigma$, we integrate out
$\sigma$ by assuming\footnote{We are assuming that $\sigma$ is much more massive than $\phi$.  This may not be valid for a truly generic configuration of a D3-brane in a compact space, but we are specifically interested in cases in which the potential for $\phi$ has been fine-tuned to be flat.  Thus, when slow-roll inflation is possible at all, the adiabatic approximation is justified.  See also \cite{Panda}.} that it evolves adiabatically while remaining in its instantaneous minimum $\sigma_\star(\phi)$, which is defined implicitly by
\beq \label{equ:sigmac} \left.
\partial_\sigma V \right|_{\sigma_\star(\phi)} = 0 \, .
\eeq
This leads to the effective single-field potential
\beq
\mathbb{V}(\phi) \equiv V(\sigma_\star(\phi),\phi)\, .
\eeq
In general, we are not able to solve equation (\ref{equ:sigmac}) analytically for
$\sigma_\star(\phi)$, so we perform this final step numerically (\S\ref{sec:numerics}).
Nevertheless, in Appendix \ref{sec:sigma} we derive useful approximate analytical solutions for the stabilized volume modulus and its dependence on $\phi$ (similar results were derived independently in \cite{Krause}). Here, we cite the basic results of that section.
First, the critical value $\sigma_F$ of the K\"ahler modulus before uplifting is determined by
$\left. D_\rho W \right|_{\phi=0,\, \sigma_F}= 0$, or equivalently \cite{KKLT},
\beq
\label{equ:KKLT}
3 \frac{|W_0|}{|A_0|} e^{a \sigma_F} = 2 a \sigma_F + 3 \quad \quad \Rightarrow \quad \quad \left.\frac{\partial V_F}{\partial \sigma} \right|_{\sigma_F}= 0\, .
\eeq
We now show how the K\"ahler modulus is shifted away from $\sigma_F$ by the inclusion of a brane-antibrane pair.
\begin{enumerate}
\item {\bf Shift induced by the uplifting}

Adding an anti-D3-brane to lift the KKLT AdS minimum to a dS minimum induces a small shift in the stabilized volume,
$\sigma_F \to \sigma_F + \delta \sigma \equiv \sigma_\star(0) \equiv \sigma_0$,
where
\beq
 \delta \sigma \approx \frac{s}{a^2 \sigma_F} \ll 1 \ll   \sigma_F\, .
\eeq
Here we found it convenient to define the ratio of the antibrane energy to the F-term energy {\it before} uplifting, {\it{i.e.}} when $\sigma=\sigma_F$,
\beq \label{equ:sDef}
s \equiv \frac{(D+ D_{\rm other}) U^{-2} (0, \sigma_F)}{|V_F(0, \sigma_F)|}\, ,
\eeq
where stability of the volume modulus in a metastable de Sitter vacuum requires $1 < s \lesssim {\cal O}(3)$.
Although $\delta \sigma$ is small, it appears in an exponent in (\ref{equ:VFTwoField}), so that its effect there has to be considered,
\beq
3 \frac{|W_0|}{|A_0|} e^{a \sigma_0} \approx 2 a \sigma_0 + 3 + 2s\, .
\eeq
When the D3-brane is near the tip, $\phi \approx 0$, the K\"ahler modulus remains at $\sigma_0$. Using this constant value even when the brane is at finite $\phi$ suffices for understanding the basic qualitative features of the potential (\ref{equ:VFTwoField}). However, important quantitative details of the potential depend sensitively on the dependence of $\sigma_\star$ on $\phi$; see Appendix D and Ref.~\cite{Panda}.

\item {\bf Shift induced by D3-brane motion}

In Appendix \ref{sec:sigma} we derive the following
analytic approximation
to the dependence of the stabilized volume on the D3-brane position:
\beq
\label{equ:sigmaApprox}
\sigma_\star(\phi) \approx \sigma_0 \left[1 + c_{3/2} \left( \frac{\phi}{\phi_\mu} \right)^{3/2}\right]\, ,
\eeq
where
\beq \label{equ:cthreehalf}
c_{3/2} \approx \frac{1}{n} \frac{1}{a \sigma_F} \left[ 1-\frac{1}{2 a \sigma_F} \right]\, .
\eeq
This expression is valid along $z_1 = - \frac{r^{3/2}}{\sqrt{2}}$, which we argue below is the interesting case in which the potential is stable in the angular directions.

\end{enumerate}

\subsubsection*{\sl Analytic single-field potential}

Along the trajectory $z_1 = - \frac{r^{3/2}}{\sqrt{2}}$, the inflaton potential is
\bea
\label{equ:VSingleField}
 \mathbb{V}(\phi) & =& \frac{\kappa^2 a |A_0|^2}{3} \frac{\exp (-2 a \sigma_\star(\phi))}{ U(\phi,\sigma_\star(\phi))^2}  g(\phi)^{2/n} \Biggl[
2 a \sigma_\star(\phi) +6- 6 \exp (a \sigma_\star(\phi) ) \frac{|W_0|}{|A_0|} \frac{1}{g(\phi)^{1/n}}  \nonumber \\
&& \hspace{2cm}
+ \frac{3c}{n}  \frac{\phi}{\phi_\mu} \frac{1}{g(\phi)^2} - \frac{3}{n} \left( \frac{\phi}{\phi_\mu} \right)^{3/2} \frac{1}{g(\phi)}
\Biggr] +  \frac{D(\phi)}{U(\phi,\sigma_\star(\phi))^2} \, ,
\eea
where $\sigma_\star(\phi)$
can be determined numerically or approximated analytically by (\ref{equ:sigmaApprox}).
Using the analytic result (\ref{equ:sigmaApprox}) in (\ref{equ:VSingleField}) captures the basic qualitative features of the potential, but is insufficient to assess detailed quantitative questions.  In particular, by using (\ref{equ:sigmaApprox}) one systematically underestimates the total number of $e$-folds supported by the potential (see Appendix D and Ref.~\cite{Panda}).

The inflaton potential (\ref{equ:VSingleField}) is one of our main results.

\section{Search for an Inflationary Trajectory}
\label{sec:Analysis}

In the preceding section, we derived the inflaton potential (\ref{equ:VSingleField}) along the angularly-stable trajectory $z_1 = - \frac{r^{3/2}}{\sqrt{2}}$.  We will now explore this potential and establish that slow roll inflation is possible for a certain range of parameters.  First, in \S\ref{sec:analytic}, we present a few analytic results about the curvature of the potential.  Then, in \S\ref{sec:micro}, we describe the constraints on the parameters of the model that are dictated by the structure of the compactification.  Finally, in \S\ref{sec:numerics}, we present the results of a numerical study of the potential.

\subsection{Analytic considerations}
\label{sec:analytic}

Let us briefly recall the reason for computing the effect of $A(\phi)$ on the inflaton potential.  Kachru {\it{et al.}} \cite{KKLMMT} derived $\eta = \frac{2}{3}$ for the case $A = const.$, and suggested that the inflaton-dependence of the nonperturbative superpotential, $A(\phi)$, could contribute corrections to the inflaton mass, which, if of the right sign, could accidentally make $\eta$ small.  This reasonable expectation hinges on the presence of quadratic corrections to the inflaton potential.

We now argue that in view of the result (\ref{equ:A}), wrapped D7-branes give {\it no} purely quadratic corrections to the inflaton potential.  To see this, we note that the holomorphic coordinates on the conifold scale as fractional powers of $\phi$, $|z_i| \propto \phi^{3/2}$, and $A(z_i)$ is a holomorphic function of the $z_i$ coordinates \cite{BDKMMM}.
This observation implies that the inflaton potential is of the form
\beq
\frac{\mathbb{V}(\phi)}{\mathbb{V}(0)} = 1 + \frac{1}{3} \frac{\phi^2}{M_P^2} + v(\phi)\, ,
\eeq
where $v(\phi)$ contains no quadratic terms.
The slow roll parameter $\eta$, which needs to be very small for sustained slow roll inflation, is
\beq
\label{equ:eta}
\eta \equiv M_P^2 \frac{\mathbb{V}_{, \phi \phi}}{\mathbb{V}} = \frac{2}{3} + M_P^2 v_{,\phi \phi}\, .
\eeq
Because $v_{,\phi\phi}$ contains no constant term, there is no possibility of cancelling the $\frac{2}{3}$ uniformly, for the entire range of $\phi$.  Instead, we can at best hope to cancel the $\frac{2}{3}$ at some special point(s) $\phi_{0}$ obeying $v_{,\phi\phi}(\phi_{0})=-\frac{2}{3}$.

Using the explicit form (\ref{equ:VSingleField}) and expanding for
small $\phi/\phi_\mu$, we find
\beq
\label{equ:etaTip}
\eta =  \frac{2}{3} - \eta_{-1/2}\, \left( \frac{\phi}{\phi_\mu} \right)^{-1/2} + \cdots \, ,
\eeq
where
\beq
\eta_{-1/2}  \approx  \frac{M_P^2}{\phi_\mu^2}  \frac{3(4s-3)}{8 n (s-1)} \frac{1}{a \sigma_0}  > 0 \, .\eeq
Hence, $\eta < 0$ sufficiently close to the tip.  On the other hand, we find that for $\phi \gg \phi_{\mu}$, $\eta > 0$. By continuity, $\eta$ must vanish at some intermediate location $\phi_0$.

The precise location of $\phi_0$ is a parameter-dependent question.
For this purpose, the most important parameter is the minimal radius $\phi_\mu$ of the D7-branes.  Notice that (\ref{equ:eta}) can be written as
\beq
\label{equ:eta2}
\eta = \frac{2}{3} + \frac{M_P^2}{\phi_\mu^2} v_{,xx}\, ,
\eeq
where $x \equiv \frac{\phi}{\phi_\mu}$ and $v_{xx}$ is insensitive to $\phi_\mu$ (see (\ref{equ:VSingleField})).
From (\ref{equ:etaTip}) we see that the second term in (\ref{equ:eta2}) dominates near the tip, giving a large negative $\eta$. This implies the opportunity for a small $\eta$ by cancellation against the positive $\frac{2}{3}$. However, only if $\phi_\mu$ is not too small can this cancellation be achieved inside the throat. Otherwise, $\eta$ remains negative throughout the regime of interest.  We conclude that for small $\phi_\mu$, $\phi_0$ is outside the throat, and hence outside the validity of our construction.

\subsection{Parameters and microscopic constraints}
\label{sec:micro}

Let us describe the microscopic parameters that determine the inflaton potential (\ref{equ:VSingleField}).  In view of (\ref{equ:sDef}), the D-term
$D_{\rm other}+2 T_3 h_0^{-1}$ and $W_0$ are represented by $s$ and by $\omega_F = a \sigma_F$, respectively.  Next, the prefactor $A_0$ only appears as an overall constant rescaling the height of the potential, so we set $A_0 \equiv 1$.  The shape of the inflaton potential is therefore determined by $n$, $\omega_F$, $s$ and $\phi_\mu$.
As we now explain, microscopic constraints lead to important restrictions on the allowed parameter ranges and induce non-trivial correlations among the above parameters.

First, the range of the radial coordinate $r$ affects the
four-dimensional Planck mass, because a longer throat makes a larger
contribution to the volume of the compact space.  In \cite{BM} two of us (D.B. and L.M.) showed that this creates a strong constraint on the allowed field range of the inflaton field $\phi$ (see also Appendix \ref{sec:reduc})
\beq
\label{equ:BMbound}
\frac{\Delta \phi}{M_P} < \frac{2}{\sqrt{N}}\, .
\eeq
Here we use the field range bound (\ref{equ:BMbound}) to constrain the microscopically viable range of $\phi_\mu$, the minimal radial extent of the D7-branes in canonical units. For this purpose we find it convenient to write the bound in the form
\beq
\label{equ:PhiMuBound}
\frac{\phi_\mu^2}{M_P^2} = \frac{1}{Q_\mu^2} \frac{1}{B_6} \frac{4}{N} \, ,
\eeq
where $B_6 \equiv \frac{V_6^w}{(V_6^w)_{\rm throat}} > 1$ parameterizes the relative contribution of the throat to the total (warped) volume of the compact space, and
\beq \label{equ:QmuDef}
Q_\mu \equiv \frac{r_{\rm UV}}{r_\mu}
\eeq
is a measure of how far into the throat the four-cycle extends.  Applicability of the results of \cite{BDKMMM} requires $Q_\mu = \frac{\phi_{\rm UV}}{\phi_\mu} > 1$.

Second, the warped volume $V_{\Sigma_4}^{w}$ of the wrapped
four-cycle $\Sigma_{4}$ is bounded below by the warped volume in the
throat region, \beq V_{\Sigma_4}^{w} =
(V_{\Sigma_4}^{w})_{\rm throat} +
(V_{\Sigma_4}^{w})_{\rm bulk} \ge
(V_{\Sigma_4}^{w})_{\rm throat}\, . \eeq
In Appendix \ref{sec:4vol} we compute
$(V_{\Sigma_4}^{w})_{\rm throat}$ for the Kuperstein embedding
\beq
T_3 (V_{\Sigma_4}^{w})_{\rm throat} = \frac{3}{2} N \log Q_{\mu}\, .
\eeq
In \S\ref{sec:potential} we explained how the unperturbed warped four-cycle volume relates to the K\"ahler modulus of the compactification (more details can be found in Refs.~\cite{BDKMMM, GM} and Appendix \ref{sec:reduc}).
If we use $B_4 \equiv \frac{V_{\Sigma_4}^w}{(V_{\Sigma_4}^w)_{\rm throat}} > 1$ to parameterize the relative contribution of the throat to the total warped volume of the four-cycle wrapped by the D7-branes, then we can relate the vev of the K\"ahler modulus to microscopic parameters of the compactification
\beq
\label{equ:OmegaBound}
\omega_F \approx \omega_0\approx \frac{3}{2} \frac{N}{n} B_4 \log Q_\mu\, .
\eeq
We require $\omega_0 < {\cal O}(30)$, because otherwise the inflation scale will be too low -- see Appendix \ref{sec:reduc}.

The constraints (\ref{equ:PhiMuBound}), (\ref{equ:OmegaBound}) will play an essential role in our
analysis.  We will find that inflationary configurations are rather
easy to find if these constraints are neglected, but imposing them
dramatically decreases the parameter space suitable for inflation.

\addtocontents{toc}{\SkipTocEntry}
\subsubsection{Bulk contributions to the volume}

We have just introduced two parameters, $B_4 = \frac{V_{\Sigma_4}^w}{(V_{\Sigma_4}^w)_{\rm throat}}$ and $B_6=\frac{V_6^w}{(V_6^w)_{\rm throat}}$, that represent ratios of total volumes to throat volumes.  In the throat, we have access to an explicit Calabi-Yau metric and can compute the volumes directly.  This metric data in the throat is one of the main reasons that warped D-brane inflation can be studied explicitly.  In contrast, we have very little data about the bulk, so we cannot compute $B_4$ and $B_6$.

Fortunately, these parameters do not directly enter the potential.  Instead, they appear in the compactification constraints (\ref{equ:PhiMuBound}) and (\ref{equ:OmegaBound}), and thereby affect the microscopically allowable ranges of the other parameters, such as $\phi_\mu$ and $Q_\mu$.  In particular, when $B_6$ is large, the range of $\phi_\mu$ is reduced, because the throat is shorter in four-dimensional Planck units.  In the numerical investigation of \S\ref{sec:numerics}, we find that inflation is possible inside the throat, and our construction is self-consistent, provided that $B_4/B_6 \gtrsim 2$.  For concreteness, we take $B_4 \sim 9$, $B_6 \sim 1.5$ in the remainder.  This means that the throat contributes a greater share of the total six-volume than it does of the wrapped four-cycle volume: in other words, the wrapped four-cycle only enters the upper reaches of the throat.  Although we expect that such a configuration can be realized, it will be valuable to find a fully explicit construction.  We note, however, that a very large value of $B_4$, implying that the D7-brane is hardly in the throat at all, would mean that the result of \cite{BDKMMM} is inapplicable, because the correction to the four-cycle volume is then dominated by the correction to the uncomputable bulk portion of the volume.

\addtocontents{toc}{\SkipTocEntry}
\subsubsection{Parameter choices}

Although a systematic study of the full multi-dimensional parameter space would undoubtedly be instructive, we here employ a simpler and more transparent strategy that we believe nevertheless accurately portrays the range of possibilities.
To this end, we set some of the discrete parameters to
reasonable values and then scan over the remaining parameters. Let
us emphasize that although the precise values chosen here are not
important, it is important that we were able to find regions in
parameter space where all our approximations are valid and all the compactification constraints are satisfied.

First, we fix $n \to 8$.  This helps to reduce the degree to which the volume shifts during inflation, as from (\ref{equ:cthreehalf}), $c_{3/2} \propto n^{-1}$.  Numerical study of the case $n=2$ yields results qualitatively similar to those we present here, but the analytical treatment is more challenging.  To ignore backreaction of the wrapped branes on the background geometry, we require that the background D3-brane charge exceeds the number of wrapped branes, $\frac{N}{n} > 1$.  For concreteness, we use $N=32$.  Finally, as previously stated, we take $B_4 \sim 5$, $B_6 \sim 2$.

This allows us to impose the microscopic constraints (\ref{equ:PhiMuBound}), (\ref{equ:OmegaBound})
on the compactification volume $\omega_F$ and the wrapped brane location $\phi_\mu$ in terms of a single parameter $Q_\mu$.
The remaining parameters that determine the potential are then $Q_\mu$ (\ref{equ:QmuDef}) and $s$ (\ref{equ:sDef}).

To search for inflationary solutions, we scanned over $Q_\mu$  and $s$, treating both as continuous
parameters, although they are in principle determined by discrete flux input.
Here, for simplicity of presentation, we will fix $Q_\mu$ to a convenient value, $Q_\mu=1.2$,
and only exhibit the scanning over $s$.  To interpret this scanning in microphysical terms,
we recall that, for fixed F-term potential and for fixed supersymmetry breaking (corresponding to the parameter $D_{\rm other}$) outside the throat, $s$ is determined by $D = 2 T_3 h_0^{-1}$, where $h_0$ is given in (\ref{warptip}) and is of order
$h_0 \approx {\rm{exp}}\left(\frac{8\pi K}{3 g_s M}\right)$.  The values of $h_0$ we will consider can be achieved for reasonable values of $K,M,g_s$.

In summary, we have arranged that all consistency conditions are satisfied,
and all parameters except for the amount of uplifting, $s$, are fixed.  As we vary the uplifting,
the shape of the potential (\ref{equ:VSingleField}) will change.  As we shall now see, for a certain range of values of $s$ the potential becomes flat enough for prolonged inflation.

\begin{figure}[h!]
    \centering
        \includegraphics[width=0.7\textwidth]{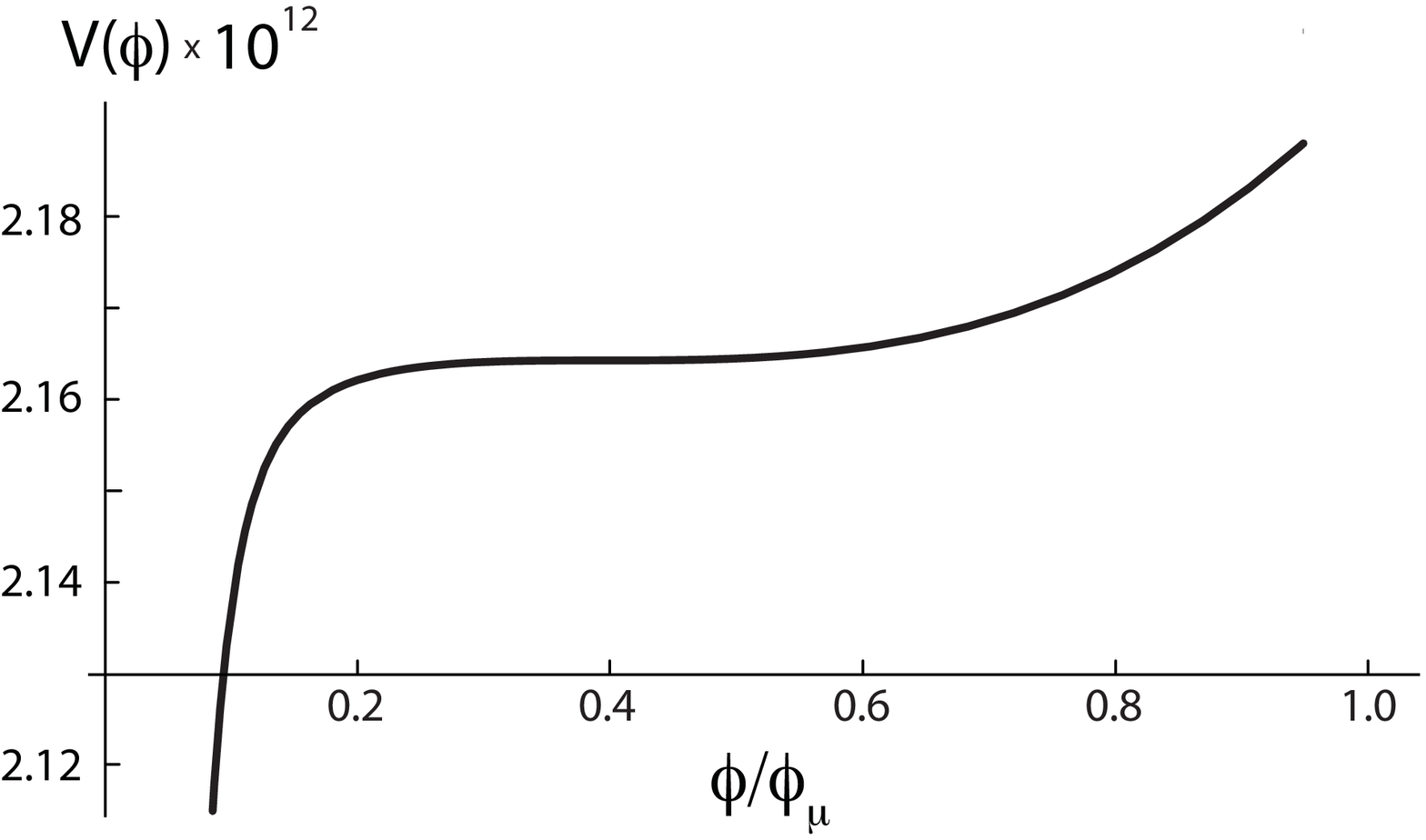}
    \caption{{\bf Inflaton potential $\mathbb{V}(\phi)$.} \newline
    {\sl Compactification data:} $n=8$, $\omega_F = 10$, $N = 32$, $Q_\mu = 1.2$,
    $B_6=1.5$, $B_4= 9$, $s=1.1$, which implies $\phi_\mu = 0.25$, $W_0 = -3.432 \times 10^{-4}$, $D+D_{\rm other}=1.2 \times 10^{-8}$,  $\omega_0 \approx 10.1$. }
    \label{fig:pot}
\end{figure}

\subsection{Numerical results}
\label{sec:numerics}

The first observation we make about (\ref{equ:VSingleField}) is that, near the parameter values we have indicated, it is generically non-monotonic.  In fact, the potential has a metastable minimum\footnote{A D3-brane located in this metastable minimum contributes to the breaking
of supersymmetry.  It would be extremely interesting to use a configuration of
a D3-brane and a moduli-stabilizing D7-brane stack to uplift to a de Sitter vacuum.
Here we have not quite accomplished this: we have, of course,
included an anti-D3-brane as well, which is well-known to accomplish
the uplifting by itself \cite{KKLT}.
If this antibrane is removed, the structure of the potential changes, and it is
not clear from our results so far that a remaining D3-brane
would suffice to uplift to a de Sitter vacuum.  We leave this as a promising direction for future work.} at some distance from the tip.  We are confident that this is a minimum and not a saddle point, because we have explicitly shown in the Appendices that the curvature of the potential in the angular directions is non-negative.  (The curvature is zero along directions protected by the unbroken $SO(3)$ symmetry of the background, and positive in the other directions.)  Moreover, we have shown that the potential is stable with respect to changes in the K\"ahler modulus.

Next, we notice that as we vary $s$, the metastable minimum grows more shallow, and the two zeroes of $V^{\prime}$, the local maximum and the local minimum, approach each other.  A zero of $V^{\prime\prime}$ is trapped in the shrinking range between these two zeroes of $V^{\prime}$.  For a critical value of $s$, the zero of $V^{\prime\prime}$ and the two zeroes of $V^{\prime}$ coincide, and the potential has an inflection point.  As $s$ changes further, the potential becomes strictly monotonic.

We therefore find that there exists a range of $s$ for which both the first and second derivatives of the potential approximately vanish.  This is an approximate inflection point.  In the next section we discuss a phenomenological model that captures the essential features of (\ref{equ:VSingleField}) in the vicinity of this inflection point.

\begin{figure}[h!]
    \centering
        \includegraphics[width=0.96\textwidth]{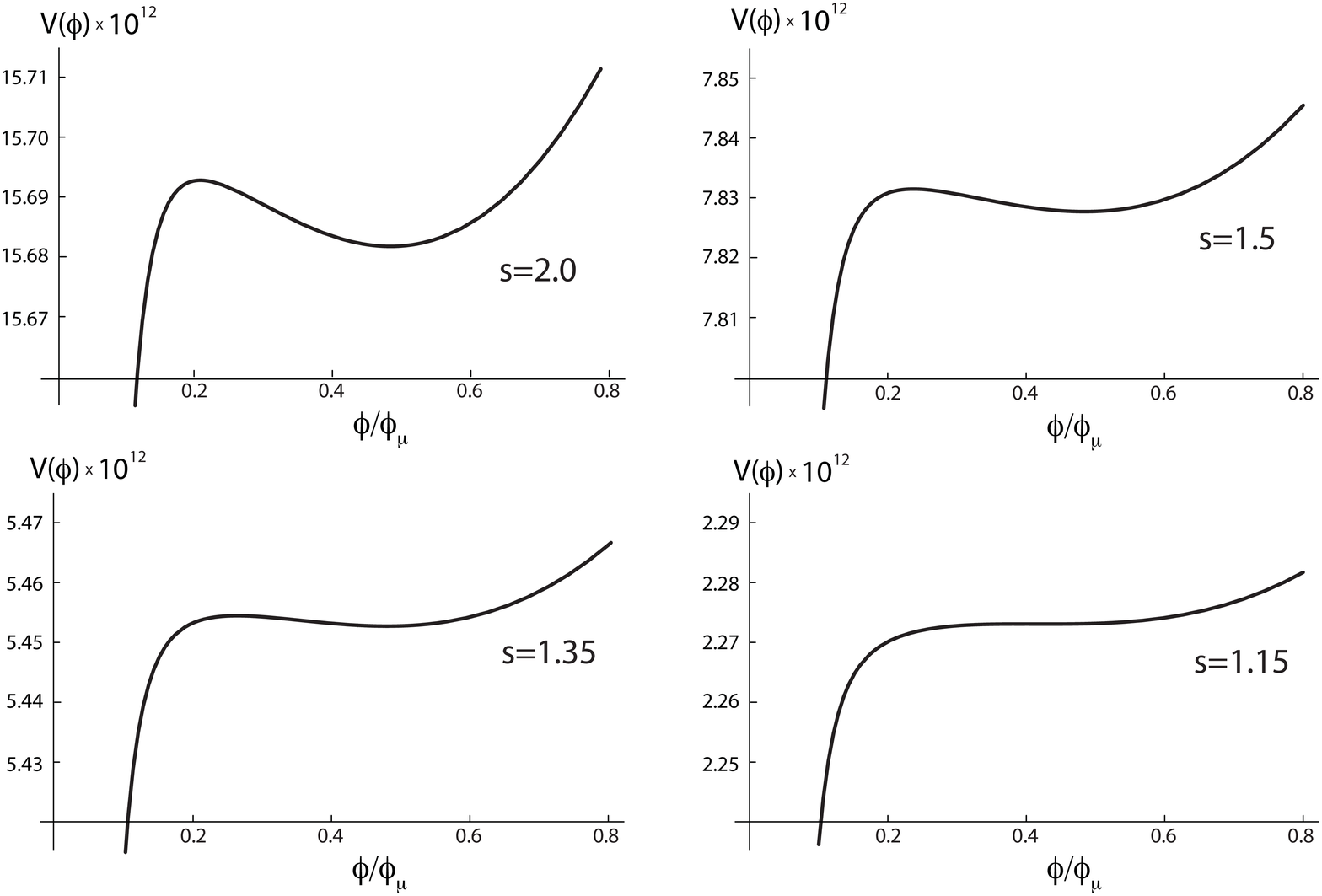}
    \caption{{\bf The inflaton potential $\mathbb{V}(\phi)$ as a function of $s$}.
    \newline
    The transition from metastability to monotonicity is shown; old inflation and new inflation are continuously connected.}
    \label{fig:sScan}
\end{figure}

\subsection{Phenomenological model: cubic inflation}

We have shown that inflationary solutions in the Kuperstein embedding arise near an inflection point at $\phi=\phi_0$, where the potential is very well approximated by the cubic form \cite{LythRiotto}\footnote{Ref.~\cite{MSSM} has developed an inflation model within the minimal supersymmetric standard model (MSSM) that has a similar cubic phenomenology. We thank Justin Khoury for bringing this model to our attention.
Inflection point inflation in the context of string theory, with the inflaton corresponding to the compactification volume, has been considered in \cite{Itzhaki}.}
\beq
\label{equ:cubic}
\mathbb{V} = V_0 + \lambda_1 (\phi - \phi_0) +  \frac{1}{3!} \lambda_3 (\phi-\phi_0)^3\, .
\eeq
Prolonged inflation requires smallness of the slow roll parameters, $\epsilon , \eta \ll 1$. From (\ref{equ:cubic}) we find\footnote{In this section we set $M_P \equiv 1$.}
\bea
\label{equ:EPS}
\epsilon &\equiv& \frac{1}{2} \left( \frac{\mathbb{V}_{,\phi}}{\mathbb{V}} \right)^2 \approx \frac{1}{2} \left( \frac{\lambda_1 + \frac{1}{2} \lambda_3 (\phi -\phi_0)^2}{V_0} \right)^2 \, , \\
\eta &\equiv&  \frac{\mathbb{V}_{,\phi \phi}}{\mathbb{V}} \approx \frac{ \lambda_3}{V_0} (\phi-\phi_0)\, .
\label{equ:ETA}
\eea
The number of $e$--folds between some value $\phi$ and the end of inflation
$\phi_{\rm end}$ is then
\beq
\label{equ:Ne}
N_e(\phi) = \int_{\phi_{\rm end}}^\phi \frac{\d \phi}{\sqrt{2 \epsilon}} = \frac{N_{\rm tot}}{\pi} \arctan \left.  \left( \frac{ \eta(\phi)}{2 \pi N_{\rm tot}^{-1}} \right) \right|_{\phi_{\rm end}}^\phi\, ,
\eeq
where
\beq \label{equ:Ntot}
N_{\rm tot} \equiv \int_{-\infty}^\infty \frac{\d \phi}{\sqrt{2 \epsilon}} = \pi \sqrt{\frac{2 V_0^2}{ \lambda_1 \lambda_3}}\, .
\eeq
In (\ref{equ:EPS}) and (\ref{equ:ETA}) we have set $\mathbb{V}(\phi) \approx V_0$ in the denominators,
while in (\ref{equ:Ntot}) we extended the integral from the range where $|\eta|<1$ to infinity.
These approximations are very good in the regime
\beq
{V_0\over \lambda_3} \ll 1\ , \qquad {V_0\over \sqrt{\lambda_1 \lambda_3} } \gg 1\, .
\eeq
The first of these conditions guarantees that inflation is of
the small-field type, while the second implies that $N_{\rm tot} \gg 1$.
We will be interested in
$N_{\rm tot} \ge N_{\rm CMB} \sim 60$.

\begin{figure}[htbp]
    \centering
        \includegraphics[width=0.60\textwidth]{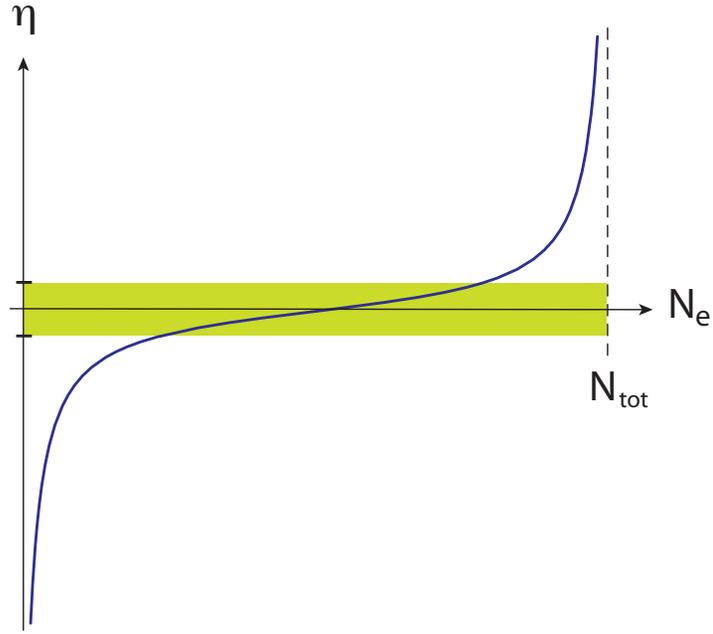}
    \caption{{\bf $\eta(\phi)$ as a function of the number of $e$-folds of inflation, $N_e$}.
    \newline
    In the green band, $|\eta| < 2\pi N_{\rm tot}^{-1}$.}
    \label{fig:Nefold}
\end{figure}

\smallskip\noindent
Equation (\ref{equ:Ne}) shows that there are $\frac{1}{2} N_{\rm tot}$ $e$--folds during which $|\eta| < 2 \pi N_{\rm tot}^{-1}$; see Figure \ref{fig:Nefold}.
For large $N_{\rm tot}$ this implies that there is a large range of $e$--folds where $\eta$ is small and the scalar perturbation spectrum is nearly scale-invariant.
Predicting the scalar spectral index in these models is
non-trivial:
\beq
n_s-1 = \left. ( 2\eta -6 \epsilon)\right|_{\phi_{\rm CMB}} \approx 2 \eta(\phi_{\rm CMB})\, .
\eeq
The scalar spectral index on CMB scales can be red, blue or even perfectly scale-invariant depending on where $\phi_{\rm CMB}$ is relative to the inflection point.
If inflation only lasts for the minimal number of $e$-folds to solve the horizon and flatness problems then the scalar spectrum is blue.  If the potential is flatter than this, so that $\epsilon$ is smaller, inflation lasts longer and $\phi_{\rm CMB}$ is more likely to be smaller than $\phi_0$. The spectrum is then red, since $\eta(\phi_{\rm CMB} < \phi_0) < 0$.

\begin{figure}[htbp]
    \centering
        \includegraphics[width=0.80\textwidth]{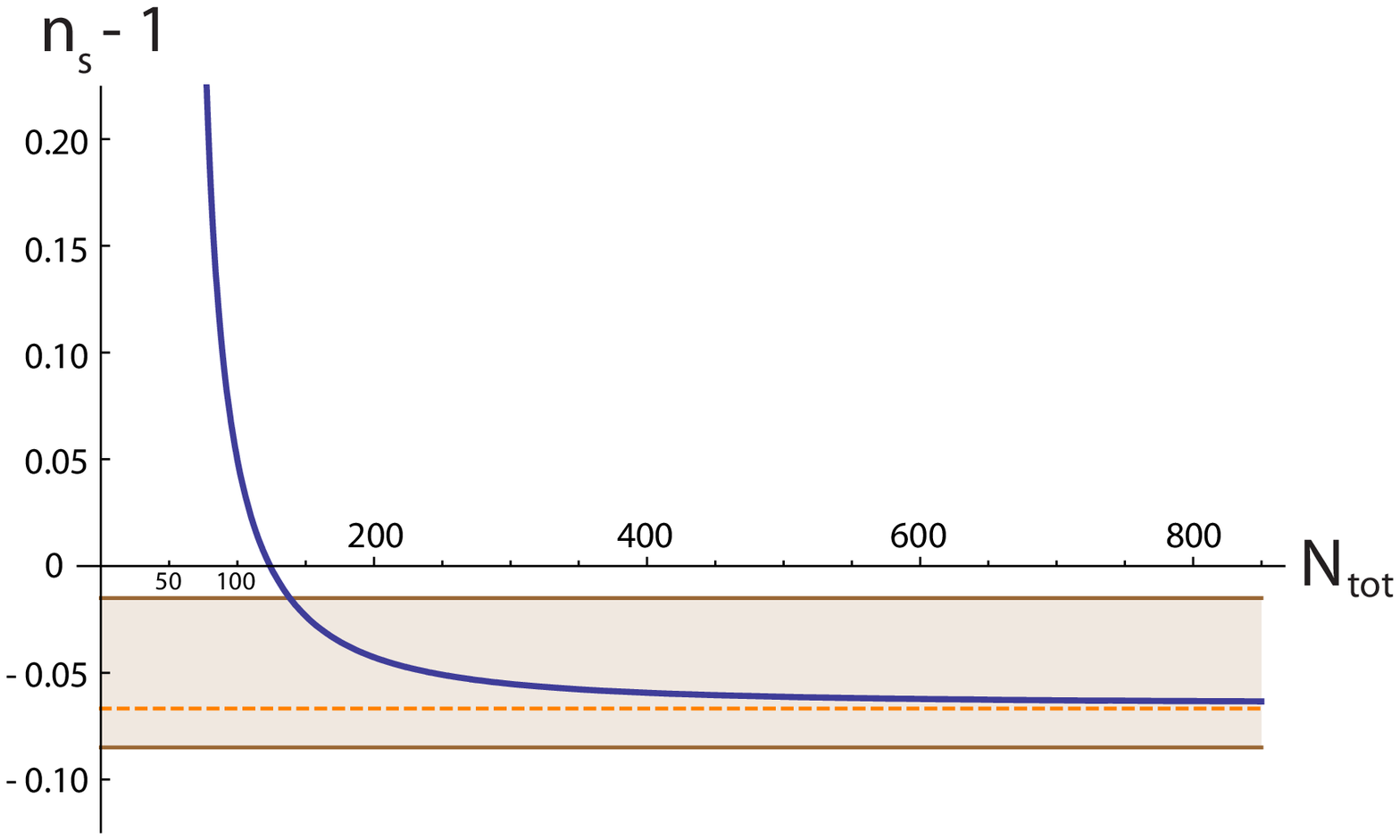}
    \caption{{\bf Spectral index $n_s$, evaluated on CMB scales, as a function of the total number of $e$-folds of inflation, $N_{\rm tot}$}.
    \newline
    The light band gives the WMAP3 2$\sigma$ limit on $n_s$ (for $r \equiv 0$) \cite{Observations}.}
    \label{fig:NsMinusOne}
\end{figure}

More concretely, we can evaluate $n_s$ by inverting (\ref{equ:Ne}) at
$\phi_{\rm CMB}$ where $N_e(\phi_{\rm CMB}) \equiv N_{\rm CMB} \sim 60$, and
using $\eta(\phi_{\rm end})=-1$. This gives
\beq
n_s -1 = \frac{4 \pi}{N_{\rm tot}} \tan \left( \pi \frac{N_{\rm CMB}}{N_{\rm tot}} -
\arctan\left( \frac{N_{\rm tot}}{2 \pi} \right)\right)\, .
\eeq
Using $\arctan x = \frac{\pi}{2} - x^{-1} + {\cal O}(x^{-3})$
we furthermore find
\beq
n_s -1 = \frac{4 \pi}{N_{\rm tot}} \tan \left( \pi \frac{N_{\rm CMB}+2}{N_{\rm tot}} - \frac{\pi}{2} + {\cal O}(N_{\rm tot}^{-3}) \right) \, .
\eeq
Using $N_{\rm CMB} + 2 \approx N_{\rm CMB} \gg 1$, we may simplify this to
\beq
\label{equ:NsMinusOne}
n_s -1 \approx - \frac{4 \pi}{N_{\rm tot}}  \cot \left( \pi \frac{N_{\rm CMB} }{N_{\rm tot}} \right)\, ,
\eeq
which has the expansion
\beq
n_s -1 \approx - \frac{4}{N_{\rm CMB}} + \frac{4 \pi^2}{3} \frac{N_{\rm CMB}}{N_{\rm tot}^2} + {\cal O}\Bigl( \frac{N_{\rm CMB}^3}{ N_{\rm tot}^{4}} \Bigr)\, .
\eeq
Equation (\ref{equ:NsMinusOne}) is plotted in Figure \ref{fig:NsMinusOne}.
We note the following properties of this result. For $N_{\rm tot} $ not much greater than $N_{\rm CMB}$
the spectrum is strongly blue and the model is hence ruled out by recent observations
\cite{Observations} (in this regime the slow-roll formulae we have used are not good approximations, but
a more exact treatment gives similar results).
For $N_{\rm tot} \approx 2 N_{\rm CMB}$ the spectrum on CMB scales is exactly scale-invariant.
For $N_{\rm tot} \gtrsim 2 N_{\rm CMB}$ the spectrum is red and asymptotes to the lower limit $n_s \to 1 - 4/N_{\rm CMB} \approx 0.93$ for $N_{\rm tot} \gg N_{\rm CMB}$.
This asymptotic limit,
which has been noted in studies of inflation near an inflection point \cite{MSSM,Itzhaki},
is more strongly red than is typical in single-field inflation models.

Given the explicit expression (\ref{equ:NsMinusOne}) for the spectral index $n_s(N_e)$ we can compute its scale-dependence or `running':
\bea
\alpha_s \equiv \frac{d n_s}{d \ln k} = -\left. \frac{d n_s}{d N_e} \right|_{N_e =N_{\rm CMB}} &=&
-\frac{4 \pi^2}{N_{\rm tot}^2} \sin^{-2} \left( \pi \frac{N_{\rm CMB}}{N_{\rm tot}} \right) \, ,\\
&\approx& -\frac{4}{N_{\rm CMB}^2} -\frac{4 \pi^2}{3} \frac{1}{N_{\rm tot}^2} + {\cal O}\Bigl( \frac{N_{\rm CMB}^2}{N_{\rm tot}^4} \Bigr)\, .
\eea
The running can be large for models with blue spectral tilt ($N_{\rm tot} \sim N_{\rm CMB}$), but is small for models with red spectra. The asymptotic value for $N_{\rm tot} \gg N_{\rm CMB}$ is $\alpha_s \to -4/N_{\rm CMB}^2 \approx -10^{-3}$.  Notice that because both the tilt $n_s-1$ and the running $\alpha_s$ are determined by $N_{\rm tot}$ alone (for fixed $N_{\rm CMB}$), this phenomenological model is predictive.

It is possible to arrange for the magnitude of scalar perturbations on CMB scales to be small,
\beq
\Delta_{\cal R}^2 = \left. \frac{1}{24 \pi^2} \frac{\mathbb{V}}{\epsilon} \right|_{\phi_{\rm CMB}} \approx \frac{1}{12 \pi^2} \frac{V_0^3}{\lambda_1^2}  \approx 2.4 \times 10^{-9}\, ,
\eeq
by adjusting the overall scale of inflation.

Let us comment briefly on some general difficulties in inflection point inflation.
Since inflation is restricted to a small region around the inflection point, an
immediate concern is the question of initial conditions. In particular,
how sensitive is the present scenario to the initial position and
velocity of the D3-brane? What fraction of initial conditions
lead to overshoot rather than to inflation?
A complete analysis of these questions is beyond the scope of the present paper;
see \cite{Krause,Itzhaki} for a discussion of some aspects of this problem.

Since this is a small-field model, it is sensitive to small corrections in the slope of the potential (see \S\ref{sec:discussion}). These corrections are important for both the background evolution, {\it i.e.} the number of $e$--folds of inflation, and for the perturbation spectrum.

Finally, we note that the appearance of the inflection point feature depends sensitively on the use of the adiabatic approximation for integrating out the volume modulus.  One might therefore be worried about cases in which the exact two-field evolution is not well captured by this approximation and a more detailed numerical study of the two-field evolution is required.\footnote{This important problem has been explored in the subsequent work \cite{Panda} where the validity of the adiabatic approximation is explicitly confirmed.}

\section{Comments on Other Embeddings}
\label{sec:ACR}

The previous two sections contained a detailed discussion of the D3-brane potential for the Kuperstein embedding.
We derived important microscopic constraints, analyzed the fine-tuning problem involved in realizing
inflationary solutions, and studied the resulting cosmological dynamics.

In this section we will make some brief remarks about other embeddings. When applicable we will emphasize
the differences and similarities to the Kuperstein case. This will illustrate the
special status of the Kuperstein embedding.\\

First, we give a simple proof that for an infinite class of D7-brane embeddings,
the ACR embeddings, there always exist trajectories for which no amount of fine-tuning
can flatten the inflaton potential.
Are\'an, Crooks and Ramallo (ACR) \cite{ACR} studied
supersymmetric four-cycles in the conifold described by the embedding equations \beq \label{equ:ACR} f(w_i) = \mu^P
- \prod_{i=1}^4 w_i^{p_i}  = 0\, , \eeq where $p_i \in
\mathbb{Z}$, $P \equiv \sum_{i=1}^4 p_i$ and $\mu^P \in
\mathbb{C}$ are constants defining the embedding.
Here $w_i \in \mathbb{C}$ are alternative
coordinates on the conifold that follow from the $z_i$ coordinates
by a linear transformation (see Appendix \ref{sec:conifold}).
The conifold constraint in these coordinates is $w_1 w_2 - w_3 w_4 = 0$.
By requiring that the $p_i$ are non-negative we can restrict attention to four-cycles that
do not reach the tip of the conifold.
Two simple special cases of the
ACR embeddings (\ref{equ:ACR}) are the {\it Ouyang embedding}
\cite{Ouyang}, $w_1 = \mu$, and the {\it Karch-Katz embedding}
\cite{Karch:2002sh}, $w_1 w_2 = \mu^2$.

To study the ACR embeddings in a unified way we define a collective coordinate $\Phi$
\beq \Phi^P
\equiv \prod_{i=1}^4 w_i^{p_i}\, ,
\eeq
such that
\beq
A(\Phi^P) = A_0
\left (1- \frac{\Phi^P}{\mu^P} \right)^{1/n} \, , \eeq and \beq \sum_i w_i \alpha_{w_i}
=  \sum_i p_i \Phi^P \alpha_{\Phi^P} = P \Phi^P \alpha_{\Phi^P} \,
, \eeq where \beq \alpha_{\Phi^P}  \equiv \frac{1}{A}
\frac{\partial A}{\partial \Phi^P} = - \frac{1}{n} \frac{1}{\mu^P -
\Phi^P}\, .
\eeq
Next, we consider the part of the F-term potential that depends on
derivatives of the superpotential with respect to the brane
coordinates
\beq \label{equ:DeltaVF} \Delta V_F = - \frac{\kappa^2 a |A|^2 e^{-2
a \sigma}}{U^2} \Bigl[ 3\, {\rm Re}(w^i \alpha_{w_i}) - \frac{1}{a
\gamma} \hat{k}_w^{i \overline{\jmath}} \alpha_{w_i}
\alpha_{\overline{w_j}} \Bigr]\,, \eeq where \beq
\label{equ:term1} {\rm Re}(w^i \alpha_{w_i})  = P\, {\rm
Re}(\Phi^P \alpha_{\Phi^P}) \eeq and \bea
\frac{1}{\gamma} \hat{k}_w^{i \overline{\jmath}} \alpha_{w_i} \alpha_{\overline{w_j}} &=& \frac{1}{ \gamma r^2} \left|  \alpha_{\Phi^P} \right|^2 \left| \Phi^P \right|^2 \times \Bigl\{ \frac{1}{2} P^2 + (p_1 - p_2)^2 +  (p_3 - p_4)^4 + Z(w_i) \Bigr\}
\label{equ:term2}\, . \eea
Here, we found it convenient to write the K\"ahler metric $\hat k_w^{i \bar \jmath}$ in $SO(4)$--invariant form (see Appendix \ref{sec:SO4}) and  defined the function
\bea
Z &\equiv& +  \left|p_1 \frac{w_4}{w_1} + p_3 \frac{w_2}{w_3}\right|^2 +  \left|p_1 \frac{w_3}{w_1} + p_4 \frac{w_2}{w_4}\right|^2  \nonumber \\
&&+  \left|p_2 \frac{w_3}{w_2} + p_4 \frac{w_1}{w_4}\right|^2 +
\left|p_2 \frac{w_4}{w_2} + p_3 \frac{w_1}{w_3}\right|^2 \, .
\eea

\addtocontents{toc}{\SkipTocEntry}
\subsection{Delta-flat directions}

We now show that there is always a radial trajectory, $\Phi =0$, along which
$\partial W/ \partial w_i$ is orthogonal to $w_i$ and lies in the null direction of $\hat k^{i \bar \jmath}_w$.
The term
$\Delta V_F$ in (\ref{equ:DeltaVF}) then vanishes and the prefactor $A$ of
the superpotential becomes independent of the brane position. We
call this a {\it delta-flat direction}. For the Ouyang embedding
this trajectory was first found by Burgess {\it et al.}
\cite{Burgess}.  As noted in \cite{Burgess}, delta-flat directions are noteworthy because they have $\Delta \eta = 0$ and therefore imply a well-known no-go result for inflation
 \cite{KKLMMT}.

More concretely, we see that for (\ref{equ:term1}) to vanish one requires
\beq
\label{equ:deltaflat}
\Phi = 0\, ,
\eeq
 {\it i.e.} at
least one of the $w_i$ that enter the embedding must vanish. In fact, we notice that
(\ref{equ:term1}) vanishes whenever (\ref{equ:term2}) does, so we
can restrict our attention to (\ref{equ:term2}). If {\it any} $p_i
> 1$, then we see immediately from the overall factor $|\Phi^P|^2
= |w_1|^{2 p_1} |w_2|^{2p_2} |w_3|^{2 p_3} |w_4|^{2 p_4}$  that
(\ref{equ:term2}) vanishes on $w_i = 0$. For $p_i \le 1$ there are only a few distinct cases: $\Phi = w_1$,
$\Phi^2 = w_1 w_2$, $\Phi^2 = w_1 w_3$, $\Phi^3 = w_1 w_2 w_3$, and
$\Phi^4= w_1 w_2 w_3 w_4$. In the next section we will illustrate the argument for the important case of the Ouyang embedding, $\Phi=w_1$; the proof is easily generalized to the remaining cases.
This completes the proof that all ACR embeddings have delta-flat trajectories.

\addtocontents{toc}{\SkipTocEntry}
\subsection{Comparison of the Ouyang and Kuperstein Embeddings}

Recently the result of \cite{BDKMMM} has been applied
\cite{Burgess, Krause} to compactifications involving the Ouyang
embedding $w_1 = \mu$.
In this case, the correction to the F-term potential is
 \beq
 \label{equ:DVFouyang} \Delta V_F =
\frac{\kappa^2}{3 U^2} \Biggl[ \frac{3}{2} (\overline{W_{,\rho}}
w_1 W_{,w_1} + c.c.) + \frac{1}{\gamma} \hat k^{1 \bar 1}_w W_{,w_1}
\overline{W_{,w_1}} \Biggr]\, , \eeq where \beq
\label{equ:kouyang}\hat{k}_w^{1 \overline{1}} = r \Bigl(1 +
\frac{|w_1|^2}{2 r^3} - \frac{|w_2|^2}{r^3} \Bigr) \, . \eeq There
are two kinds of radial extremal trajectories: the delta-flat
trajectory $w_1=w_3=w_4=0$, for which $\Delta V_F =0$
\cite{Burgess}, and also a trajectory $w_2=w_3=w_4=0$, $w_1 \in
\mathbb{R}$ used in \cite{Krause}.

The Kuperstein scenario is closely related to the Ouyang scenario
except for two subtle differences which we now discuss:
\begin{enumerate}
\item There exists {\it no} delta-flat direction for the
Kuperstein embedding. \item  The single-field potential along the non-delta-flat direction for the
Ouyang embedding is identical in shape to that along the corresponding
Kuperstein trajectory. However, the angular stability is different (see Appendix
\ref{sec:stability}).  This trajectory in the Kuperstein embedding is stable for small $r$, while in the Ouyang embedding it is unstable in that regime.
\end{enumerate}
To see this compare the correction to the F-term potential for the
Ouyang embedding, (\ref{equ:DVFouyang}), with the corresponding
term for the Kuperstein embedding,
 \beq \Delta V_F =
\frac{\kappa^2}{3 U^2} \Biggl[ \frac{3}{2} (\overline{W_{,\rho}}
z_1 W_{,z_1} + c.c.) + \frac{1}{\gamma} k^{1 \bar 1} W_{,z_1}
\overline{W_{,z_1}}  \Biggr]\, , \eeq where \beq \label{equ:kkup}
k^{1\bar 1}= r \Bigl(1- \frac{1}{2} \frac{|z_1|^2}{r^3}\Bigr)\, .
\eeq This shows immediately that the Kuperstein embedding does not
have a delta-flat direction, since $ k^{1 \bar 1}$
 cannot vanish for $r>0$.
This is to be viewed in contrast to the Ouyang case for
which (\ref{equ:kouyang}) vanishes on $w_1=0$.

Considering $ k^{1\bar 1},\hat k^{1\bar 1}_w$ for each case one may
further show that the trajectories $2|z_1|^2 = r^3$ and $|w_1|^2 =
r^3$ lead to identical shapes for the single-field potential.
However, `off-shell', {\it i.e.} away from the extremal path, $
\hat{k}^{1 \bar 1}_w$ for the Ouyang embedding is of a different
form from $k^{1 \bar 1}$ for the Kuperstein embedding. It is for
this reason that the angular stability of the two scenarios is
different (see Appendix \ref{sec:stability} for details). In
particular, while for the Kuperstein embedding the trajectory is
stable for the regime of interest, for the Ouyang embedding it is
unstable.

To discuss the issue of stability in simple terms, we consider
$\theta_1=\theta_2=\theta$ and $\tilde \psi \equiv \psi-\phi_1-\phi_2$, so that
$w_1= e^{\frac{i}{2} \tilde \psi} r^{3/2} \sin^2 (\theta/2)$. Then, as shown in
\cite{Burgess} for $n=1$, \beq V_F(\theta) = V_1 \sin^2 (\theta/2) + V_2
\sin^4 (\theta/2) + const.\ , \eeq where \bea V_1 &=& {\kappa^2 |A_0|^2
e^{-2 a\sigma}\over 3 U^2} \frac{r}{\gamma \mu^2} \bigg (2 - a\mu \gamma
\sqrt{r} \cos \frac{\tilde \psi}{2}
\left (9+ 4 a\sigma + 6 W_0 {e^{a\sigma}\over |A_0|}\right )\bigg )\ , \\
V_2 &=&  \frac{1}{4}
{\kappa^2 |A_0|^2
e^{-2 a\sigma}\over 3 U^2} \frac{r}{\gamma \mu^2}
\bigg (-2 + a \gamma r^2 (12+ 8 a\sigma ) \bigg )\ . \eea We
see that ${\partial V_F\over \partial \theta}$ vanishes for
$\theta=0$ or $\pi$.

The $\theta=0$ trajectory is delta-flat \cite{Burgess}. For this
trajectory, ${\partial^2  V_F\over \partial \theta^2}={1 \over 2}
V_1$ which is clearly positive for small $r$ and stays positive up
to some critical radius $r_c$. To compute $r_c$ we evaluate $V_1$
at $\sigma = \sigma_0$ using \beq 4 a \sigma_0 + 9 - 6 W_0
{e^{a\sigma_0}\over |A_0|} \approx 3-4s\, . \eeq We find \beq
\label{equ:V1approx} V_1 \approx  {\kappa^2 |A_0|^2 e^{-2
a\sigma_0}\over 3 U^2} \frac{r}{\gamma \mu^2} \bigg (2 + a\mu
\gamma (4s-3) \sqrt{r}  \cos \frac{\tilde \psi}{2} \bigg )\, .
\eeq
For $s>3/4$ and for real positive $\mu$,
the potential is minimized at $\tilde \psi = 2 \pi$, where
\beq
\label{equ:V1approx2}
V_1 \approx  {\kappa^2 |A_0|^2
e^{-2 a\sigma_0}\over 3 U^2} \frac{r}{\gamma \mu^2} \bigg (2 - a\mu \gamma (4s-3)
\sqrt{r}
\bigg )\ ,
\eeq
This is positive as long as $r$ is less than $r_c$, where
\beq
\label{equ:BurgessRC}
\frac{r_c}{r_\mu} = \frac{1}{(4s-3)^2} \left(\frac{9}{a \sigma_0} \right)^2 \frac{M_P^4}{\phi_\mu^4}\, .
\eeq
Applying the field range bound in the form $\frac{\phi_\mu^2}{M_P^2} < \frac{4}{N}$ one finds
\beq
\frac{r_c}{r_\mu} > \frac{N^2}{4^2} \frac{1}{(4s-3)^2}  \left(\frac{9}{a \sigma_0} \right)^2\, .
\eeq
For typical parameters we therefore conclude that $r_c \ge r_\mu$ and the delta-flat direction is hence stable from the tip to at least the location $r_{\mu}$ of the D7-branes.

For the $\theta=\pi$ trajectory, ${\partial^2 V_F\over
\partial \theta^2}=-{1\over 2} V_1$. This is negative for $r<r_c$ and the $\theta = \pi$ trajectory is therefore unstable in this regime.
This analysis was carried out for $n=1$ but it illustrates
the essential qualitative point (for a more general analysis with the same conclusion, see Appendix C.2).

In Appendix \ref{sec:HigherACR} we show that for all ACR
embeddings there are alternative trajectories with $\Phi \ne 0$
that are not delta-flat. This is important because it implies
that, for a D3-brane moving along such a trajectory, $\eta$ can be
different from $\frac{2}{3}$. We postpone a more general treatment of such
trajectories for the future (but see Appendix \ref{sec:HigherACR} for some preliminary remarks).

\section{Discussion}
\label{sec:discussion}

In this section we will briefly take stock of our progress towards an explicit model of D-brane inflation.  For this purpose, it is useful to consider the more general problem of deriving a low-energy Lagrangian from the data of a string compactification.

In principle, the data of a compactification -- such as the background geometry, brane positions, and fluxes -- determine the low-energy effective Lagrangian in full.  In practice, one typically begins by deriving the leading-order effective Lagrangian, which follows from dimensional reduction of the classical ten-dimensional supergravity action, including the effect of fluxes but treating D-branes as probes.  Then, one can include corrections to this action, including such things as nonperturbative terms in the superpotential, D-brane backreaction effects, string loop corrections to the K\"ahler potential, and $\alpha^{\prime}$ corrections to the K\"ahler potential.  Except in cases with extended supersymmetry, it is typically impossible to obtain results beyond leading order in either series of corrections to the K\"ahler potential.

For the present purpose, an instructive way to organize these corrections is according to their effects on the slow-roll parameter $\eta$, as follows.  The leading order classical four-dimensional Lagrangian we denote ${\cal L}_0$.  Correction terms are well-known to give rise to inflaton masses of order $H$, and hence corrections of order unity to $\eta$.  When all such effects, from any source whatsoever, have been added to ${\cal L}_0$, we denote this corrected Lagrangian ${\cal L}_1$.  By definition, this is the Lagrangian whose inflaton mass term is a good approximation to the `true' mass term that would follow from a dimensional reduction incorporating corrections of arbitrarily high degree.  (Notice that the leading-order classical Lagrangian ${\cal L}_0$ may itself contain large inflaton mass terms.)
Finally, if to ${\cal L}_1$ we add the leading terms that give corrections to $\eta$ that are parametrically small compared to unity, we call the resulting Lagrangian ${\cal L}_2$.  In sum, we propose to organize corrections to the Lagrangian according to the degree of their effects on $\eta$, even though such an organization does not correspond to a literal expansion parameter such as the string coupling.

To determine whether a given model gives rise to prolonged slow-roll inflation, one needs to know ${\cal L}_1$.  However, it is rarely true that all the required results are available.  For example, in the warped brane inflation model of \cite{KKLMMT}, the inflaton-dependence of the threshold factor $A(\phi)$ of the nonperturbative superpotential was not known until recently \cite{BHK,BDKMMM}.  Similarly, in K\"ahler moduli inflation ({\it{cf.}} the second reference in \cite{otherinflation}), a particular term in the K\"ahler potential that could give $\Delta \eta \sim 1$ has not yet been computed (though it has been conjectured that this term might vanish.)  Although such partial data is generally insufficient to determine whether a model is successful, even this degree of detail is relatively rare: a fair fraction of proposed models of string inflation include only the ${\cal L}_0$ data, without any corrections at all.

In this paper, we have progressed towards a full understanding of the Lagrangian ${\cal L}_1$ for the warped brane inflation models of \cite{KKLMMT}.  However, as we will now explain, further work is necessary.

First, let us briefly recall the best-understood correction terms.  The D3-brane potential receives contributions from the mixing between the volume and the D3-brane position in the DeWolfe-Giddings K\"ahler potential (\ref{equ:KKK}).  Moreover, the nonperturbative superpotential receives the correction (\ref{equ:A}) sourced by the backreaction of the D3-brane on the warp factor.  Holomorphy of the gauge kinetic function ensures that this correction, which corresponds to a one-loop threshold factor, is the only perturbative correction to the superpotential.  The only additional contributions to the superpotential come from multi-instantons, which give a negligibly small effect.

The K\"ahler potential, however, is not protected by holomorphy, and in general receives $\alpha'$ and $g_{s}$ corrections.  In the
large-volume, weak-coupling limit, these corrections are suppressed relative to the leading terms in the DeWolfe-Giddings
K\"ahler potential, and so generate mass terms that are generically smaller than $H$ by powers of the inverse volume or
powers of the string coupling.  Hence, by our above definition, these corrections correspond to terms in ${\cal L}_2$.
Although complete results for these terms are not available for a general compactification, the work of Berg, Haack,
and K\"ors \cite{BHKKahler,BHKKahlerTwo} in the toroidal orientifold case gives substantial guidance.  These authors found
that the leading D3-brane-dependent corrections to ${\cal K}$ are of two types, one suppressed by an additional power of $\rho$
compared to the DeWolfe-Giddings result, and the other suppressed by one power of $g_{s}$.  For $\rho \gg 1$, $g_{s} \ll 1$, these
terms give a parametrically small correction to the D3-brane potential, and so belong to what we have called ${\cal L}_2$.  Even better, in some cases \cite{BHKKahler} the expectations of naive dimensional analysis are borne out, and the numerical prefactors of these higher-order terms are moderately small.\footnote{We thank M. Berg for helpful discussions of this point.}

Corrections due to the fluxes are a further possibility. The best-understood $\alpha'$ corrections arise from the term
$(\alpha')^3 R^4$ in ten dimensions, with $R^4$ standing for an appropriate contraction of four powers of the Riemann tensor.  In
the presence of three-form flux $G_3$, there are additional terms from $(G_{3})^{2} R^{3}$, and it is less clear how these correct
the K\"ahler potential \cite{BBHL}.  However, it has been argued that at large volume, hence low flux density, this effect is subleading \cite{BBCQ}.

Finally, and perhaps most importantly, perturbations of the background fields in the throat, which can arise from sources in the bulk, can give substantial corrections to the D3-brane potential.\footnote{We thank S. Kachru for explanations of this point.}
It was argued in \cite{DMSU} (see also \cite{Krause}) that in special cases these effects may be small compared to the forces from the D7-brane.  However, lacking a more complete understanding of these effects, we do not claim that the potential we have presented is completely general.  Instead, our construction is representative of a particularly tractable class of situations in which the bulk effects are small.

\section{Conclusions}
\label{sec:Conclusion}

In this paper we have systematically studied the potential for a D3-brane in a warped throat containing holomorphically-embedded D7-branes.  As explained in \cite{KKLMMT}, this system is a promising candidate for an explicit model of inflation in string theory.  However, the warped brane inflation model of \cite{KKLMMT} is well-known to suffer from an inflaton mass problem, in that corrections from moduli stabilization tend to curve the potential and make slow-roll inflation impossible.  The true severity of this problem -- and, correspondingly, the status of the model -- have remained unclear, because the functional form of one particular correction term, arising from a threshold correction to the nonperturbative superpotential, was unavailable before the recent result of \cite{BDKMMM}.  In this work, building on \cite{BDKMMM}, we have studied the corrected potential in detail.  This equipped us to assess the true status of the warped brane inflation model \cite{KKLMMT} and to ascertain whether prolonged slow-roll inflation is indeed possible.

Our method for analyzing the potential involved several nontrivial improvements over existing approximations.  First, we systematically identified stable minima in the angular directions of the conifold, and showed how the radial potential depends on the choice of angular minimum.
Second, we showed that the common assumption that the compactification volume remains stabilized at its minimum during inflation is inadequate: the volume shrinks slightly as the D3-brane falls down the throat, and this leads to non-negligible corrections to the effective D3-brane potential.  We gave an analytic expression for the volume as a function of the D3-brane position and showed that this is an excellent approximation to the full result.

For a large class of embeddings of the wrapped branes, the ACR embeddings, we showed that the radial potential for a D3-brane at a particular angular extremum is necessarily too steep to support inflation, because the contributions computed in \cite{BDKMMM} vanish along the trajectory, and so the no-go result of \cite{KKLMMT} applies.  This was first explained in \cite{Burgess} for a special ACR embedding, the Ouyang embedding \cite{Ouyang}. Our results here generalize this to the full ACR class. It follows that these trajectories in throats containing wrapped branes with ACR embeddings do not permit slow roll D-brane inflation, even if one allows an arbitrary degree of fine-tuning: there is simply no parameter that can be varied to flatten the potential in such a case. However, we also showed that all ACR embeddings have alternative trajectories, corresponding to other choices of the angular extremum, for which this no-go result does not apply.

Our main result was an analysis of a very simple and symmetric embedding, the Kuperstein embedding, that does allow a flat D3-brane potential.
We found that for certain fine-tuned ranges of the compactification parameters, the potential is flat enough to allow prolonged inflation.
However, the resulting potential is not as simple as that conjectured in \cite{KKLMMT} and further elaborated in \cite{CloserTowards}: moduli stabilization gives rise to a potential that is much more complicated than a mass term for radial motion. Furthermore, adjusting the potential by varying microscopic parameters changes features in the potential instead of just rescaling the mass term.  In particular, one can fine-tune to arrange for a flat region suitable for inflation, but we found that this will occur around an inflection point away from the origin. Hence, when inflation occurs, it does so near an approximate inflection point, rather than in a shallow quadratic potential centered on the origin.  In short, we found that a low-order Taylor expansion of the potential around the minimum at the tip of the throat does not properly describe those regions of the potential where inflation is possible.  Instead, one is obliged to use the complete potential presented here.  Our result implies that the phenomenology of some classes of warped D-brane inflation models is well-described by an effective single-field potential with a constant and cubic term.  As we explained, models of this sort (see \cite{MSSM} for an analogous example in the MSSM) are particularly sensitive to the initial conditions.  Moreover, the tilt of the scalar spectrum is exquisitely sensitive to the slope of the potential near the approximate inflection point.

One important, general lesson of our work is that there is considerably less freedom to adjust the parameters of this system than one might expect from the low-energy effective action.  First, the functional form of $A(\phi)$ derived in \cite{BDKMMM} is rather special: most importantly, $A(\phi)$ contains no quadratic term, and does not lead to any new quadratic terms in the inflaton potential.  This implies that the potential cannot be flattened uniformly, and $\eta$ can only be small in a limited region.  Second, the range of $\phi$ is limited by the microscopic constraint of \cite{BM}.  Finally, the field range is linked to the scale of inflation, because of a new geometric constraint linking the size of the wrapped four-cycle and the length of the throat.  Using these results, we found that although this system depends on many microscopic parameters, in many cases it is nevertheless impossible to choose these parameters in such a way that the slow-roll parameter $\eta$ is fine-tuned to vanish.  This occurs because constraints originating in the geometry of the compactification correlate the microscopic quantities, so that the true number of adjustable parameters is much smaller than a naive estimate would suggest.

An important direction for future work is a more comprehensive understanding of any additional corrections to the potential, such as $\alpha'$ corrections, string loop corrections, and perturbations of the throat metric due to bulk sources.  We have argued that in some cases the presence of such effects is not fatal for inflation, but precise observational predictions will certainly depend on these effects.

Finally, there is a pressing need for a more natural model of string inflation than the one we have presented here.

\section*{Acknowledgments}
We are grateful to Paul Steinhardt for collaboration on the
companion paper \cite{ShortPaper} and for many useful discussions
of this paper. It is a pleasure to thank Marcus Berg, Cliff
Burgess, Jim Cline, Joe Conlon, Oliver DeWolfe, Steve Giddings, Shamit Kachru, Renata
Kallosh, Justin Khoury, Juan Maldacena, Arvind Murugan, Enrico Pajer, Sudhakar Panda, Joe
Polchinski, Gary Shiu, Eva Silverstein, Shinji Tsujikawa, Henry Tye, Bret Underwood,
and Herman Verlinde for many useful discussions and insightful
comments.  We especially thank Shamit Kachru for detailed comments
on a draft. D.B. thanks Julian Baumann for assistance with Figure
1.
D.B. thanks Dragan Huterer, Hiranya Peiris, and the KICP, Chicago for their hospitality while this work was completed.
A.D. would like to thank the Galileo Galilei Institute for
Theoretical Physics, the Perimeter Institute, and the University of
Chicago for their hospitality. The research of A.D. was supported
in part by Grant RFBR 07-02-00878, and Grant for Support of
Scientific Schools NSh-8004.2006.2.  L.M. is grateful to the KITP
Santa Barbara and the theory groups at the Perimeter Institute,
the University of Pennsylvania, the University of Wisconsin,
Cornell, Stanford, the University of Texas, the University of
Michigan, Texas A\&M, Harvard, and Yale for their hospitality. This
research was supported in part by the National Science Foundation
under Grant No. PHY-0243680 and by the Department of Energy under
grant DE-FG02-90ER40542. Any opinions, findings, and conclusions
or recommendations expressed in this material are those of the
authors and do not necessarily reflect the views of these funding
agencies.


\newpage
\appendix

\section{Background Geometry}
\label{sec:conifold}

\addtocontents{toc}{\SkipTocEntry}
\subsection{The singular conifold}

The singular conifold is a non-compact Calabi-Yau threefold
defined by the constraint equation \cite{Candelas}
\beq
\label{equ:constraint} \sum_{i=1}^4 z_i^2 = 0\, , \eeq
where
$\{z_i; i=1,2,3,4\}$ are complex coordinates in $\mathbb{C}^4$.
The conifold constraint (\ref{equ:constraint}) may be written as $\det(W) = 0$ where
\beq W
= \frac{1}{\sqrt{2}} \left( \begin{array}{cc}
z_3+iz_4 & z_1 -i z_2 \\ z_1 + i z_2 & -z_3 + i z_4
\end{array} \right) \equiv
\left( \begin{array}{cc} w_3 & w_2 \\ w_1 & w_4 \end{array}
\right) \, . \eeq
The K\"ahler potential (\ref{Kpot}) is chosen such that the K\"ahler metric on the conifold,
$k_{\alpha \overline{\beta}} \equiv \partial_\alpha
\partial_{\overline{\beta}} k$, is Calabi-Yau (Ricci-flat)
\bea
\d s^2 &=& \partial_\alpha \partial_{\overline{\beta}}k \, \d u^\alpha \d \overline{u^{\beta}} \nonumber \\
&=& k''\,  |{\rm Tr}(W^\dagger \d W)|^2 + k' \, {\rm Tr}(\d W \d
W^\dagger) \label{equ:cmetric} \, , \eea where $(\dots)' \equiv
\frac{d(\dots)}{d r^3}$ and $\{u_\alpha; \alpha = 1,2,3 \}$ are three
complex coordinates on the conifold, {\it e.g.} $u_\alpha =
z_\alpha$.
The metric on the conifold (\ref{equ:cmetric}) may be
cast in the form (\ref{conicform})
where \beq \d s_{T^{1,1}}^2 = \frac{1}{9} \Bigl( \d \psi +
\sum_{i=1}^2 \cos \theta_i \d \phi_i \Bigr)^2 + \frac{1}{6}
\sum_{i=1}^2 (\d \theta_i + \sin^2 \theta_i \d \phi_i^2)\, , \eeq is
the metric on the Einstein space $T^{1,1}$. The complex
coordinates $z_i$ are related to the real coordinates $\{ r \in [0, \infty],
\theta_i \in [0,\pi], \phi_i \in [0,2\pi],\psi \in [0,4\pi] \}$
via
{\small
\begin{eqnarray}
\label{equ:z1}
z_1 &=& \frac{r^{3/2}}{\sqrt{2}} e^{\frac{i}{2} \psi} \left[ \cos \left(\frac{\theta_1+\theta_2}{2}\right)   \cos \left(\frac{\phi_1+\phi_2}{2}\right) + i   \cos \left(\frac{\theta_1-\theta_2}{2}\right)   \sin \left(\frac{\phi_1+\phi_2}{2}\right) \right],\\
z_2 &=& \frac{r^{3/2}}{\sqrt{2}} e^{\frac{i}{2} \psi} \left[ -\cos \left(\frac{\theta_1+\theta_2}{2}\right)   \sin \left(\frac{\phi_1+\phi_2}{2}\right) + i   \cos \left(\frac{\theta_1-\theta_2}{2}\right)   \cos \left(\frac{\phi_1+\phi_2}{2}\right) \right],\\
z_3 &=& \frac{r^{3/2}}{\sqrt{2}} e^{\frac{i}{2} \psi} \left[ -\sin \left(\frac{\theta_1+\theta_2}{2}\right)   \cos \left(\frac{\phi_1-\phi_2}{2}\right) + i   \sin \left(\frac{\theta_1-\theta_2}{2}\right)   \sin \left(\frac{\phi_1-\phi_2}{2}\right) \right],\\
z_4 &=& \frac{r^{3/2}}{\sqrt{2}} e^{\frac{i}{2} \psi} \left[ -\sin
\left(\frac{\theta_1+\theta_2}{2}\right)   \sin
\left(\frac{\phi_1-\phi_2}{2}\right) - i   \sin
\left(\frac{\theta_1-\theta_2}{2}\right)   \cos
\left(\frac{\phi_1-\phi_2}{2}\right) \right].\label{equ:z4}
\end{eqnarray} }
\noindent
The complex coordinates $w_i$ are related to the real
coordinates $\{r, \theta_i,\phi_i,\psi\}$ via
\begin{eqnarray}
w_1 &=& r^{3/2} e^{\frac{i}{2}(\psi-\phi_1-\phi_2)} \sin \frac{\theta_1}{2} \sin \frac{\theta_2}{2} \label{equ:w1}\, ,\\
w_2 &=& r^{3/2} e^{\frac{i}{2}(\psi+\phi_1+\phi_2)} \cos \frac{\theta_1}{2} \cos \frac{\theta_2}{2}\, ,\\
w_3 &=& r^{3/2} e^{\frac{i}{2}(\psi+\phi_1-\phi_2)} \cos \frac{\theta_1}{2} \sin \frac{\theta_2}{2}\, ,\\
w_4 &=& r^{3/2} e^{\frac{i}{2}(\psi-\phi_1+\phi_2)} \sin
\frac{\theta_1}{2} \cos \frac{\theta_2}{2} \label{equ:w4}\, .
\end{eqnarray}

\addtocontents{toc}{\SkipTocEntry}
\subsection{Making $SO(4)$ symmetry manifest}
\label{sec:SO4}

It is sometimes convenient to write the
F-term potential (\ref{equ:Fterm}) in terms of the
four
homogeneous
coordinates $z_i$ of the embedding space $\mathbb{C}^4$ which makes the $SO(4)$ symmetry of the
conifold transparent.
For that reason we define a new metric $\widehat{\cal K}^{A \overline{B}}$ which depends on $z_i$ in such a way that
for any function $W(z_i)$ the following identity is satisfied \beq
\label{eau:constraint_} D_A W
\widehat{\cal K}^{A \overline{B}} \overline{D_B W} \equiv D_\Sigma W
{\cal K}^{\Sigma \overline{\Omega}} \overline{D_\Omega W}\, , \eeq
where $\{ Z^A\} \equiv \{ \rho, z_i; i = 1,2,3,4\}$ and $\{Z^\Sigma \} \equiv \{\rho, z_\alpha; \alpha = 1,2, 3 \}$. In this equation the
conifold constraint, $z_4^2=z_4^2(z_\alpha) = - \sum_{\alpha=1}^3 z_\alpha^2 $, is
substituted {\it after} differentiation on the left-hand side and
 {\it before} differentiation on the right-hand side.
The metric $\widehat{\cal K}^{A \overline{B}}(z_i)$ defined through (\ref{eau:constraint_})
is not unique and the choice of one over another is a matter of convenience.
 We construct $\widehat{\cal K}^{A \overline{B}}$ with the help of the auxiliary matrix $J^A_{\ \ \Sigma}$
  \beq \widehat{\cal K}^{A
\overline{B}} = J^A_{\ \ \Sigma} \, {\cal K}^{\Sigma
\overline{\Omega}} \, J^{\overline{B}}_{\ \ \overline{\Omega}}\, ,
\eeq where $J^A_{\ \ \Sigma}$ is defined as follows
\beq D_\Sigma W = \frac{\partial Z^A}{ \partial
Z^\Sigma} D_A W \equiv J^A_{\ \ \Sigma}\, D_A W\, , \quad \quad \quad
J^A_{\ \ \Sigma}=\left(
\begin{array}{c|c} 1 & 0 \\ \hline
0 & \delta_{i\alpha}\\
0& { - z_\alpha\over \sqrt{ - \sum_{\gamma = 1}^3 z_\gamma^2}}\\
\end{array}\right). \eeq
This gives ${\cal \widehat{K}}^{A \overline{B}}$ as a function of $z_\alpha$.
To find it as a function of $z_i$ guess a ${\cal \widehat{K}}^{A \overline{B}}(z_i)$
such that it reduces to ${\cal \widehat{K}}^{A \overline{B}}(z_\alpha)$ after substituting the conifold constraint.
This step and hence ${\cal \widehat{K}}^{A \overline{B}}(z_i)$  is not unique.
Nevertheless finding an $SO(4)$--invariant ${\cal \widehat{K}}^{A \overline{B}}(z_i)$
is not difficult, {\it e.g.} replacing  $\Bigl(-\sum_{\gamma=1}^3 z_\gamma^2 \Bigr)^{1/2}$ by $z_4$ everywhere in
${\cal \widehat{K}}^{A \overline{B}}(z_\alpha)$ and $J^A_{\ \ \Sigma}$ we find
\beq  {\cal \widehat{K}}^{A \overline{B}}  =   \frac{\kappa^2 U}{3} \left(
\begin{array}{c|c} U + \gamma k_{l} \hat{k}^{l \overline{m}}
k_{\overline{m}} & k_l \hat{k}^{l \overline{\jmath}}  \\ \hline
\hat{k}^{i \overline{m}} k_{\overline{m}}
& \frac{1}{\gamma} \hat{k}^{i \overline{\jmath}}\\
\end{array}\right)\, ,
\eeq
where \beq \label{equ:ki} k_{i} = \frac{\overline{z_i}}{r}\, , \eeq and
\beq
\label{equ:kij} \hat{k}^{i \overline{\jmath}} = J^i_{\ \alpha} \,
k^{\alpha \overline{\beta}} \, J^{\overline{\jmath}}_{\
\overline{\beta}} = r\, \left[ \delta^{i \overline{\jmath}} +
\frac{1}{2} \frac{z_i \overline{z_j}}{r^3} - \frac{\overline{z_i}
z_j}{r^3}\right]\, . \eeq Notice that $\hat{k}^{i \overline{\jmath}}$
is {\it not} the inverse of
$k_{i \overline{\jmath}} = \frac{1}{r}\,
\Bigl[ \delta_{i \bar \jmath} - \frac{1}{3} \frac{\overline{z_i} z_j}{r^3}
\Bigr]$, which is $ k^{i \overline{\jmath}} = r \, \Bigl[
\delta^{i \overline{\jmath}} + \frac{1}{2} \frac{z_i \overline{z_j}}{r^3} \Bigr]\,
. $
From (\ref{equ:ki}) and (\ref{equ:kij}) one then finds
\beq
k_{l} \hat{k}^{l \bar \jmath} = \frac{3}{2} \overline{z^j}\, ,
\quad \quad
k_{l} \hat{k}^{l \overline{m}}
k_{\overline{m}} = \frac{3}{2} r^2 = \hat r^2 = k \, , \eeq and hence,
\beq {\cal \widehat{K}}^{A \overline{B}}  = \frac{\kappa^2 U}{3} \left( \begin{array}{c|c}
\rho + \overline{\rho} &  \frac{3}{2} \overline{z_j} \\ \hline
\frac{3}{2} z_i
 & \frac{r}{ \gamma}\, \Bigl[ \delta^{i \overline{\jmath}} + \frac{1}{2} \frac{z_i \overline{z_j}}{r^3} - \frac{\overline{z_i} z_j}{r^3} \Bigr] \\
\end{array}\right)\, .
\eeq Using the above results we arrive at the F-term potential
\beq \label{equ:VVVF} V_F = \frac{\kappa^2}{3 U^2} \left[ (\rho +
\overline{\rho}) |W_{,\rho}|^2 - 3 (\overline{W} W_{, \rho} +
c.c.) + \frac{3}{2} (\overline{W_{,\rho}} z^i W_{,i} + c.c.) +
\frac{1}{\gamma} \hat{k}^{i \overline{\jmath}} W_{,i}
\overline{W_{,j}}  \right]\, . \eeq
The result (\ref{equ:VVVF}) is essential
for the main analysis presented in this paper. In terms of the
$w_i$-coordinates the F-term potential (\ref{equ:VVVF}) is \beq
\label{equ:VVVVF} V_F = \frac{\kappa^2}{3 U^2} \left[ (\rho +
\overline{\rho}) |W_{,\rho}|^2 - 3 (\overline{W} W_{, \rho} +
c.c.) + \frac{3}{2} (\overline{W_{,\rho}} w^i W_{,i} + c.c.) +
\frac{1}{\gamma} \hat{k}_w^{i \overline{\jmath}} W_{,i}
\overline{W_{,j}}  \right]\, , \eeq where \beq \hat{k}_w^{i
\overline{\jmath}}  = r\, \left[ \delta^{i \overline{\jmath}} +
\frac{1}{2} \frac{w_{i} \overline{w_{j}}}{r^3} -
\frac{c_i^{i'}c_j^{j'}\overline{w_{i'}} w_{j'}}{r^3}\right]\, .
\eeq The matrix $c^{i'}_j$ has only four non-zero elements
$c^1_2=c^2_1=1$ and $c^3_4=c^4_3=-1$.

\section{Dimensional Reduction}
\label{sec:reduc}

As explained in \cite{BDKMMM} and reviewed in \S\ref{sec:potential} of this paper, D7-branes wrapping certain four-cycles in the compactification source nonperturbative effects that stabilize the volume modulus.
In \S\ref{sec:rho} we identify the real part of the K\"ahler modulus with the warped volume of the four-cycle and give a detailed derivation of the DeWolfe-Giddings K\"ahler potential.
These results are well known, but our goal here is to fix notation with enough care to permit precise discussions elsewhere in the paper.
\S\ref{sec:4vol} computes the throat contribution to the warped four-cycle volume for our setup and relates it to microscopic compactification data.
We also explain how this relates microscopic input to the energy scale of inflation.
Finally, in \S\ref{sec:bound}, we present an improved derivation of the field range bound of \cite{BM} that includes a non-trivial breathing mode of the compactification.

\addtocontents{toc}{\SkipTocEntry}
\subsection{K\"ahler modulus and K\"ahler potential}
\label{sec:rho}

We consider the line element
\beq
\label{equ:back}
\d s^2 = h^{-1/2}(y) e^{-6u} g_{\mu \nu} \d x^\mu \d x^\nu + h^{1/2}(y) e^{2u} g_{\alpha \bar{\beta}}
 \d y^\alpha \d \overline{y^\beta}\, ,
\eeq
where $g_{\mu \nu}$ is the four-dimensional Einstein frame metric, $h$ is the warp factor, $g_{\alpha \bar{\beta}}$ is a fiducial metric on the internal space, and the factor $e^{u}$ extracts the breathing mode.

Before proceeding, let us explain the division of the metric $\tilde g_{\alpha \bar{\beta}} \equiv e^{2u} g_{\alpha \bar{\beta}}$ into a fiducial metric $g$ and a breathing mode $e^{2u}$.  These two objects do play different roles: note in particular that although $g$ affects the four-dimensional Planck mass
\beq M_P^2 = \frac{1}{\pi}  (T_3)^2 \int \d^6 y \sqrt{ g}\, h  \equiv   \frac{1}{\pi}  (T_3)^2 V_6^w\, , \eeq the breathing mode
does not, because of the factor $e^{-6u}$ in the spacetime term of (\ref{equ:back}).

For any fixed location $Y \equiv Y_{\alpha}$ of the D3-brane, there is a minimum $\rho_\star(Y)$ of the nonperturbatively-generated potential for the K\"ahler modulus $\rho$.  (We will soon come to a precise definition of $\rho$ in terms of the fields in (\ref{equ:back}).)  We argued in \S\ref{sec:Kuperstein} for an adiabatic approximation in which, as the D3-brane moves and hence the location $\rho_\star(Y)$ of the instantaneous minimum changes, $\rho$ moves to remain in this instantaneous minimum $\rho_\star(Y)$.  In terms of the fields in (\ref{equ:back}), this is most conveniently represented\footnote{At the end of this section we will verify that choosing a different normalization of the fiducial metric does not affect any physical quantities.} by fixing the fiducial metric once and for all, but allowing the breathing mode $e^{u(Y,\bar{Y})}$ to keep track of the change in the volume that is due to the displacement of the D3-brane.

To this end, we normalize the fiducial metric $g$ to correspond to $\rho_\star(0)$, {\it{i.e.}} so that the volume computed with $g$ is precisely the physical volume of the internal space when the D3-brane sits at the tip of the throat, $Y \approx 0$.  Then, $e^{u(Y,\bar{Y})}$ represents the change in the volume that is due to the displacement of the D3-brane away from the tip, and by this definition, $u(0)=0$.
In the throat region, the fiducial metric takes the form
\beq
\label{equ:tm}
 g_{\alpha \bar{\beta}}
 \d y^\alpha \d \overline{y^\beta} = \d \hat r^2 + \hat r^2 \d s^{2}_{T^{1,1}}\, .
\eeq

The nonperturbative superpotential arising from strong gauge dynamics on $n$ D7-branes (or Euclidean D3-instantons) wrapping a four-cycle $\Sigma_4$ in the background (\ref{equ:back}) is \cite{BDKMMM}
\beq |W_{\rm np}|^2 \propto \exp \left[ - \frac{2 T_3
V_{\Sigma_4}^w e^{4u}}{n}\right]\, , \eeq
where $ V_{\Sigma_4}^w \equiv \int \d^4 \xi \sqrt{ g^{ind}} \, h$.
Before the mobile D3-brane enters the throat region, the (warped) four-cycle volume is $( V_{\Sigma_4}^w)_0$.
When the D3-brane is in the throat, it induces a perturbation in the warp factor, $\delta h$, which sources a change in the warped four-cycle volume, $\delta V_{\Sigma_4}^w(Y)$. (This change in the warped volume is logically distinct from the shift in the stabilized value of the K\"ahler modulus studied in Appendix \ref{sec:sigma}.)
Hence,
\bea
|W_{\rm np}|^2 &\propto& \exp \left[ - \frac{2 T_3 \delta
V_{\Sigma_4}^w(Y) e^{4u(Y, \bar Y)}}{n}\right]  \exp \left[ - \frac{2 T_3 (V_{\Sigma_4}^w)_0 e^{4u(Y,\bar Y)}}{n}\right]  \\
&\equiv& |A(Y)|^2 e^{-a(\rho + \bar \rho)}\, .
\eea
We now define $a \equiv {2 \pi \over n}$ and identify the K\"ahler modulus \cite{BDKMMM}
\beq \label{equ:defineGamma}
\rho + \bar \rho \equiv \Gamma e^{4 u (Y, \bar Y)} + \gamma k(Y,\bar Y)\, , \quad \quad \Gamma \equiv  \frac{T_3 (V_{\Sigma_4}^w)_0}{\pi} \, .
\eeq
Here $k$ is the little K\"ahler potential
for the fiducial metric on the conifold,
$\partial_\alpha \partial_{\bar \beta}  k \equiv  g_{\alpha \bar \beta}$.
The term proportional to $k$ in (\ref{equ:defineGamma}) has to be added to make $\rho$ holomorphic; see \cite{GM,BDKMMM} for extensive discussions of this issue.  Because $u(0)= k(0)=0$, we may relate the parameter $ \Gamma$ to the value of the stabilized K\"ahler modulus $\rho$ when the D3-brane is at the tip of the throat,
\beq
\label{equ:Gamma}
\Gamma \equiv \rho_\star(0)+\bar{\rho}_\star(0) = 2 \sigma_\star(0) \equiv 2 \sigma_0\, .
\eeq
The range of allowed values for $\sigma_0$ is constrained by the throat contribution to the warped four-cycle volume, which we will compute in the next section.

With these definitions the DeWolfe-Giddings K\"ahler potential (see footnote 5 for comments on possible corrections to this K\"ahler potential) is
\beq
\label{equ:K}
{\cal K} = - 3 M_P^2 \log \left[\rho + \bar \rho - \gamma k(Y, \bar Y) \right] = - 3 M_P^2 \log \left[ \Gamma e^{4 u(Y, \bar Y)} \right] \, .
\eeq
To determine the constant $ \gamma$, we compare the kinetic terms derived from the DBI action,
 \beq
{\cal L}_{\rm kin}^{\rm DBI} = - \frac{1}{2} T_3 e^{-4u} g_{\alpha \bar \beta} \partial_\mu Y^\alpha
\partial^\mu \overline{Y^\beta}\, ,  \label{equ:DBIkin} \eeq
with the kinetic terms\footnote{Of course, the complete kinetic terms for the D3-brane coordinates and the volume modulus are
\bea
{\cal L}_{\rm kin} &\equiv& - {\cal K}_{\rho \overline{\rho}} \partial_\mu \rho \partial^\mu \overline{\rho} -{\cal K}_{Y_\alpha \overline{Y_\beta}} \partial_\mu Y^\alpha \partial^\mu \overline{Y^\beta}
-  \left( {\cal K}_{\rho \overline{Y_\beta}} \partial_\mu \rho \partial^\mu \overline{Y^\beta} + c.c. \right) \nonumber \\
&=&
3 M_P^2 \frac{\gamma k_{\alpha \overline{\beta}} \partial_\mu Y^\alpha \partial^\mu \overline{Y^\beta}}{U} \nonumber\\
&& - 3 M_P^2 \left| \frac{\partial_\mu \rho}{U} \right|^2
 +3 M_P^2  \left|\frac{\gamma k_{\alpha} \partial_\mu \overline{Y^\alpha}}{U} \right|^2 +  3 M_P^2 \left( \frac{\partial_\mu \rho}{U} \frac{\gamma k_{\overline{\beta}} \partial^\mu \overline{Y^\beta}}{U} + c.c. \right) \nonumber \\
&\approx& \frac{3 M_P^2 \gamma k_{\alpha \overline{\beta}} \partial_\mu Y^\alpha \partial^\mu \overline{Y^\beta}}{U}\, , \nonumber
\eea
In the final relation we have focused attention on a subset of the kinetic terms for the D3-brane coordinates.  This is justified for several reasons.  First, in this paper we are specifically searching for (fine-tuned) configurations in which the D3-brane potential is unusually flat.  In such a case, it is consistent to use an adiabatic approximation for the motion of the volume modulus, and omit the kinetic terms for $\rho$.  Next, the term involving $|k_{\alpha} \partial_\mu \overline{Y^\alpha}|^2$ is suppressed, relative to the term we have retained, by $U^{-1} \ll 1$.  Furthermore, $k_{\alpha}$ vanishes at the tip of a singular conifold, and is correspondingly very small at the tip of the deformed conifold we are considering.  Finally, from the explicit form of $k$ we learn that the term $|k_{\alpha} \partial_\mu \overline{Y^\alpha}|^2$ is suppressed by the small quantity $(\phi/M_P)^2$ relative to the term we have retained.} derived from (\ref{equ:K}),
\bea
{\cal L}_{\rm kin}^{\cal K} &=& - {\cal K}_{\alpha \bar \beta} \partial_\mu Y^\alpha
\partial^\mu \overline{Y^\beta} \nonumber \\
&\approx&  3 M_P^2 e^{-4u} \partial_\alpha \partial_{\bar \beta} e^{4u} \partial_\mu Y^\alpha
\partial^\mu \overline{Y^\beta} \label{equ:Kkin} \, ,
\eea
where
\beq
\partial_\alpha \partial_{\bar \beta} e^{4u} = - \frac{\gamma}{\Gamma} \partial_\alpha \partial_{\bar \beta}  k = - \frac{\gamma}{\Gamma} g_{\alpha \bar \beta}\, .
\eeq
Equations (\ref{equ:DBIkin}) and (\ref{equ:Kkin}) are consistent if we define
\beq
\gamma \equiv \frac{\Gamma}{6} \frac{T_3}{M_P^2} = \frac{1}{6} \frac{({V}_{\Sigma_4}^w)_0}{{V}_6^w}  \, .
\eeq
Using equation (\ref{equ:Gamma}) this may be written in the useful form
\beq
\gamma = \frac{\sigma_0}{3} \frac{T_3}{M_P^2}\, .
\eeq
We now notice that the physical quantities of interest, such as $\rho$ and $ \gamma {k}$, are independent of the split into breathing mode $e^{u(0)}$ and fiducial metric $g$ in (\ref{equ:back}).
For example,
\beq
 \gamma  k = \frac{1}{6} \frac{T_3}{ [M_P^2 e^{6 u}]} \frac{T_3 [(V_{\Sigma_4}^w)_0 e^{4u}]}{\pi} [e^{2 u} k] = \frac{1}{6} \frac{(\tilde V_{\Sigma_4}^w)_0}{\tilde V_6^w} \tilde k \equiv \tilde \gamma \tilde k \, ,
\eeq
where we have defined
\beq
\tilde k \equiv e^{2u} k = e^{2u} r^2 = \tilde r^2 \, ,
\eeq
and
\beq
\tilde \gamma \equiv e^{-2u} \gamma =  \frac{\tilde \Gamma}{6} \frac{T_3}{ [M_P^2 e^{6 u}]} = \frac{1}{6} \frac{(\tilde V_{\Sigma_4}^w)_0}{\tilde V_6^w} \, ,
\eeq
where \beq \tilde \Gamma \equiv \Gamma e^{4u} = 2\sigma_0 e^{4u} \, . \eeq
This shows that $\tilde \gamma \tilde k$ is invariant under the change of conventions
\beq
e^{2u(0)} \to \lambda e^{2u(0)}\, , \qquad  g_{\alpha\bar{\beta}} \to \lambda^{-1} g_{\alpha\bar{\beta}}\, .
\eeq
One easily sees that $\rho+\bar{\rho}$ is likewise invariant.  We have therefore justified our original choice of the convenient value $u(0)=0$.

\addtocontents{toc}{\SkipTocEntry}
\subsection{Warped four-cycle volume}
\label{sec:4vol}

The four-cycle $\Sigma_4$ wrapped by the D7-branes is, in principle, compact.
However, we have access to explicit metric information only in the throat region, but can still make progress by deriving results that are largely independent of the unknown bulk region.  The four-cycle volume receives
contributions both from the throat region and the bulk, $(V_{\Sigma_4}^{w})_0 =  (V_{\Sigma_4}^{w})_{0, \rm throat} +  (V_{\Sigma_4}^{w})_{0, \rm bulk}$.  In the remainder of this section we compute
$(V_{\Sigma_4}^{w})_{0, \rm throat}$ for various embeddings and
use  $(V_{\Sigma_4}^{w})_0 \equiv B_4
(V_{\Sigma_4}^{w})_{0, \rm throat}$, with $B_4 > 1$, to
parameterize the unknown bulk contribution.
In the non-compact limit, $(V_{\Sigma_4}^{w})_{0, \rm throat}$ diverges.  Here we identify the leading order divergence and regularize the throat volume by introducing the UV cutoff $\hr_{\rm UV}$, where the throat attaches to the bulk and $h(\hr_{\rm UV}) \approx 1$.
We use the following approximation to the warp factor \cite{KT, HKO},
\beq
\label{equ:hKT}
h \approx \frac{L^4}{\hr^4} \log \frac{\hr}{\varepsilon^{2/3} }\, ,
\eeq
where
\beq
\label{equ:L}
2 T_3 L^4 = \frac{27}{16 \pi^2} \left( \frac{3 g_s M}{2 \pi K} \right) N\, ,
\quad \quad \log Q_0 \equiv \log \frac{\hr_{\rm UV}}{\varepsilon^{2/3}} \approx \frac{2 \pi K}{3 g_s M}\, .
\eeq
At large radius $\hr$ the D7-brane wraps an $S^3$ and the metric on the four-cycle is
\beq
\d s^2_{\Sigma_4} \to \d \hr^2 + \hr^2 \d s^2_{S^3}\, .
\eeq
The warped four-cycle volume then becomes
\bea
(V_{\Sigma_4}^{w})_0 \equiv \int \d^4 \xi \sqrt{g^{ind}} h &=& B_4 L^4 {\rm Vol}(S^3)
\int_{\hr_\mu}^{\hr_{\rm UV}} \frac{\d \hr}{\hr} \log \frac{\hr}{\varepsilon^{2/3}} \\
&=& B_4 L^4 {\rm Vol}(S^3)  \log Q_\mu \left[ \log Q_0 - \frac{1}{2} \log Q_\mu \right]\, ,
\eea
where $Q_\mu \equiv \frac{\hr_{\rm UV}}{\hr_\mu} > 1$.
The numerical value of ${\rm Vol}(S^3)$ depends on the specific embedding.
For the Kuperstein embedding we have \cite{Kuperstein}
\beq
{\rm Vol}(S^3) = \frac{16 \pi^2}{9}\, ,
\eeq
and hence, by (\ref{equ:defineGamma}),
\beq
\Gamma = \frac{3}{2 \pi} B_4 N \log Q_\mu \left[ 1 - \frac{1}{2} \frac{\log Q_\mu}{\log Q_0} \right]\, .
\eeq
This implies
\bea
\omega_0 \equiv a \sigma_0 &=&
\frac{3}{2} B_4 \left(\frac{N}{n} \right)
\log Q_\mu \left[ 1 - \frac{1}{2} \frac{\log Q_\mu}{\log Q_0} \right]\, , \\
&\approx& \frac{3}{2} B_4 \left(\frac{N}{n} \right)
\log Q_\mu\, .
\eea
Since, $B_4 > 1$, $N/n > 1$ and $Q_\mu > 1$, this can assume a range of values, with $\omega_0 = {\cal O}(10)$ being easily achievable.
The value of $\omega_0$ is important as it determines the scale of inflation
\beq
\frac{V_{dS}}{M_P^{4}} \sim \frac{e^{-2 \omega_0}}{2 \omega_0} \, .
\eeq
In particular, for $B_4 = {\cal O}(1)$ and $\log Q_\mu = {\cal O}(1)$, constraints on the minimal phenomenologically viable  inflation scale, $V_{dS} > {\cal O}({\rm TeV}^4) \sim 10^{-60} M_P^4$ or $\omega_0 \lesssim 150 $, translate into an upper limit on the background five form flux
\beq
\frac{N}{n} < {\cal O}(10^2)\, .
\eeq
This can be a serious constraint on nonperturbative volume stabilization by the KKLT mechanism.

Finally, the ACR \cite{ACR} embeddings, $\prod_{i=1}^4 w_i^{p_i} = \mu^P$, satisfy
\beq
{\rm Vol}(S^3) = \frac{16 \pi^2}{9} P\, ,
\eeq
and hence
\beq
\label{equ:ACRSigma}
\omega_0 \approx \frac{3}{2} P B_4 \left(\frac{N}{n} \right)
\log Q_\mu\, .
\eeq
Possible corrections to these results coming from induced magnetic fields on the D7-branes are discussed in Appendix \ref{sec:magnetic}.
There we show that the result for the Kuperstein four-cycle is in fact unchanged, while ACR four-cycles may receive a correction. However, even for the ACR case this would only affect numerical details, and in any event this uncertainty in the precise ACR result may be absorbed in the uncertainty of the value of $B_4$.

\addtocontents{toc}{\SkipTocEntry}
\subsection{Canonical field range}
\label{sec:bound}

\addtocontents{toc}{\SkipTocEntry}
\subsubsection{Canonical inflaton}

The inflaton action includes the kinetic term
\beq
{\cal L}_{\rm kin} = - \frac{1}{2} T_3 e^{-4u(\hat r)} (\partial_\mu \hat r)^2  \equiv - \frac{1}{2} (\partial_\mu \varphi)^2\, ,
\eeq
so
the canonical inflaton field is
\beq
\label{equ:CanPhi}
\varphi = \int  \left( T_3 \frac{\Gamma}{\tilde \Gamma(\hat r)} \right)^{1/2} \d \hat r\, .
\eeq
For analytical considerations the following approximation is often sufficient (see Appendix \ref{sec:sigma} for more accurate analytical results)
\beq
\varphi^2 \approx \frac{\Gamma}{\tilde \Gamma(\hat r)} T_3 \hat r^2  \approx \frac{2 \sigma_0}{\rho+ \bar \rho} T_3 \hat r^2 \approx T_3 \hat r^2 \equiv \phi^2\, .
\eeq
This is independent of the split into breathing mode and fiducial metric in  (\ref{equ:back}):
\beq
\frac{\varphi^2}{M_P^2} = \frac{[\Gamma e^{4u}]}{\tilde \Gamma} \frac{T_3 [e^{2u} \hr^2]}{[M_P^2 e^{6u}]} = \frac{ \tilde r^2}{\frac{1}{\pi} T_3 \widetilde V^w_6}\, .
\eeq
This implies
\beq
\frac{\gamma k}{\rho + \bar \rho} = \frac{\Gamma}{\rho + \bar \rho} \frac{T_3 \hat r^2}{ 6 M_P^2} \approx
\frac{\Gamma}{\tilde \Gamma} \frac{T_3 \hat r^2}{ 6 M_P^2} =
\frac{1}{6} \frac{\varphi^2}{M_P^2}
\eeq
and the DeWolfe-Giddings K\"ahler potential becomes separable
\beq
\kappa^2 {\cal K} \approx - 3 \log(\rho + \bar \rho) - 3 \log u\, , \quad \quad u \equiv 1 - \frac{1}{6} \frac{\varphi^2}{M_P^2}\, .
\eeq

\addtocontents{toc}{\SkipTocEntry}
\subsubsection{Bound on the field range}

We now derive the microscopic bound on the inflaton field range in four-dimensional Planck units \cite{BM}. Recall that the Planck mass depends on the warped volume of the internal space as
\beq
M_P^2 = \frac{1}{\pi} (T_3)^2 V_6^w\, ,
\eeq
where $V_6^w$ is computed from the fiducial metric excluding the breathing mode.
The warped volume of the internal
space receives contributions both from the throat and from the bulk,
 $V_6^w =  (V_6^w)_{\rm throat} +  (V_6^w)_{\rm bulk}$.
Since the bulk metric is not known we use the parametrization
$V_6^w \equiv B_6 (V_6^w)_{\rm throat}$, with
$B_6 > 1$, to characterize the unknown bulk contribution.  Using the warp factor (\ref{equ:hKT}), we find that the leading contribution to the throat volume is
 \beq (V_6^w)_{\rm throat} = \frac{\pi}{T_3} \frac{N}{4} \hr_{\rm UV}^2
 \, . \eeq This implies that the range of the inflaton is bounded by \cite{BM}
\beq \label{equ:OurBound}
\frac{\varphi^2}{M_P^2} \le \frac{\varphi_{\rm UV}^2 }{M_P^2}  =
\frac{4}{e^{4 u(r_{\rm UV})} N B_6} \, , \eeq
where
\beq
e^{4 u(r_{\rm UV})} = \frac{\sigma(r_{\rm UV})}{\sigma(0)}\, .
\eeq
In \cite{BM} two of us (D.B. and L.M.) explained that this relation sharply limits the amount of gravitational waves that can be expected in D-brane inflation models.  One of the main results of the present paper is that the bound (\ref{equ:OurBound}) also provides a surprisingly strong constraint on the possibility of fine-tuning the inflaton potential, even in cases where the energy scale of inflation is too low for gravitational waves to be relevant.

The bound (\ref{equ:OurBound}) is written in a slightly stronger and more general form than the bound given in \cite{BM},
\beq \label{equ:OurOldBound} \frac{\varphi^2}{M_P^2} \le \frac{4}{N} \, . \eeq  The factor $B_6 > 1$ is simply an explicit representation of the unknown bulk contribution to $V_6^w$ and hence to $M_P^2$.  The factor of the breathing mode requires further explanation.  Recall that we have used the convention $e^{u(0)}=1$, so that the breathing mode factor is unity when the D3-brane is at the tip of the throat.  We will now argue that in the configurations of interest, {\it fluctuations} of the breathing mode as the D3-brane is displaced from the tip are characterized by the condition $e^{4 u(r)} = \frac{\sigma(r)}{\sigma(0)} \ge 1$.  The D3-brane potential from moduli stabilization is minimized when the D3-brane is at the tip of the throat, in all the configurations studied in this paper (see also \cite{DMSU}).  Displacing the D3-brane from the tip increases the potential, and so tends to increase the compactification volume, because positive energy causes a runaway potential for the volume.  (We have confirmed this expectation by explicit numerical analysis in \S\ref{sec:numerics}, and by analytical arguments in Appendix \ref{sec:sigma}.)  Thus, moving the D3-brane up the throat dilates the space, and leads to $e^{4 u(r)} = \frac{\sigma(r)}{\sigma(0)} > 1 $ for $r>0$.  This effect goes in the direction of making the bound (\ref{equ:OurBound}) {\it{stronger}} than (\ref{equ:OurOldBound}), but the effect is very small in practice: in the cases we considered, $1< \frac{\sigma(r_{\rm UV})}{\sigma(0)} < 1.1$.  Such a factor is a negligible correction in comparison to the uncertainties in $B_6$, so it is very reasonable to omit it, as in \cite{BM}.

\section{Stability in the Angular Directions}
\label{sec:stability}

\addtocontents{toc}{\SkipTocEntry}
\subsection{Kuperstein embedding}
The following analysis complements our Kuperstein case study of \S\ref{sec:Kuperstein} and \S\ref{sec:Analysis}.
In \S\ref{sec:min} we derive the condition for extremal trajectories $z_1 = \pm \frac{r^{3/2}}{\sqrt{2}}$
whose angular stability we investigate in \S\ref{sec:stab}.

\addtocontents{toc}{\SkipTocEntry}
\subsubsection{Extremal trajectories}
\label{sec:min}

Recall that the Kuperstein potential (\ref{equ:FKup}) depends only on  the following dynamical fields: $|z_1|^2$,
$z_1 + \overline{z_1}$,
$r$, and $\sigma = {\rm Re}(\rho)$.  Along a trajectory that extremizes the potential in the angular directions we must have
$\frac{\partial V}{\partial \Psi_i}=0$ for all $r$, so we aim to find points
in $T^{1,1}$ that satisfy \beq \frac{\partial |z_1|^2}{
\partial \Psi_i}=0= \frac{\partial (z_1+ \overline{z_1})}{\partial
\Psi_i}\, . \label{d|z|}\eeq We examine (\ref{d|z|}) by
introducing local coordinates in the vicinity of a fiducial point
$\mathbf{z}_0 \equiv ( z_1', z_2', z_3', z_4')$. The coordinates
around this point are given by the five generators of $SO(4)$
acting nontrivially on $\mathbf{z}_0$ \beq \label{equ:coord}
\mathbf{z}(r, \Psi_i) = \exp(\mathsf{T}) \, \mathbf{z}_0\, . \eeq
The Kuperstein embedding, $z_1 = \mu$, breaks the global $SO(4)$
symmetry of the conifold down to $SO(3)$, and the D3-brane
potential preserves this $SO(3)$ symmetry. We will find that the
actual trajectory breaks this $SO(3)$ down to $SO(2)$,
 which we take to act on $z_3$ and $z_4$.
The coordinates that make this $SO(2)$ stability group manifest
are given by
\beq
\label{equ:T}
\mathsf{T} \equiv \left(
\begin{array}{cc|cc}
  0 & \alpha_2 & \alpha_3 & \alpha_4 \\
  -\alpha_2 & 0 & \beta_3 & \beta_4 \\
  \hline
  -\alpha_3 & -\beta_3& 0 & 0 \\
  -\alpha_4 & -\beta_4  & 0 & 0 \\
\end{array}
\right)\, , \eeq
where $\Psi_i \equiv \{ \alpha_i, \beta_i \} \in \mathbb{R}$
are the local coordinates of the base of the
cone.
We aim to find $\mathbf{z}_0$ such that the potential $V(z_1 + \overline{ z_1}, |z_1|^2)$ is extremal along $\mathbf{z}_0$. We here find trajectories along which the linear variation of $z_1 + \overline{ z_1}$ and $|z_1|^2$ vanishes.
First, we observe from (\ref{equ:coord}) and (\ref{equ:T}) that for arbitrary $\mathbf{z}_0$ we have
\beq
\delta z_1 = \sum_{i=2}^4 \alpha_i z_i'\, , \qquad \alpha_i \in \mathbb{R}\, .
\eeq
and, hence,
\beq
\label{equ:z12}
\delta |z_1|^2 = \sum_{i=2}^4 \alpha_i (z_i' \bar z_1' + z_1' \bar z_i') \equiv 0\, .
 \eeq
To satisfy (\ref{equ:z12}) for all $\alpha_i$ one requires
\beq
z_i' = i \rho_i z_1'\, , \qquad \rho_i \in \mathbb{R}\, .
\eeq
We may use $SO(3)$ to set $\rho_3 = \rho_4 =0$, while keeping $\rho_2$ finite.
The conifold constraint, $z_1^2 + z_2^2 = 0$, then implies $\rho_2 = \pm 1$, while
the requirement
\beq
\delta(z_1' + \bar z_1') = a_2 (z_2' + \bar z_2') = 0\, ,
\eeq
makes $z_2'$ purely imaginary and $z_1'$ real.
This proves that the following is an extremal trajectory of the brane potential for the Kuperstein potential
\beq
\label{equ:z0}
z_1' = \pm \frac{1}{\sqrt{2}} r^{3/2}\, , \qquad z_2' = \pm i z_1'\, .
\eeq

\addtocontents{toc}{\SkipTocEntry}
\subsubsection{Stability}
\label{sec:stab}

Let us now study the stability of the potential along the path (\ref{equ:z0}). From (\ref{equ:coord}) one finds
\beq
z_1 = z_1' \left[ 1 - \frac{1}{2} (\alpha_2^2 + \alpha_3^2 + \alpha_4^2) +    \frac{i}{2} \rho_2 (2 \alpha_2 - \alpha_3 \beta_3 - \alpha_4 \beta_4)  + \cdots \right]\, ,
\eeq
and
\bea
z_1 + \bar z_1 &=& 2 z_1' \left[ 1 - \frac{1}{2} (\alpha_2^2 + \alpha_3^2 + \alpha_4^2) + \cdots \right] \, ,\\
|z_1|^2 &=& (z_1')^2 \left[ 1 - (\alpha_3^2 + \alpha_4^2) + \cdots \right]\, .
\eea
Since the Kuperstein potential (\ref{equ:FKup}) depends only on $r$, $z_1 + \overline{z_1}$,
and $z_1 \overline{z_1}$, and because \beq \left. \frac{\partial
|z_1|^2}{\partial \Psi_i}\right|_0 = \left. \frac{\partial (z_1 +
\overline{z_1})}{\partial \Psi_i}\right|_0 = 0 \eeq where $\left. \bigl( \dots \bigr) \right|_{0}$
denotes evaluation at ${\bf z}_0$, we find
\beq
\left. \frac{\partial V}{\partial \Psi_i} \right|_0 = 0\, , \eeq
and
\beq \left. \frac{\partial^2 V}{\partial \Psi_i
\partial \Psi_j}\right|_{0} = \left. \left[ \frac{\partial V}{\partial |z_1|^2}
\frac{\partial^2 |z_1|^2}{\partial \Psi_i
\partial \Psi_j} + \frac{\partial V}{\partial(z_1 +
\overline{z_1})} \frac{\partial^2 (z_1 +\overline{z_1}) }{\partial
\Psi_i \partial \Psi_j} \right] \right|_0 \, , \eeq where \beq \left. \partial_i
\partial_j |z_1|^2 \right|_0 = \pm {\frac{r^{3/2}}{\sqrt{2}}}
\left. \partial_i \partial_j (z_1 + \overline{z_1}) \right|_0 +
r^3 \delta_{i 2} \delta_{j 2} = - r^3 \delta_{i 3}
\delta_{j 3}   - r^3 \delta_{i 4} \delta_{j 4}\, . \eeq
Hence, the angular mass matrix at ${\bf z}_0$ has the  form \beq
\left. \frac{\partial^2 V}{\partial \Psi_i
\partial \Psi_j}\right|_{0}  = \left. \left(%
\begin{array}{ccccc}
  X & 0 & 0 & 0 & 0 \\
  0 & X+Y & 0 & 0 & 0 \\
  0 & 0 & X+Y & 0 & 0\\
  0 & 0  & 0 & 0 & 0\\
  0  & 0 & 0 &  0 & 0
\end{array}%
\right) \right| \begin{array}{c} \alpha_2 \\ \alpha_3 \\ \alpha_4 \\ \beta_3 \\
\beta_4 \end{array} \eeq where \bea
X &\equiv& \left.  \mp \sqrt{2} r^{3/2}\frac{\partial V}{\partial (z_1 + \overline{z_1})} \right|_0 \, , \\
Y &\equiv&  \left. - r^3 \frac{\partial V}{\partial |z_1|^2}
\right|_0\, . \eea The flat directions in the $\beta_3, \beta_4$ angles
parameterize the symmetry group $SO(3)/SO(2)$ that leaves the Kuperstein embedding $z_1
= \mu$ invariant. The angles $\alpha_3$ and $\alpha_4$ have degenerate
eigenvalues.

To calculate the eigenvalues $X$ and $Y$ we note that
the F-term potential (\ref{equ:FKup}) may be written as
\bea \label{equ:fwa}
&&  V_F =  {\cal C}(r,\sigma) {\cal G}^{1/n} \Biggl[ (2 a \sigma + 6) - \frac{6 \,e^{a\sigma} |W_0| }{|A_0|} {\cal G}^{-1/2n}  \nonumber \\
&& \hspace{1cm}  +  \ \frac{3}{2n} \Bigl({1 \over \mu} (z_1 +
\overline{z_1}) - \frac{2}{\mu^2}  |z_1|^2 \Bigr) {\cal G}^{-1} +
\frac{4 c}{n} \frac{r}{r_\mu}
\Bigl(
1 - \frac{|z_1|^2}{2 r^3}\Bigr) {\cal G}^{-1} \Biggr]\, , \eea
and the D-term potential (\ref{equ:Dterm}) depends only on $r$ and is independent of the angles (at least far from the tip \cite{DMSU}).
In (\ref{equ:fwa}) we have defined the  functions
\bea {\cal C}(r,\sigma) &\equiv&  \frac{\kappa^2 a |A_0|^2 e^{-2a \sigma}}{3 U^2}> 0\, , \quad \quad \partial_{z_1 + \overline{z_1}} {\cal C}= \partial_{|z_1|^2} {\cal C} = 0\, ,  \\
{\cal G} &\equiv& \frac{|A|}{|A_0|} = \left|1-\frac{z_1}{\mu} \right|^2 \, , \quad \quad \partial_{z_1 + \overline{z_1}} {\cal G} = -\frac{1}{\mu}\, , \quad \partial_{|z_1|^2} {\cal G} = \frac{1}{\mu^2} \, , \\
{\cal G}_0 &\equiv& \left. {\cal G} \right|_0 = |1 \mp x^{3/2}|^2 = g(x)^2\, ,
 \eea
 and the variable
 \beq
 x \equiv
 \frac{r}{r_\mu} = \frac{\phi}{\phi_\mu}\, .
 \eeq
After a lengthy but straightforward computation we find
\bea
X &=& \pm \frac{2 {\cal C}}{n} \frac{x^{3/2}}{|1 \mp x^{3/2}|^{2(1-1/n)}} \Biggl[ 2 a \sigma + \frac{9}{2} - \frac{3 \,e^{a\sigma} |W_0| }{|A_0|} \frac{1}{|1\mp x^{3/2}|^{1/n}}  \nonumber \\
&& \hspace{1cm}  \mp \ 3 \Bigl( 1-\frac{1}{n}\Bigr) \frac{x^{3/2} (1\mp x^{3/2})}{|1 \mp x^{3/2}|^2}  - 3c  \Bigl( 1-\frac{1}{n}\Bigr) \frac{x}{|1 \mp x^{3/2}|^2} \Biggr]\, ,
\label{equ:X}
\eea
and
\beq \label{equ:XY} X+Y = \left( 1  \mp
x^{3/2} \right) X + \frac{2 {\cal C} c }{n}  \,  \frac{x}{|1 \mp x^{3/2}|^{2(1- 1/n)}} \left( 1 + \frac{3}{2 c} x^2 \right)
\, , \eeq
where
\beq
c \equiv \frac{9}{4 n \, a \sigma_0 \frac{\phi_\mu^2}{M_P^2}} \, .
\eeq
Notice that stability of the trajectory $z_1 = \pm \frac{1}{\sqrt{2}} r^{3/2}$ for both
positive and negative real $z_1$ only requires that $X > 0$ (from
(\ref{equ:XY}) this automatically implies $X+Y > 0$, at least for
the regime of interest, $r < r_\mu$). Hence, from equation
(\ref{equ:X}) a simple numerical check can decide whether a
specific scenario is stable in the angular directions. For all
potential inflationary trajectories we have performed
this stability test.

\subsubsection*{\sl Analytical approximation}

\noindent
In \S\ref{sec:Kuperstein} and Appendix \ref{sec:sigma} we explain how the position of the K\"ahler modulus in the AdS vacuum, $\sigma_F$, shifts to $\sigma_0$ when the minimum gets uplifted to dS,
\beq
a \sigma_0 \approx a \sigma_F +\frac{s}{a \sigma_F}\, ,
\eeq
where $\frac{3 |W_0|}{|A_0|} e^{a \sigma_F} = 2 a \sigma_F + 3$ and $\frac{3 |W_0|}{|A_0|} e^{a \sigma_0} \approx 2 a \sigma_0 + 3 + 2 s$. Substituting this into (\ref{equ:X}) we find
\bea
X &=& \pm \frac{2 {\cal C}}{n} \frac{x^{3/2}}{|1 \mp x^{3/2}|^{2(1-1/n)}} \Biggl[ \frac{3}{2} - \frac{2s}{|1 \mp x^{3/2}|^{1/n}}+ 2 a \sigma_0 \Bigl( 1- \frac{1}{|1 \mp x^{3/2}|^{1/n}}\Bigr)\nonumber \\
&& \hspace{4cm} + \ 3 \Bigl( 1- \frac{1}{|1 \mp x^{3/2}|^{1/n}}\Bigr)
 \mp 3 \Bigl( 1-\frac{1}{n}\Bigr) \frac{x^{3/2} (1 \mp x^{3/2})}{|1 \mp x^{3/2}|^2} \nonumber \\
&& \hspace{4cm} - 3c \Bigl( 1-\frac{1}{n}\Bigr) \frac{x}{|1 \mp x^{3/2}|^2} \Biggr]\, .
\label{equ:XX}
\eea
In the limit $x \to 0$ this becomes
\beq
\lim_{x \to 0} X = \mp \frac{{\cal C}}{n} x^{3/2}  \left[ 4 s - 3 \right]\, .
\eeq
Hence, the near-tip region is stable on the negative axis and unstable on the positive axis. Notice that it was essential to include the shift from uplifting to realize this.

\addtocontents{toc}{\SkipTocEntry}
\subsection{Ouyang embedding}
\label{sec:Ouyang}

\addtocontents{toc}{\SkipTocEntry}
\subsubsection{Extremal trajectories}

For the Ouyang embedding, $w_1 = \mu$, the brane potential depends on $w_1 + \overline{ w_1} $, $|w_1|^2$ and $|w_2|^2$.
To find extremal trajectories of the potential we therefore require
\beq \frac{\partial |w_1|^2}{
\partial \Psi_i}= \frac{\partial |w_2|^2}{\partial
\Psi_i} = \frac{\partial (w_1+ \overline{ w_1})}{\partial \Psi_i}
= 0\, . \label{d|w|}\eeq We introduce local coordinates by
applying generators of $SU(2)$ to the generic point $W_0$ \beq
\label{equ:Wtrans} W = e^{i \mathsf{T}_1} W_0 e^{-i
\mathsf{T}_2}\, , \qquad W_0 \equiv \left( \begin{array}{cc} w_3'
& w_2' \\ w_1' & w_4' \end{array} \right) \, , \eeq where \beq
\label{equ:Tmatrix} \mathsf{T}_i \equiv  \left( \begin{array}{cc}
\alpha_i & \beta_i + i \gamma_i \\ \beta_i - i \gamma_i & -
\alpha_i \end{array} \right) \, . \eeq This implies \beq \delta
w_1 = - i (\alpha_1 + \alpha_2) w_1' + (- \beta_1 + i \gamma_1)
w_3' + (\beta_2 - i \gamma_2) w_4' + \cdots \eeq and $\delta (w_1
+ \bar w_1) = 0$ gives $w_1' \in \mathbb{R}$, $w_3' = w_4' = 0$.
We find that $\delta |w_1|^2 = 0$ and  $\delta |w_2|^2 = 0$ if
$w_2' \in \mathbb{R}$. The conifold constraint $w_1' w_2' = 0$
then restricts the solution to the following two options: \beq
w_1' = 0\, , \qquad |w_2'| = r^{3/2}\, , \qquad \Leftrightarrow
\qquad \theta_1 = \theta_2  = 0\, , \eeq or \beq w_1' = \pm
r^{3/2}\, , \qquad w_2' = 0\, , \qquad \Leftrightarrow \qquad
\theta_1 = \theta_2 = \pi\, . \eeq

\subsubsection*{\sl Delta-flat direction}

For $w_1' = 0$ the superpotential correction to the potential vanishes and inflation is impossible, as noted in \cite{Burgess} and reviewed in \S\ref{sec:ACR}.

\subsubsection*{\sl Non-delta-flat direction}

For $w_2' = 0$ the superpotential correction to the potential does not vanish. In fact, along this extremal trajectory the potential can be shown to be identical to the potential for the Kuperstein case.
However, angular stability for the Ouyang and Kuperstein cases is different, as we now discuss.

\addtocontents{toc}{\SkipTocEntry}
\subsubsection{Stability}

\subsubsection*{\sl Delta-flat direction}

Near $w_1' = 0$, $w_2' \ne 0$ we have
\bea
w_1 &=& w_2' \left[ \beta_1 \beta_2 - \gamma_1 \gamma_2 - i (\beta_1 \gamma_2 + \beta_2 \gamma_1)\right] \\
w_2 &=& w_2' \left[ 1 + i(\alpha_1 + \alpha_2) - \frac{1}{2}(\alpha_1 + \alpha_2)^2 - \frac{1}{2} (\beta_1^2+ \beta_2^2 + \gamma_1^2 + \gamma_2^2) \right]
\eea
and (to second order)
\bea \label{equ:woneequalszero}
w_1 + \overline{ w_1} &=& 2 w_2' \left[ \beta_1 \beta_2 - \gamma_1 \gamma_2 \right] \\
|w_1|^2 &=& 0 \\
|w_2|^2 &=& (w_2')^2 \left[1 -  (\beta_1^2+ \beta_2^2 + \gamma_1^2 + \gamma_2^2) \right]\, .
\eea
The mass matrix in these coordinates in non-diagonal, but may easily be diagonalized by the  transformation
\beq
\beta_{1,2} = \frac{u_1 \pm v_1}{\sqrt{2}}\, , \quad \gamma_{1,2} = \frac{v_2 \pm u_2}{\sqrt{2}}\, , \quad U^2 \equiv u_1^2 + u_2^2\, , \quad V^2 \equiv v_1^2 + v_2^2\, .
\eeq
This gives
\bea
w_1 + \overline{w_1} &=&  w_2' \left[ U^2-V^2 \right] \\
|w_2|^2 &=& (w_2')^2 \left[1 -  (U^2 + V^2) \right]\, .
\eea
A lengthy but straightforward computation gives the eigenvalues of the angular mass matrix of the potential along the delta-flat direction
\bea
\label{equ:XU}
X_{U} &=& \frac{2{\cal C} c}{n} \, x \left[  1 \mp \frac{1}{4 c} \,x^{1/2}  \left( 4 a \sigma + 9 - \frac{6 e^{a \sigma} |W_0|}{|A_0|}\right)\right]\, , \\
X_{V} &=& \frac{2{\cal C} c}{n} \, x \left[  1 \pm \frac{1}{4 c} \, x^{1/2}  \left( 4 a \sigma + 9 - \frac{6 e^{a \sigma} |W_0|}{|A_0|}\right)\right]\, .
\label{equ:XV}
\eea
We have confirmed that these stability criteria are precisely what was found for the delta-flat direction in \cite{Burgess} ({\it cf.} their equation (3.15)).
Equation (\ref{equ:XU}) and (\ref{equ:XV}) show that the delta-flat direction is angularly stable if $x < x_c$ and unstable if $x > x_c$, where
\beq
\label{equ:xc}
x_c \equiv \left( \frac{4c}{4s-3} \right)^2 = \frac{1}{(4s-3)^2} \left( \frac{9}{a \sigma_0} \right)^2 \frac{1}{n^2} \frac{M_P^4}{\phi_\mu^4}\, .
\eeq
Equation (\ref{equ:xc}) is the generalization of (\ref{equ:BurgessRC}) to general $n$.
Applying the field range bound in the form $\frac{\phi_\mu^2}{M_P^2} < \frac{4}{N}$ one finds
\beq
x_c > \frac{1}{(4s-3)^2} \left( \frac{9}{a \sigma_0} \right)^2 \frac{N^2}{(4 n)^2} \ge 1\, .
\eeq
For typical parameters the delta-flat direction is hence stable from the tip to at least the location of the D7-branes.

\subsubsection*{\sl Non-delta-flat direction}

The non-delta-flat trajectory of the Ouyang embedding is very closely related to the extremal trajectories of the Kuperstein embedding.
In fact, the shape of the potential is identical for the two cases. However, the stability analysis reveals subtle, but important differences.

Near $w_2' = 0$, $w_1' \ne 0$ we have
\bea
w_1 + \overline{w_1} &=& 2 w_1' \left[ 1 - \frac{1}{2}(U^2 + V^2 )  \right] \, ,\\
|w_1|^2 &=& (w_1')^2 \left[1 - V^2\right]\, ,
\eea
where we have defined
\bea
U^2 &\equiv& (\alpha_1 + \alpha_2)^2\, , \\
V^2 &\equiv& \beta_1^2+ \beta_2^2 + \gamma_1^2 + \gamma_2^2\, .
\eea
Computing the eigenvalues of the angular mass matrix we find results that are almost identical to the Kuperstein results (\ref{equ:X}) and (\ref{equ:XY}), except for one crucial sign difference:
\bea
X_U  &=&  X\, , \\
 X_V &=& \left( 1  \mp
x^{3/2} \right) X\,  + \,\frac{2 {\cal C} c }{n} \,  \frac{x}{|1 \mp x^{3/2}|^{2(1- 1/n)}} \left( \fbox{$\displaystyle - $} \ 1 + \frac{3}{2 c} x^2 \right)
\label{equ:XVO}
\, , \eea
where $X$ is the Kuperstein result (\ref{equ:X}). Since the leading term in (\ref{equ:XVO}) now comes with the opposite sign ({\it cf.} (\ref{equ:XY})), the non-delta-flat trajectory for Ouyang is typically unstable for small $x$ whereas the corresponding trajectory for Kuperstein is stable,
\beq
 X_V \approx   -  \,\frac{2 {\cal C} c }{n}  \,  \frac{x}{|1 \mp x^{3/2}|^{2(1- 1/n)}} < 0
\, . \eeq
This is consistent with the results of \cite{Burgess}.

\addtocontents{toc}{\SkipTocEntry}
\subsection{Stability for small $r$}

To summarize our discussion of stability for the Kuperstein
scenario and the Ouyang scenario we now give an intuitive
explanation of angular stability in the limit of small $r$.

For either the Kuperstein or the Ouyang embedding, stability near
the tip $r\rightarrow 0$ is controlled by the term in the
potential proportional to $k^{1\bar{1}}$.  We first focus on the
Kuperstein embedding. Since $k^{1\bar{1}}$ contains a term
proportional to $r^{-3}$, its contribution to the second
derivative of the potential with respect to an angular variable
$\Psi_i$, ${\partial^2 V\over \partial \Psi_i^2}$, grows as $r$.
All other terms grow at least as $r^{3/2}$ (this follows from
${\partial\  \over \partial \Psi} ={\partial z_i\over \partial
\Psi}{\partial\ \over
\partial z_i}+c.c.$ and ${\partial z_i \over
\partial \Psi} \sim r^{3/2}$). A parallel consideration confirms that $\hat k_w^{1\bar{1}}$
is responsible for the leading contribution to the stability
analysis in the case of the Ouyang embedding as well.

Now, the trajectories $z_1=\pm {r^{3/2}\over \sqrt{2}}$ maximize
$|z_1|^2$ for a given $r$, and any variation of angles may only
increase  $k^{1\bar{1}}=r\left(1-{|z_1|^2\over 2 r^3}\right)$.
Hence the  trajectories in question are stable at small $r$ under
fluctuations of any angles that affect $|z_1|^2$. So far, this
analysis does not include the phase of $z_1$, which of course
leaves $|z_1|^2$ invariant.  The leading correction to the
potential from fluctuations of this phase comes not through
$k^{1\bar{1}}$ but through terms in $V$ proportional to
$z_1+\overline{z_1}$.  These terms change sign when $z_1$ does;
thus, one of the signs in $z_1=\pm {r^{3/2}\over \sqrt{2}}$
corresponds to the stable trajectory, while the other sign
corresponds to an unstable trajectory.  We showed above that the
stable trajectory in fact corresponds to $z_1<0$ (careful
consideration of the shift of the volume, {\it{cf.}} Appendix D,
is required for that argument.)

The analysis for the Ouyang embedding is very similar. The
delta-flat trajectory $|w_2|^2=r^{3}$ ($\theta_1=\theta_2=0$)
maximizes the ratio ${|w_2|^2\over r^3}$. Thus, any angular
fluctuation can only decrease the ratio ${|w_2|^2\over r^3}$,
without affecting ${|w_1|^2\over r^3}$ to second order ({\it cf.}
(\ref{equ:woneequalszero})). This is easily checked with the help
of the angular coordinates $\theta_i$ (\ref{equ:w1}). On the other
hand, the trajectory $w_1=\pm r^{3/2}$ ($\theta_1 =\theta_2 =
\pi$) maximizes ${|w_1|^2\over r^3}$, and angular fluctuations
away from this trajectory decrease the ratio ${|w_1|^2\over r^3}$,
without affecting ${|w_2|^2\over r^3}$ to second order.
 As a result, $\hat k_w^{1\bar{1}}=r\left(1+{|w_1|^2\over 2
r^3}-{|w_2|^2\over r^3}\right)$ cannot decrease in the case of the
delta-flat trajectory $|w_2|^2 =r^{3}$, but necessarily has a
negative mode along the non-delta-flat trajectory $w_1 =\pm
r^{3/2}$. Hence, the non-delta-flat trajectory is unstable for
small $r$. No further consideration is needed to show that the
delta-flat trajectory $|w_2|^2=r^{3}$ is stable. Since angular
fluctuations around $w_1=0$ cannot affect the term involving
$w_1+\overline{w_1}$, the leading contribution always comes from
$\hat k_w^{1\bar{1}}$.

We have therefore demonstrated that near the tip, the trajectory
$z_1 = -\frac{r^{3/2}}{\sqrt{2}}$ is stable for the Kuperstein embedding, whereas
the trajectory $w_1 = \pm r^{3/2}$ in the Ouyang embedding is
unstable.

\addtocontents{toc}{\SkipTocEntry}
\subsection{Higher-degree ACR embeddings}
\label{sec:HigherACR}

We conclude this Appendix with a derivation of the non-delta flat trajectory for higher degree ACR embeddings. As before,
to find an extremal radial direction we need to satisfy the equations  ${\partial V\over \partial \Psi_i}=0$ for any $r$. Since the potential for a general ACR embedding depends only on $|\Phi|^{2P}$, ${\rm Re}(\Phi^P)$
and
\bea
Z &\equiv&  \left|p_1 \frac{w_4}{w_1} + p_3 \frac{w_2}{w_3}\right|^2 +  \left|p_1 \frac{w_3}{w_1} + p_4 \frac{w_2}{w_4}\right|^2  \nonumber \\
&&+  \left|p_2 \frac{w_3}{w_2} + p_4 \frac{w_1}{w_4}\right|^2 +
\left|p_2 \frac{w_4}{w_2} + p_3 \frac{w_1}{w_3}\right|^2\
,\label{equ:2term}\eea we will consider these terms separately.
Extremizing ${\rm Re}(\Phi^P)$ with respect to the phase of $\Phi$
selects real $\Phi^P$. For any {\it non-zero} $\Phi$ we can use equations (\ref{equ:Wtrans}) and (\ref{equ:Tmatrix}) to show that
\beq
\label{equ:master}
{\rm Re} \left[\frac{\delta \Phi^P}{\Phi^P} \right] = {\rm Re} \left[ \sum_{i} p_i \frac{\delta w_i}{w_i} \right]= 0
\eeq
for
\bea
\label{equ:ACR1}
\left| \frac{w_1}{w_3} \right|^2 &=& \tan^2{\theta_1\over 2} = {p_1+p_4\over p_2+p_3}\, ,\\
\left| \frac{w_1}{w_4} \right|^2 &=& \tan^2{\theta_2\over 2} = {p_1+p_3\over p_2+p_4}\, ,
\label{equ:ACR2} \eea and \beq
\label{equ:ACR3}\Phi^P = \pm \lambda^P r^{3P/2}\, , \eeq where
\beq \lambda^P \equiv {(p_1+p_3)^{{1\over
2}(p_1+p_3)}(p_1+p_4)^{{1\over 2}(p_1+p_4)}(p_2+p_3)^{{1\over
2}(p_2+p_3)}(p_2+p_4)^{{1\over 2}(p_2+p_4)}\over P^P}\ . \eeq
Notice that this
point trivially extremizes $|\Phi|^{2P}$ since
\beq
\delta |\Phi|^{2P} = |\Phi|^{2P} {\rm Re} \left[\frac{\delta \Phi^P}{\Phi^P} \right] = 0\, .
\eeq
In addition, since $\Phi^P$ is real, equation (\ref{equ:master}) also implies $\delta {\rm Re}(\Phi^P) = 0$.
Finally, surprisingly enough, (\ref{equ:ACR1}, \ref{equ:ACR2}) also
extremizes (\ref{equ:2term}) with respect to the angular directions.
Therefore (\ref{equ:ACR3}) is an
extremal radial trajectory which exists for a generic case in
addition to the delta-flat direction, $\Phi = 0$.
For $P=1$ this reproduces the non-delta-flat direction for the Ouyang embedding, {\it cf}.~Appendix \ref{sec:Ouyang}.
For $P=2$, $p_1 = p_2 = 1$ we find that the Karch-Katz embedding has a non-delta-flat direction given by $\Phi^2 = w_1 w_2 = w_3 w_4 = \pm \frac{1}{4} r^3$.

Although every ACR embedding contains non-delta flat directions (\ref{equ:ACR3}), we now argue that these higher degree embeddings ($P > 1$) are {\it not} promising settings for flat potentials.
Recall that
for the $P=1$ Kuperstein an inflection point could be arranged along $z_1 \in \mathbb{R}^{-}$ (modulo consistency with  microscopic constraints), basically because an $x^{3/2}$ term in the potential led to a negative divergence of $\eta$ for small $x$, which could be balanced against the $x^2$ term familiar from the $\eta$-problem for intermediate $x$ (see \S\ref{sec:analytic}).
For $P > 1$ embeddings this possibility disappears, since there are then no terms of order lower than quadratic in $x$. Hence,
\beq
\lim_{x \to 0} \eta = \frac{2}{3}\, .
\eeq
This is suggestive evidence that fine tuning in higher degree ACR embeddings cannot give inflation, at least not at $x < 1$ ({\it cf.} \cite{Krause}).

\newpage
\section{Stabilization of the Volume}
\label{sec:sigma}

In the AdS minimum of a KKLT compactification, the stabilized value of the K\"ahler modulus $\omega_F  \equiv a \sigma_F$ is given by the SUSY condition
\beq
\label{equ:KKLT1}
\left. D_\rho W \right|_{r=0,\, \omega_F}= 0 \quad \quad \Rightarrow \quad \quad \left.\frac{\partial V_F}{\partial \omega} \right|_{\omega_F}= 0\, ,
\eeq
which in terms of the flux superpotential is \cite{KKLT}
\beq
\label{equ:KKLT}
3 \frac{|W_0|}{|A_0|} e^{\omega_F} = 2 \omega_F + 3\, .
\eeq
Adding an antibrane to lift the KKLT AdS minimum to a dS minimum induces a small shift in the stabilized volume, $\omega_0 = \omega_F + \delta \omega$. We compute this shift in \S\ref{sec:Dshift}. This gives the value of the K\"ahler modulus in the absence of a mobile D3-brane (or when the brane is near the tip of the throat). The presence of a D3-brane away from the tip induces a further shift of the volume that depends on the brane position, which we denote $\omega_\star(r)$. We compute this dependence of the K\"ahler modulus on the D3-brane position in \S\ref{sec:D3shift}.  (See also \cite{Krause}.)

\addtocontents{toc}{\SkipTocEntry}
\subsection{Shift induced by uplifting}
\label{sec:Dshift}

The stabilized value of the K\"ahler after uplifting, $\omega_0 = \omega_F + \delta \omega$, is determined from
\beq
\label{equ:sig1}
\left. \frac{\partial V}{\partial \omega}\right|_{\omega_0}  = 0 \approx \left. \frac{\partial^2 V_F}{\partial \omega^2} \right|_{\omega_F} \delta \omega + \left. \frac{\partial V_D}{\partial \omega} \right|_{\omega_0}\, ,
\eeq
where, from (\ref{equ:Dterm}),
\beq
\label{equ:sig2}
 \left. \frac{\partial V_D}{\partial \omega} \right|_{\omega_0} = \left. \frac{- 2V_D}{\omega} \right|_{\omega_0} \approx \left. \frac{-2 V_D}{ \omega_F} \right|_{\omega_F} \left[ 1- 3 \frac{\delta \omega}{\omega_F}\right]\, .
\eeq
Solving (\ref{equ:sig1}) and (\ref{equ:sig2}) for $\delta \omega$ we find
\beq
\label{equ:sig3}
 \delta \omega = \frac{\omega_F}{3 + \frac{ \omega^2}{2 V_D} \left. \frac{\partial^2 V_F}{\partial \omega^2}\right|_{\omega_F}} \, .
\eeq
From equation (\ref{equ:VFTwoField}) we have
\bea
V_F(0, \omega) &=& C \frac{e^{-2 \omega}}{(2 \omega)^2} \left[ (2 \omega + 6) - 6 e^{\omega} \frac{|W_0|}{|A_0|}\right]\\
V_D(0, \omega)&=& (D+D_{\rm other})\frac{1}{(2 \omega)^2}
\eea
and
\beq
\label{equ:dVFds}
\frac{\partial V_F}{\partial \omega} = - \frac{\omega +2}{\omega} \left[V_F + C\frac{e^{-2 \omega}}{2 \omega} \right] \, .
\eeq
Since
\beq
V_F(0,\omega_F )= - C \frac{ e^{ -2 \omega_F} } {2 \omega_F } \, ,
\eeq
equation (\ref{equ:dVFds}) vanishes at $\omega_F$,
confirming equation (\ref{equ:KKLT1}).
The second derivative of the F-term potential at $\omega_F$ is
\bea
\left. \frac{\partial^2 V_F}{\partial \omega^2}\right|_{\omega_F} &=& - \frac{2}{(\omega_F)^2}
\left[   (\omega_F)^2 + \frac{5}{2} \omega_F + 1 \right] V_F(0,\omega_F) \nonumber \\
&\approx& + 2 \, |V_F(0, \omega_F)| \, ,
\eea
where $\omega_F \gg 1$.  Recalling the definition (\ref{equ:sDef})
and using $\frac{(\omega_F)^2}{s} \gg 1$ in equation (\ref{equ:sig3}), we find
\beq
\label{equ:wCzero}
\omega_0 \approx \omega_F + \frac{s}{\omega_F}\, .
\eeq
Although $\delta \omega$ is small, it appears in an exponent in the potential (\ref{equ:VFTwoField}), so that its effect there has to be considered:
\bea
3 \frac{|W_0|}{|A_0|} e^{\omega_0} = 3 \frac{|W_0|}{|A_0|} e^{\omega_F} e^{\delta \omega}&\approx& \left(2 \omega_F + 3 \right) \left(1 + \frac{s}{\omega_F} \right) \nonumber \\ &\approx& 2 \omega_F + 3 + 2s\, , \nonumber \\
&\approx& 2 \omega_0 + 3 + 2s\, .
\eea

\addtocontents{toc}{\SkipTocEntry}
\subsection{Shift induced by brane motion}
\label{sec:D3shift}

Adding a mobile D3-brane to the compactification induces a further shift of the K\"ahler modulus that depends on the radial position of the brane, $\omega_\star(r)$. The function $\omega_\star(r)$ is determined by the solution of a transcendental equation,
\beq
\label{equ:dVdw}
\left. \partial_\omega V \right|_{\omega_\star(r)}= 0\, .
\eeq
Although equation (\ref{equ:dVdw}) does not have an exact analytic solution,
here we derive a simple approximate solution. (See also \cite{Krause}.)
The precise form of the solution we give here is only valid for the trajectory $z_1 = - \frac{r^{3/2}}{\sqrt{2}}$ of the Kuperstein potential (see \S\ref{sec:potential}). While these results can easily be generalized to the trajectory $z_1 = \frac{r^{3/2}}{\sqrt{2}}$ and to more general embeddings, we have argued in the main body of the paper that the trajectory $z_1 = - \frac{r^{3/2}}{\sqrt{2}}$ in the Kuperstein embedding is of primary interest as far as the possibility of inflationary solutions is concerned.

\begin{figure}[htbp]
    \centering
        \includegraphics[width=1.0\textwidth]{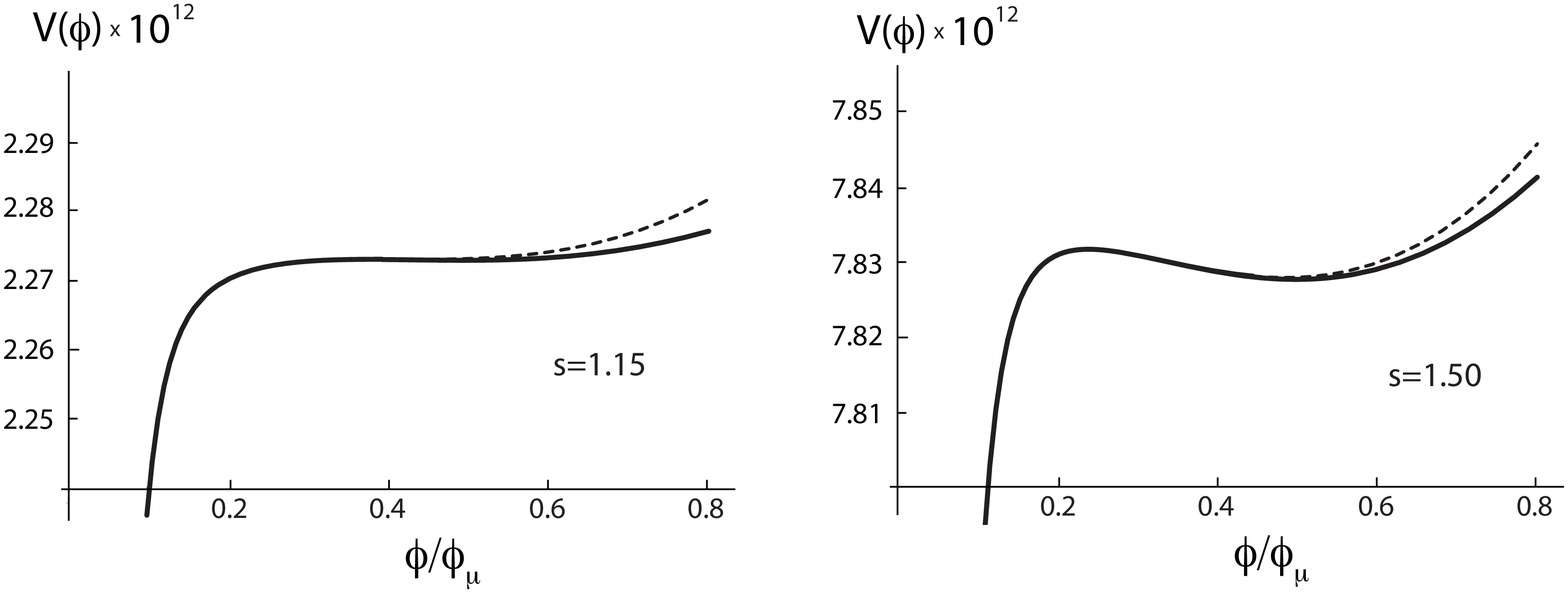}
    \caption{{\bf D3-brane potential $\mathbb{V}(\phi)$.} \newline
    Shown are the analytic potential derived from (\ref{equ:wCApprox})(dashed line) and the exact numerical result (solid line).}
    \label{fig:Pot}
\end{figure}

First, we notice that $\omega_\star$ appears both in polynomial terms and exponential terms in (\ref{equ:dVdw}). Letting $\omega_\star \to \omega_0 = \omega_F + \frac{s}{\omega_F}$ (see equation (\ref{equ:wCzero})) in all polynomial terms, but {\it not} in the exponentials, we transform the transcendental equation (\ref{equ:dVdw}) into a quadratic equation for $e^{\omega_\star(r)}$.
Solving this we obtain the dependence of the K\"ahler modulus on the brane position as an expansion in $x$
\beq
\label{equ:omegaA}
\omega_\star(r) = \omega_0 \left[ 1 + c_1 x + c_{3/2} x^{3/2} + c_2 x^2 + \dots \right]\, ,
\eeq
where
\beq
\omega_0 = \omega_F+ \frac{s}{\omega_F} + \frac{s(3s-5)}{2\omega_F^2} + {\cal O}(\omega_F^{-3}) \, ,
\eeq
and
\bea
c_1 &=& \left(\frac{27}{8n^2} \frac{M_P^2}{\phi_\mu^2} \right) \frac{1}{\omega_F^3} + {\cal O}(\omega_F^{-4})\, ,\\
c_{3/2} &=&  \frac{1}{n} \frac{1}{ \omega_F} \left[1 - \frac{1}{2 \omega_F} \right] +{\cal O}(\omega_F^{-3})\, , \\
c_2 &=& \left( \frac{s-1}{6} \frac{\phi_\mu^2}{M_P^2} \right) \frac{1}{\omega_F^2} + {\cal O}(\omega_F^{-3}) \, .
\eea
As we can see from this, typically $c_{3/2} \gg c_1, c_2$, so that the following approximation can be tried
\beq
\label{equ:wCApprox}
\omega_\star(r) \approx \omega_0 \left[1 + c_{3/2} x^{3/2} \right]\, .
\eeq
Notice also
that $c_1$, $c_{3/2}$, $c_2$ are all positive -- the volume shrinks as the D3-brane moves towards the tip.
Fig.~\ref{fig:Pot} shows the D3-brane potential derived from (D.20) (dashed line) and the exact numerical result (solid line). Notice that the analytic approximation is always steeper than the more precise numerical result and therefore systematically underestimates the total number of $e$-folds \cite{Panda}.

\addtocontents{toc}{\SkipTocEntry}
\subsection{Canonical inflaton revisited}

Equipped with (\ref{equ:wCApprox}) for the evolution of the volume modulus, we may obtain a more accurate analytical result for the canonical inflaton field (\ref{equ:CanPhi}),
\beq
\label{equ:varphi}
\varphi(r) = \int \Bigl( \frac{3 \sigma_0 T_3}{U(r, \sigma_\star(r))} \Bigr)^{1/2}\d r \, ,
\eeq
where
 \beq
 \frac{3 \sigma_0 T_3}{U(r, \sigma_\star(r))} = \frac{3}{2} T_3 \left( \frac{\omega_\star(r)}{\omega_\star(0)}  - \frac{\frac{3}{2} T_3 r^2}{6 M_P^2}\right)^{-1}\, .
 \eeq
This implies
 \bea
 \varphi(r) &=& \int \left[ 1 + c_{3/2} \left(\frac{\phi}{\phi_\mu} \right)^{3/2} - \frac{\phi^2}{6 M_P^2} \right]^{-1/2} \d \phi \\
 &\approx& \phi \left[ 1 - \frac{c_{3/2}}{5} \left(\frac{\phi}{\phi_\mu} \right)^{3/2} + \frac{\phi^2}{36 M_P^2} \right] \, ,\\
 &\approx& \phi\, .
 \eea

\newpage
\section{Effects of Induced Magnetic Fields}
\label{sec:magnetic}

One may be concerned that magnetic fields induced on the wrapped D7-branes
could give rise to an additional contribution to the D3-brane potential.
In this appendix we address this problem and show that if
supersymmetry is broken by a brane-antibrane pair in a
(conformally) Calabi-Yau throat, the induced magnetic fields do
not affect our considerations.

Let us consider a D7-brane with the worldvolume action $S_{D_7}=
S_{\rm DBI}+S_{\rm CS}$, \beq \label{actionD7} S_{D_7} = - T_7
\int_{{\cal M}_{8}} \d^8 \xi \, e^{-\Phi} \sqrt{- \det \left( \hat
G_{ab} + \hat B_{ab} + 2 \pi \alpha' F_{ab} \right) } - T_7
\int_{{\cal M}_8}  \sum_q \hat C_{[q] } \wedge e^{\hat B_{[2]} + 2
\pi \alpha' F}\, , \eeq where hats denote pullbacks of bulk
fields onto the worldvolume of the D7-brane, $C_{[p]}$ are the
RR potentials, and $F$ is the worldvolume gauge field strength.
The indices $a,b$ run through the eight dimensions of the D7-brane
worldvolume, while Greek indices $\mu,\nu$ and Roman indices
$i,j$ will denote directions along Minkowski spacetime and along
the internal Calabi-Yau space, respectively. Let the warped
background metric be \beq \label{met}\d s^2=h^{-1/2}
\eta_{\mu\nu}\d x^\mu \d x^\mu+h^{1/2}\sum_{i=1}^6 G_i^2\ , \eeq
where $G_i$ are six-dimensional vielbeins.
 We assume that the D7-brane wraps some four-cycle $\Sigma_4$ in
the compact space, and that there is a magnetic field $B$ in the
bulk and an induced gauge field $F$ on the D7-brane.
The gauge field can be
decomposed into  components $\F$ in the internal dimensions and
components $\FF$ in spacetime. We absorb a dimensionful coefficient, $2\pi\alpha'$,
in the definition of $\F$ and $\FF$. The action of the D7-brane
depends on $\F$ through the combination $M \equiv \hat{B}+\F$.  We are
interested in the Born-Infeld and Chern-Simons actions for the
D7-brane in the presence of a D3-brane and a non-trivial gauge
field $\FF$ in the spacetime directions.
Let us first consider the D7-brane alone,
without any D3-branes.  We do not
yet assume that the compactification manifold is conformally
Calabi-Yau. The kappa-symmetry condition for a supersymmetric
embedding of the D7-brane can be reformulated purely in geometrical
terms without use of the $\Gamma$-algebra
\cite{GMS}. First, supersymmetry requires the four-cycle $\Sigma_4$ to be
holomorphic. Then the following identities for the forms on
$\Sigma_4$ should be satisfied
 \beq
 \label{kpa}
 h^{1/2}\hat{J}\wedge
M=\tanh\theta\left({h\over 2}\hat{J}\wedge \hat{J}-{1\over
2}M\wedge M\right)\, , \quad \quad M^{2,0}=0\, .
\eeq
 Here $J$
is a (pseudo)-K\"ahler form of  the internal manifold $J={i\over
2}\sum_{i=1}^3 G_{2i-1}\wedge G_{2i}$ and the angle $\theta$
characterizes the geometry of the supergravity solution as explained in
\cite{GMS}. For the warped Calabi-Yau case of the warped deformed conifold \cite{KS} solution,
$\theta=0$, while the solutions on the baryonic branch
\cite{Butti} satisfy $\cosh\theta=e^{-\Phi}$, where $\Phi$ is the dilaton.
We use the block-diagonal form of the matrices in (\ref{actionD7})
to evaluate the DBI action  \beq \det \left( \hat G_{ab} + \hat
B_{ab} + 2 \pi \alpha' F_{ab} \right)={\rm
det}\left(h^{-1/2}\eta_{\mu\nu}+\FF_{\mu\nu}\right)\times {\rm
det}\left(h^{1/2} \hat{G}_{ij}+M_{ij}\right)\, , \eeq and express
the answer as a power series in $\FF$ \beq \sqrt{-{\rm
det}\left(h^{-1/2}\eta+\FF\right)}=\left[h^{-1}-{1\over 4}{\rm Tr}
\, \FF^2+{\cal{O}}(\FF^4)\right]\, . \eeq Since $\hat{G}_{ij}$
(and $\hat{J}$) has only $(1,1)$ components, equation (\ref{kpa})
implies
 \beq \label{factor}
  {\rm
det}\left(h^{1/2}\hat{G}+M\right) = \left(h\  {\rm
det}^{1/2}\hat{G}-{\rm Pf}M\right)^2+{h}\left(\hat{J}\wedge
M\right)_{1234}^2\, , \eeq where ${\rm Pf} X$ of a 2-form $X$ is
defined as ${\rm Pf} X={1\over 2}(X\wedge X)_{1234}$.  Expression
(\ref{factor}) is algebraic because
 ${\rm det}\hat{G}$ also factorizes: ${\rm Pf} \hat{J}={\rm
det}^{1/2}\hat{G}$.  Combining (\ref{factor}) with (\ref{kpa}) we get
\beq {\rm det}(h^{1/2}\hat{G}+M)= \left(h\ {\rm
det}^{1/2}\hat{G}-{\rm Pf} M\right)^2\left(1+\tanh^2\theta\right)
\, , \eeq and \beq S_{\rm DBI}=-T_7\int_{{\cal M}_{8}} \d^8 \xi \,
e^{-\Phi}\left[h^{-1}-{1\over 4}{\rm Tr} \,
\FF^2+{\cal{O}}(\FF^4)\right]\left(h \, {\rm
det}^{1/2}\hat{G}-{\rm Pf}M\right) {1\over \cosh\theta}\, . \eeq
From now on we focus on the warped deformed conifold solution, although the calculation
below is the same for any background with imaginary self-dual flux.  Since the dilaton is
constant, $\Phi= \theta=0$, the leading terms of the $\FF$-expansion
are \bea S_{\rm DBI} &=&- \frac{1}{2} T_7{\rm Vol}_4\int_\Sigma
\left(\hat{J}\wedge \hat{J}-h^{-1}M\wedge M\right)+ \nonumber \\
&+& T_7\int_{\mathbb{R}^{1,3}} \d^4x\ {\rm Tr}\,
\FF^2\int_{\Sigma_4} \left(h\ {\rm det}^{1/2}\hat{G}\, \d^4\xi
-{1\over 2} {M\wedge M}\right)+\mathcal{O}(\FF^4)\, .
\label{equ:DBIterm} \eea To compute the Chern-Simons term, we note
that the only relevant RR flux in the warped deformed conifold case is \beq C_4=h^{-1}
{\rm Vol}_4\, , \eeq where ${\rm Vol}_4=\d x^{1}\wedge \dots \wedge
\d x^{4}$ is a volume form in the Minkowski space
 and
$\left. e^{\hat{B}+2\pi\alpha'F} \right|_{\Sigma_4}= \left.
e^{M+\FF} \right|_{\Sigma_4}= \left. {1\over 2}M\wedge M
\right|_{\Sigma_4}$. Hence,  \beq S_{\rm CS}= -\frac{1}{2} T_7{\rm
Vol}_4\int_{\Sigma_4} h^{-1} M\wedge M\, . \label{equ:CSterm} \eeq
Adding the DBI and CS actions, we have \bea \label{ten0}
S_{D_7} &=& -{1\over 2}T_7{\rm Vol}_4\int_{\Sigma_4} \hat{J}\wedge \hat{J}+ \nonumber \\
&+& {T_7}\int_{\mathbb{R}^{1,3}}\d^4x \, {\rm Tr}\, \FF^2
\int_{\Sigma_4} \left(h\ {\rm det}^{1/2}\hat{G}\, \d^4\xi -{1\over
2} {M\wedge M}\right)+\mathcal{O}(\FF^4)\ . \label{ten1} \eea
Finally, we are ready to add the D3-brane to our considerations.
As was discussed in \cite{BDKMMM}, to leading order the D3-brane
only changes the warp factor $h$, but does not affect the metric
$G$ or the RR or NSNS forms. The $h$ (and $M$) dependence of the leading
term of the DBI action (\ref{equ:DBIterm}) is precisely cancelled
by the CS term (\ref{equ:CSterm}). Since the leading term in
(\ref{ten0}) is independent of $h$, it is not affected by the
D3-brane.

For the solutions in question the K\"ahler form $J$ is closed. Therefore the leading term in (\ref{ten0}),
$\int_{\Sigma_4} \hat{J}\wedge\hat{J}$,
does not depend on $\Sigma_4$ itself but only on its homology class. This also means that the D7-brane tension does not
vary under small perturbations of $\Sigma_4$, as expected for the tension of supersymmetric branes.
This is only true when $\Sigma_4$ is compact. In the non-compact case this term diverges as the fourth power of a cut-off radius $r^4_{\rm UV}$.

In the subleading term in (\ref{ten1}) the warp factor $h$ and magnetic field $M$ do not mix.
The first term is proportional to $h$ and is responsible for the factor in the nonperturbative superpotential discussed in detail in \cite{BDKMMM}. For the validity of the considerations of \cite{BDKMMM} it is important that this term is independent of $M$.
The second term involves $M$, but is $h$-independent. To leading order it is not affected by
the backreaction of the D3-brane.

\addtocontents{toc}{\SkipTocEntry}
\subsection{Kuperstein embedding}

After deriving the general formula (\ref{ten0}) we are ready to
apply it to the particular case of the Kuperstein embedding. We
are interested in calculating the unperturbed value of the gauge
coupling \bea \label{equ:cc} \frac{1}{g^2_0} = 2 T_7 \int_{\Sigma_4}
\left(h\ {\rm det}^{1/2}\hat{G}\, \d^4\xi -{1\over 2} {M\wedge
M}\right) \eea before the D3-brane is added. In \S\ref{sec:4vol} we computed the warped four-cycle volume (the first
term in (\ref{equ:cc})), but ignored the effects of induced
magnetic fields (the second term in (\ref{equ:cc})). Here we
justify this approximation. For the Kuperstein embedding this is a
straightforward calculation, since in this case $F=0$ and hence
$M=\hat{B}$ \cite{Kuperstein}. The integral (\ref{equ:cc}) is
divergent in the non-compact case. To estimate its lower limit in
the compact case we evaluate its dependence on the cutoff
parameter $r_{\rm UV}$, defined by $h(r_{\rm UV}) \approx 1$. For
large $r_{\rm UV}$ we use the KT \cite{KT} approximation for the metric
(\ref{equ:KTh}) and the NSNS fields \cite{KT} and drop all
corrections in $1/r$. In this limit we find  \beq \label{equ:BB}
-{1\over 2}B\wedge B ={9\over 4}\log(r)^2\  \d \theta_1 \wedge
\sin\theta_1 \d \phi_1\wedge \d \theta_2 \wedge \sin\theta_2 \d \phi_2\,
. \eeq

In the large $r$ region, where the conifold can be approximated by a
cone over $T^{1,1}$, the four-cycle $\Sigma_4$ specified by the
Kuperstein embedding $z_1=\mu$ approaches a cone over an
$S^3$ given by \beq \label{equ:embedding}
\phi_2=\phi_1\, , \quad \theta_2=\theta_1\, . \eeq We choose the variables $\phi_1,\theta_1,\psi$ and $r$
to parameterize the internal part of the D7-brane world-volume
$\Sigma_4$ with $\theta_2,\phi_2$ given by (\ref{equ:embedding}).
It is clear from (\ref{equ:embedding}) that (\ref{equ:BB}) gives
no contribution in the large $r$ limit. Therefore its contribution
is finite when $r_{\rm UV} \rightarrow \infty$ and is small compared to
the divergent warped
four-cycle volume computed in (\ref{sec:4vol})
if $r_{\rm UV}$ is large enough.  Hence the contribution of the magnetic field in (\ref{equ:cc})
is negligible for the Kuperstein embedding.

\addtocontents{toc}{\SkipTocEntry}
\subsection{ACR embeddings}

For the case of the general ACR embedding (\ref{equ:ACR}), we note
that in the large $r$ region, where $\Sigma_4$ can be approximated
by a cone over a certain three-sphere (or a sum of several three-spheres),
(\ref{equ:BB}) vanishes as well. This follows immediately from the
equations specifying the base of the cone, \beq \theta_i=0,  \pi\, .\eeq
This is not sufficient to say that the second term in (\ref{equ:cc}) gives
no contribution, because the gauge field $F$ is not zero in this case.
Nevertheless one would not expect the contribution of the induced
gauge field to be dominant in the DBI action, as the value of the induced
gauge field is such that the DBI action is minimized. Therefore
the second term in (\ref{equ:cc}) can only change the coefficient in
front of the leading divergence in (\ref{equ:ACRSigma}) but will not
change the power of the divergence itself.


\newpage
\section{Symbols Used in the Paper}
\label{sec:symbols}

\begin{table*}[!h]
\caption{Definitions of symbols used in this paper}
\begin{center}
\begin{tabular}{cll}
\hline
Variable &  Description & Definition\\
\hline \hline
$z_i$ & complex conifold coordinates & $\sum_i z_i^2 = 0$, (\ref{equ:z1})--(\ref{equ:z4})\\
$w_i$ & complex conifold coordinates & $w_1 w_2 - w_3 w_4 =0$, (\ref{equ:w1})--(\ref{equ:w4})\\
$\varepsilon$ & conifold deformation parameter & $\sum_i z_i^2 = \varepsilon$, equation (\ref{equ:Cconstraint}) \\
$r$ & radial coordinate on the conifold & $r^3 = \sum_i |z_i|^2$ \\
$\hat r$ & radial coordinate on the conifold & $\hat r^2 = \frac{3}{2} r^2$, $\d s^2 = \d \hat r^2 + \dots $ \\
$\tilde r$ & radial coordinate on the conifold & $\tilde r = e^u \hat r$ \\
$e^u$ & breathing mode & equation (\ref{equ:back})\\
$g_{\alpha \bar \beta}$ & fiducial metric & \\
$\tilde g_{\alpha \bar \beta}$ & physical metric & $\tilde g_{\alpha \bar \beta} = e^{2u} g_{\alpha \bar \beta}$ \\
$k$ & K\"ahler potential & $g_{\alpha \bar \beta} = k_{, \alpha \bar \beta}$, $k = \frac{3}{2} (\sum_i z_i^2)^{2/3}$\\
$\phi_i$, $\theta_i$, $\psi$ & angular coordinates on $T^{1,1}$ & equations (\ref{equ:z1})--(\ref{equ:z4})\\
$h(r)$ & warp factor & equation (\ref{warpedansatz})\\
$r_{\rm UV}$ & radius at the UV end & $\ln r_{\rm UV}/\varepsilon^{2/3} = 2 \pi K/(3 g_s M) $\\
$\phi$ & inflaton field & $\phi^2 = T_3 \hat r^2$\\
$\varphi$ & canonical inflaton field &  Appendix \ref{sec:reduc}, $\varphi^2 \approx \frac{\sigma_\star(0)}{\sigma(\phi)} \phi^2$ \\
\hline
${\cal K}$ & K\"ahler potential & $\kappa^2 {\cal K} = - 3 \log U$ \\
$W$ & superpotential & \\
$U$ & argument of K\"ahler potential \cite{DeWG} & $U = \rho + \bar \rho - \gamma k$ \\
$\rho$ & complex K\"ahler modulus & \\
$\sigma$ & real part of $\rho$ & $2 \sigma = \rho + \bar \rho$\\
$\tau$ & imaginary part of $\rho$ & $2 i\, \tau = \rho - \bar \rho$\\
$\sigma_\star$ & stabilized volume modulus & $\left. \partial_\sigma V \right|_{\sigma_\star}  = 0$ \\
$\omega_\star$ & rescaled volume modulus& $\omega_\star \equiv a \sigma_\star$ \\
$W_0$ & GVW-flux superpotential & $W_0 = \int G \wedge \Omega$\\
$W_{\rm np}$ & non-perturbative superpotential & $W_{\rm np} = A(z_i) e^{-a \rho}$\\
$f(z_i)$ & embedding equation & $A(z_i)\propto (f(z_i))^{1/n}$\\
$g(z_i)$ &embedding equation & $g(z_i) = f(z_i)/f(0)$\\
$A_0$ & prefactor of $W_{\rm np}$ & $A_0 = A(z_i =0)$ \\
$V_F$ & F-term potential & equation (\ref{equ:VF}) \\
$V_D$ & D-term potential & $V_D = D(r) U^{-2}(\rho,r)$; equation (\ref{equ:Dterm}) \\
$D$ & scale of D-term energy & $D \equiv 2 h_0^{-1} T_3$ \\
\hline
\end{tabular}
\end{center}
\end{table*}

\begin{table*}[!h]
\caption{Definitions of symbols used in this paper (continued)}
\begin{center}
\begin{tabular}{cll}
\hline
Variable &  Description & Definition\\
\hline \hline
$\epsilon$, $\eta$ &slow-roll parameters & $\epsilon = \frac{1}{2} (V'/V)^2$, $\eta = V''/V$\\
$g_s$ & string coupling & \\
$T_3$ & D3-brane tension & $T_3^{-1} = (2\pi)^3 g_s (\alpha')^2$\\
$\mu$ & embedding parameter & $z_1= \mu$ \\
$n$ & \# of embedded D7's & \\
$a$ & parameter in $W_{\rm np} $ & $a = 2\pi/n$ \\
$r_\mu$ & minimal radius of D7 & $r_\mu^3 = 2 \mu^2$ \\
$\phi_\mu$ & rescaled $r_\mu$ & $\phi_\mu^2 = \frac{3}{2} T_3 r_\mu^2$ \\
$M$, $K$ & flux on the $A$ and $B$ cycle & \\
$N$ & five-form flux & $N \equiv M K$ \\
$L$ & AdS radius & equation (\ref{equ:L})\\
$M_P$ & 4d Planck mass & $M_P^2 = \frac{1}{\pi} (T_3)^2 V_6^w$ \\
$V_6^w$ & warped 6-volume & \\
$V_{\Sigma_4}^w$ & warped 4-cycle volume & Appendix \ref{sec:reduc}\\
$s$ & ratio of D- and F-term & equation (\ref{equ:sDef})  \\
$x$ & ratio of $r$ and $r_\mu$ &  $x = r/r_\mu$\\
$x_0$ & location of $\eta = 0$ & \\
$Q_\mu$ & ratio of $r_{\rm UV}$ and $r_\mu$ & $Q_\mu = r_{\rm UV}/r_\mu > 1$\\
$B_6$ & bulk contribution to $V_6^w$& $(V_6^w)_{\rm bulk} = B_6 (V_6^w)_{\rm throat}$\\
$B_4$ & bulk contribution to $V_{\Sigma_4}^w$& $(V_{\Sigma_4}^w)_{\rm bulk} = B_4 (V_{\Sigma_4}^w)_{\rm throat}$ \\
$\omega_F$ & K\"ahler modulus {\it before} uplifting & equation (\ref{equ:KKLT})  \\
$\omega_0$ & K\"ahler modulus {\it after} uplifting & $\omega_0 \approx \omega_F  + s /\omega_F $\\
$\Gamma$ & factor in ${\cal K}$ &  $\Gamma = 2 \sigma_0$ \\
$\tilde \Gamma$ & factor in ${\cal K}$ & $\tilde \Gamma \equiv \Gamma e^{4u} = U$ \\
$\gamma$ & prefactor in the K\"ahler potential & $\gamma = \frac{\Gamma}{6} \frac{T_3}{M_P^2} = \frac{\sigma_0}{3} \frac{T_3}{M_P^2}$\\
$c$ & factor in $V_F$& $c^{-1} = 4\pi \gamma r_\mu^2$\\
$c_{3/2}$ &factor in volume shift & equation (\ref{equ:cthreehalf})\\
$X$, $X+Y$ & eigenvalues of Hessian & Appendix \ref{sec:stability}\\
$P$ & degree of ACR embeddings & $\prod_i w_i^{p_i} = \mu^P$\\
$p_i$ & embedding parameter& $p_i \in \mathbb{Z}$ \\
$\Phi$ & collective coordinate for ACR & $\Phi^P \equiv \prod_i w_i^{p_i}$\\
\hline
&&\\
\end{tabular}
\end{center}
\end{table*}

\newpage
\begingroup\raggedright\endgroup

\end{document}